\documentclass[superscriptaddress,
bibnotes,amsmath,amssymb,aps,pra,floatfix,twocolumn]{revtex4-2}

\usepackage[margin=1in]{geometry}
\usepackage[english]{babel}
\usepackage{xcolor}
\usepackage{ucs}
\usepackage[utf8]{inputenc}
\usepackage{amsmath}
\usepackage{amssymb}
\usepackage{amsthm}
\usepackage{graphicx}
\usepackage{bm}
\usepackage{bbm}
\usepackage{braket}
\usepackage{empheq}
\usepackage{hyperref}
\usepackage{lipsum}
\usepackage{verbatim}
\usepackage[normalem]{ulem}

\newcommand{\gammam}{\gamma_-}
\newcommand{\gammaz}{\gamma_z}

\newcommand{\cave}[1]{\langle #1 \rangle}

\begin{document}

\title{Non-Gaussian generalized two-mode squeezing: applications to two-ensemble spin squeezing and beyond}

\date{\today}

\author{Mikhail Mamaev}
\email{mamaev@uchicago.edu}
\affiliation{Pritzker School of Molecular Engineering, University of Chicago, Chicago, Illinois, USA}

\author{Martin Koppenh\"{o}fer}
\affiliation{Fraunhofer Institute for Applied Solid State Physics IAF, Tullastr.~72, 79108 Freiburg, Germany}

\author{Andrew Pocklington}
\affiliation{Department of Physics, University of Chicago, Chicago, Illinois, USA }
\affiliation{Pritzker School of Molecular Engineering, University of Chicago, Chicago, Illinois, USA}

\author{Aashish A. Clerk}
\affiliation{Pritzker School of Molecular Engineering, University of Chicago, Chicago, Illinois, USA}

\begin{abstract}
Bosonic two-mode squeezed states are paradigmatic entangled Gaussian states that have wide utility in quantum information and metrology.  Here, we show that the basic structure of these states can be generalized to arbitrary bipartite quantum systems in a manner that allows simultaneous, Heisenberg-limited estimation of two independent parameters for finite-dimensional systems.  Further, we show that these general states can always be stabilized by a relatively simple Markovian dissipative process.  In the specific case where the two subsystems are ensembles of two-level atoms or spins, our generalized states define a notion of two-mode spin squeezing that is valid beyond the Gaussian limit and that enables true multi-parameter estimation.  We discuss how generalized Ramsey measurements allow one to reach the two-parameter quantum Cramer-Rao bound, and how the dissipative preparation scheme is compatible with current experiments.
\end{abstract}

\maketitle

\textit{Introduction.}
The central goal of quantum metrology is to harness many-body entanglement to improve the precision of a sensor beyond the bound set by the projection noise of non-entangled particles~\cite{giovannetti2006,degen2017,pezze2018squeezingReview,ma2011squeezingReview}. 
Spin-squeezed states~\cite{kitagawa1993squeezing} are particularly attractive given their relative robustness e.g., to undesired dissipation like dephasing. Given the experimental success in producing such states~\cite{leroux2010, gross2010, Riedel2010, Strobel2014, Muessel2015, hosten2016squeezing, bohnet2016squeezing, braverman2019, colombo2022heisenbergScaledSqz, hines2023squeezing, franke2023squeezing, eckner2023squeezing, bornet2023squeezing, robinson2024squeezing}, interest has naturally turned to 
entangling two distinct ensembles. Can this be done in a way that enables quantum-enhanced multi-parameter estimation~\cite{genoni2013multiParameter, humphreys2013multiParameter, zhang2014multiParameter, gao2014multiParameter, baumgratz2016multiParameter, gessner2020multiparameter, baamara2023quantum}? 
In the simpler case of harmonic oscillators, it is easy to generalize squeezing to two modes via a Gaussian two-mode squeezed state (TMSS), which enables simultaneous entanglement-enhanced estimation of two orthogonal parameters (see e.g.~\cite{fabre2020bosonicReview}).  A fundamental question is how to generalize this structure to more arbitrary subsystems, e.g. two spin ensembles for which only non-Gaussian states allow one to reach the ultimate sensitivity limits given the finite-dimensional Hilbert space.
Further, what are the fundamental sensitivity limits of these generalized states?  Is there a way to produce and utilize them using existing experimental resources?

\begin{figure}[t!]
\center
\includegraphics[width=1\columnwidth]{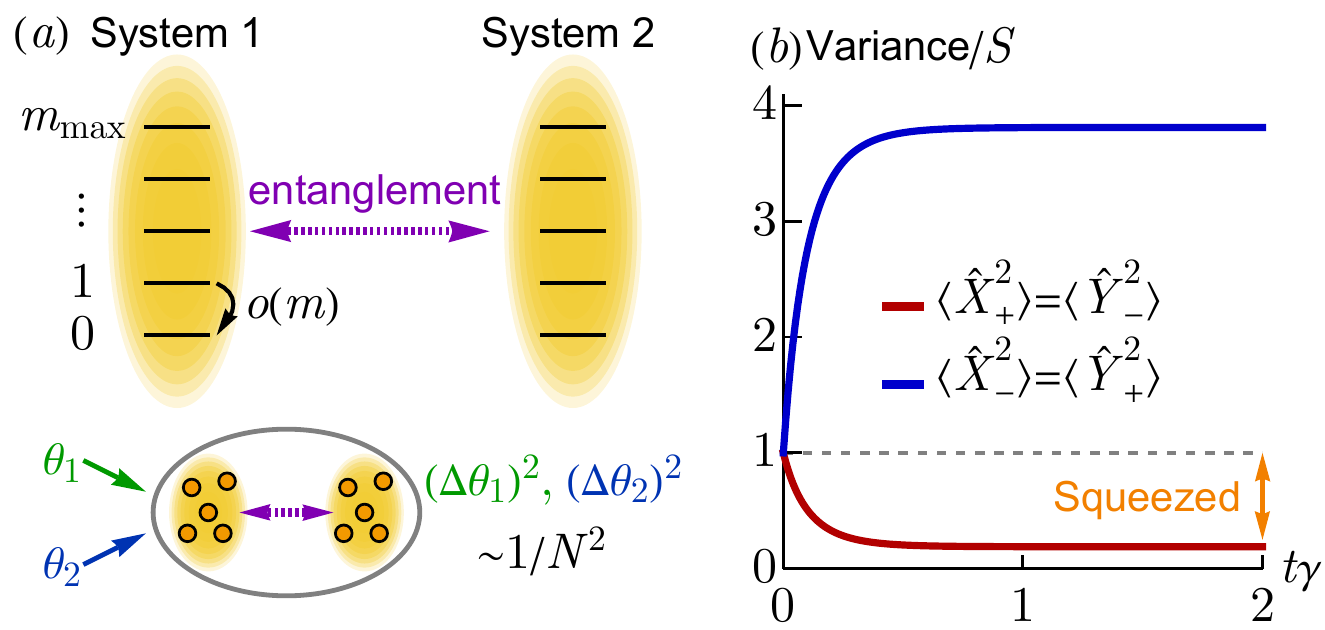}
\caption{
(a) Schematic of a generic bipartite system, with each subsystem $i=1,2$ having basis states $\ket{m}_i$
($m \in \{0, \dots, m_{\mathrm{max}} \}$), 
and a lowering operator with matrix elements $o(m)$.  We introduce an entangled state that generalizes bosonic two-mode squeezing and enables simultaneous estimation of two independent parameters $\theta_{1,2}$ with Heisenberg-limited sensitivity. (b) Time evolution of the variance for squeezed and anti-squeezed operators $\langle\hat{X}_{+}^2\rangle = \langle\hat{Y}_{-}^2\rangle$ and $\langle\hat{X}_{-}^2\rangle = \langle\hat{Y}_{+}^2\rangle$ respectively, under dissipative dynamics that stabilizes our generalized TMSS. The subsystems here are large spins each of size $S = 9/2$ (hence $m_{\mathrm{max}}$ = $2S$ = $9$), and set the squeezing parameter to $r= 0.75$.}
\label{fig_Schematic}
\end{figure}

In this work, we provide an answer to all of the above questions.  We introduce a generalization of a bosonic TMSS that can be defined for {\it any} (possibly finite-dimensional) bipartite system.  
We show that these states generically enable {\it simultaneous} estimation of two independent parameters $\theta_1,\theta_2$ with Heisenberg-limited scaling in finite-dimensional systems; for the application of two ensembles of $N/2$ two-level atoms each, we find a sensitivity of $\Delta \theta_{1,2} \sim \sqrt{3} / N$, which can be attained for both parameters simultaneously with a Ramsey-style measurement. Crucially, this sensitivity persists even in the presence of correlated noise sources such as common-mode noise, which would be impossible if measuring the two components of a bipartite system independently. Further, there is a generic Markovian dissipative evolution that can be used to prepare and stabilize these states (an evolution that closely mimics a two-mode squeezing photonic environment), and our dissipative scheme is a natural generalization of dissipative single-ensemble spin squeezing~\cite{agarwal1989squeezedDissipator, dalla2013dissipatorImplementation, groszkowski2022reservoir}. We discuss experimental implementation in cavity QED setups, and also analyze the robustness against typical imperfections (finding an advantageous scaling with the collective cooperativity $C$ that characterizes the setup).  
In contrast to prior work~\cite{gessner2018sensitivity, gessner2020multiparameter,Fadel2023}, here we explore the beyond-Gaussian limit of multi-mode squeezed metrology, including encoding the signals via non-commuting, nonlocal generators and the prospect of simultaneous non-commuting measurements.
An advantage of such a nonlocal encoding is the ability to convert noncommuting observables of one subensemble (e.g., $\hat{X}_1$ and $\hat{Y}_1$) into commuting nonlocal observables $\hat{X}_+$ and $\hat{Y}_-$ that allow one to reach the quantum Cramer-Rao bound (QCRB).

\textit{Multi-mode squeezed state.}
We consider a bipartite system composed of two identical, independent subsystems (labeled $i \in \{ 1,2 \}$).  Each subsystem has a local Hilbert space with basis states $\ket{m}_{i}$, $m \in \{0, 1, \dots, m_{\mathrm{max}} \}$ [see Fig.~\ref{fig_Schematic}(a)].  We introduce a one-parameter family of  entangled states of this system that directly mimics the construction of a bosonic TMSS:
\begin{equation}
    \label{eq_steadyState}
    \ket{\psi_G(r)} = \mathcal{N} \sum_{m=0}^{m_{\mathrm{max}}}\left[-\tanh(r)\right]^{m} 
    \ket{m}_{1}\otimes \ket{m}_{2},
\end{equation}
where 
$\mathcal{N} = (\cosh^2(r)[ 1- \tanh^{2m_{\mathrm{max}}+2}(r)])^{-1/2}$ and $r$ is a generalized squeezing strength.
We refer to this state as a generalized two-mode squeezed state (GTMSS).  
It reduces to a conventional TMSS in the case where each subsystem is a bosonic mode, $\ket{m}_{i}$ is a Fock state, and $m_{\mathrm{max}} = \infty$.  Much in the way a bosonic TMSS can be generated by driving two modes with squeezed light, we show below that our GTMSS can be prepared using a relatively simple form of engineered dissipation.

To harness our state for metrology, we will also mirror the bosonic construction, and will introduce generalized Hermitian ``quadrature" operators that will be used both to encode parameter dependence, and also for readout.  The first step is to chose a generalized lowering operator $\hat{O}_{i}$ for each subsystem:
\begin{equation}
    \begin{aligned}
    \hat{O}_{i} \ket{m}_{i} &= o(m) \ket{m-1}_{i}.
   \end{aligned}
\end{equation}
These lowering operators must annihilate the va\-cu\-um state $\ket{m=0}$, hence we require 
$o(0) = 0$.  We also assume without loss of generality $o(m) \in \mathbb{R}$.   As concrete examples, if the subsystems are bosonic modes and $\hat{O}_i$ are standard annihilation operators, then $o(m) = \sqrt{m}$ and $m_{\mathrm{max}}=\infty$.  
If the subsystems are spins of size $S$, $\ket{m}_i$ can be chosen as states of fixed angular momentum projection along the quantization axis, and $\hat{O}_i$ as standard collective spin lowering operators, with $o(m) = \sqrt{S(S+1) - (m-S)(m-S-1)}$ and $m_{\mathrm{max}} = 2S$.

We now introduce a set of Hermitian operators that mimic the construction of bosonic quadrature operators.  We first introduce local generalized quadratures via 
$\hat{X}_{i} = (\hat{O}_{i}^{\dagger} + \hat{O}_{i})/2$, $\hat{Y}_{i} =-i(\hat{O}_{i}^{\dagger} - \hat{O}_{i})/2$.  Generalized joint quadrature operators are then defined as
\begin{align}
    \label{eq_CollectiveQuads}
	\hat{X}_{\pm} &= \hat{X}_{1} \pm \hat{X}_{2}, &
	\hat{Y}_{\pm} &= \hat{Y}_{1} \pm \hat{Y}_{2}.
\end{align}
For bosonic modes, these are conventional two-mode collective quadratures whereas, for spins, they correspond to sums and differences of the $x$ and $y$ collective spin projection.
Note that in our general case (and in stark contrast to bosons), we cannot find pairs of these operators that commute with one another. However, we will in general be able to find pairs where the commutator vanishes when applied to our GTMSS. This will be crucial in enabling optimal sensing properties.

\textit{Quantum Fisher information.}
Consider the multi-parameter estimation problem where we apply the unitary $\hat{U} = \exp(-i \vec{\theta} \cdot \vec{W})$ to our state, with
$\vec{W} = \{\hat{X}_{+}, \hat{X}_{-}, \hat{Y}_{+}, 
\hat{Y}_{-}\}$, where $\vec{\theta} = (\theta_{X+},\theta_{X-},\theta_{Y+},\theta_{Y-})$ are the infinitesimal parameters of interest.  
As our state is pure, the quantum Fisher information matrix $\mathcal{Q}$ is 
proportional to the covariance matrix:  
$\mathcal{Q}_{ij} = 
    4 (\langle\hat{W}_i\hat{W}_j \rangle-\langle\hat{W}_i \rangle\langle\hat{W}_j \rangle)$.
Recall that if we only care about a single parameter $\theta_j$, the QCRB tells us that, asymptotically, the minimal achievable estimation error is $\Delta \theta_j = 1 / \sqrt{M \mathcal{Q}_{jj}}$, where $M$ is the number of repetitions of the experiment.  

For the state $\ket{\psi_{G}(r)}$, we find simply:
\begin{equation}
\label{eq_QFIM}
\begin{aligned}
\mathcal{Q} &= 
    \mathcal{N}_Q \left(\begin{array}{cccc} 
e^{-2r} & 0 & 0 & 0\\
0 & e^{2r} & 0 & 0\\
0 & 0 & e^{2r} & 0 \\
0 & 0 & 0 & e^{-2r}
\end{array}\right),\\
\mathcal{N}_Q &= \frac{2\sum_{m=0}^{m_{\mathrm{max}}}\tanh^{2m}(r) o^2(m)}{\cosh^2(r) \sinh^2(r)\left[1- \tanh^{2m_{\mathrm{max}}+2}(r)\right]}.
\end{aligned}
\end{equation}
Remarkably, the covariance matrix directly mirrors a bosonic TMSS:  the generalized quadratures 
$\hat{X}_{+}, \hat{Y}_{-}$ have reduced fluctuations (i.e.~they are ``squeezed"), whereas 
$\hat{X}_{-}, \hat{Y}_{+}$ are ``anti-squeezed".  
We plot the corresponding variances in Fig.~\ref{fig_Schematic}(b). Note that all system-specific details enter only through the overall prefactor $\mathcal{N}_Q$.  Further, from the QCRB, the optimal sensitivity for the parameters $\theta_{X-}, \theta_{Y+}$  is always $e^{4r}$
better than that of $\theta_{X+}, \theta_{Y-}$

Our state also enables one to simultaneously achieve the QCRB on both $\theta_{X-}, \theta_{Y+}$ through a simple pair of measurements.  
We establish this by computing the symmetric logarithmic derivative (SLD) operators, which provide a means to find an optimal measurement scheme~\cite{liu2020multiparReview, szczykulska2016multiparReview, giovannetti2011multiparReview}. The SLD operators for our state are (see \cite{Supplement})
\begin{align}
\hat{L}_{\hat{X}_{\pm}} &= 2 e^{\mp 2r} \hat{Y}_{{\pm}}, &
\hat{L}_{\hat{Y}_{\pm}} &= -2 e^{\pm 2r} \hat{X}_{\pm}.
\label{eq_SLDs}
\end{align}
These SLDs are simply proportional to our generalized quadrature operators.
One finds that 
the commutator of the two sensitive operators is 
$[\hat{X}_{-}, \hat{Y}_{+}] = i (\hat{Z}_{1}-\hat{Z}_{2})$ with $\hat{Z}_{i}=(\hat{O}_{i}^{\dagger}\hat{O}_{i} - \hat{O}_{i}\hat{O}_{i}^{\dagger})/2$.  While the commutator is non-zero, it vanishes when applied to the state $\ket{\psi_{G}(r)}$ due to its symmetric ``paired" form.
A general result then implies that it is possible to simultaneously achieve the QCRB for both $\theta_{X-}, \theta_{Y+}$
by finding new SLDs that commute, and measuring these operators~\cite{liu2020multiparReview,matsumoto2002,Ragy2016,Pezze2017}.  
We provide a simple example for very small spin ensembles in \cite{Supplement}.
In general, however, the new SLDs correspond to complicated high-weight operators that can be challenging to measure.
In our case, as the commutator vanishes when acting on $\ket{\psi_{G}(r)}$ (and not just in expectation), we have a stronger result that allows us to pursue a different strategy: As shown in \cite{Supplement}, simple sequential measurements of $\hat{X}_+$ and $\hat{Y}_-$ (i.e.~generalized Ramsey measurements corresponding to the original SLDs) allows one to {\it simultaneously} extract the two parameters $\theta_{X-}$ and $\theta_{Y+}$ with an estimation error that exhibits Heisenberg limited $1/N$ scaling (with a prefactor that is only a factor of $\simeq 3$ larger than the QCRB bound derived below).  

We now finally ask about the ultimate sensitivity that our state allows for the estimation of the two parameters $\theta_{X-}, \theta_{Y+}$.  To do this, we consider the large elements of the QFIM in the infinite squeezing limit, $\mathcal{Q}_{\rm max} \equiv \text{lim}_{r\to\infty}\mathcal{N}_Q e^{2r}$
\footnote{Note that one can construct examples where the QFI does not monotonically increase with $r$.  However, for most cases of interest, including the crucial case of two spin ensembles, we do have a monotonic increase.}. We find:
\begin{equation}
\label{eq_asymptoticQFI}
\mathcal{Q}_{\rm max}= \frac{8}{1+m_{\mathrm{max}}} \sum_{m=0}^{m_{\mathrm{max}}}o^2(m) = \frac{8 || \hat{O} ||^2}{1+m_{\mathrm{max}}} ,
\end{equation}
where the norm here is the Frobenius norm.  We thus have a very general kind of Heisenberg limit on the maximal QFI (which not surprisingly depends on the overall scale of our generalized lowering operator, which we have not fixed).  In the concrete case where each subsystem corresponds to the collective manifold of $N/2$ two-level atoms, and $\hat{O}$ is a standard collective angular momentum lowering operator for one of the ensembles, one finds:
\begin{equation}
    \mathcal{Q}_{\rm max} = \frac{4}{3}m_{\mathrm{max}}(m_{\mathrm{max}}+2) \approx \frac{N^2}{3}.
\end{equation}
The QFI for both optimal parameters yields a Heisenberg-limit like scaling ($\propto N^2$), with a corresponding estimation error scaling like $\sqrt{3}/N$.  Even in more general settings, we can argue this scaling is generic, as long as one normalizes the operator $\hat{O}$ so that it also grows with system size the same way as a collective spin lowering operator, i.e. $||\hat{O}||^2 \sim N^3$.

\textit{Metrological utility for spin ensembles.}
We now consider the relevant case where each subsystem is a collective spin of size $S = N/4$, and $\hat{O}$ is a collective angular momentum lowering operator.  Our general SLD calculation tells us that, e.g., to optimally estimate $\theta_{Y+}$ we should measure the squeezed variable $\hat{X}_+$, which is just the sum of the $x$ collective spin components of each ensemble. The estimation error for such a measurement reduces to a signal-to-noise ratio, which can be expressed in terms of a generalized version of the Wineland squeezing parameter.  We have $(\Delta \theta_{Y+})^2 = \xi^2 / N$, with  
$\xi^2 = N \langle\hat{X}_{+}^2\rangle/|\langle \hat{Z}_{1}\rangle + \langle \hat{Z}_{2}\rangle|^2$.  Defining $f_{\pm} =\tanh^{4S+2}(r)\pm 1$, we find
\begin{equation}
\label{eq_squeezing}
\xi^2 =-\frac{e^{-2 r} S  f_{-} \left[(2 S+1) f_{+}+\cosh (2 r) f_{-}\right]}{2 \left[\left(\cosh ^2(r)+S\right)f_{+}-\cosh (2 r)\right]^2}.
\end{equation}
One obtains an analogous result for the estimation of $\theta_{X-}$ (which involves measuring $\hat{Y}_-$).

\begin{figure}
\center
\includegraphics[width=0.95\columnwidth]{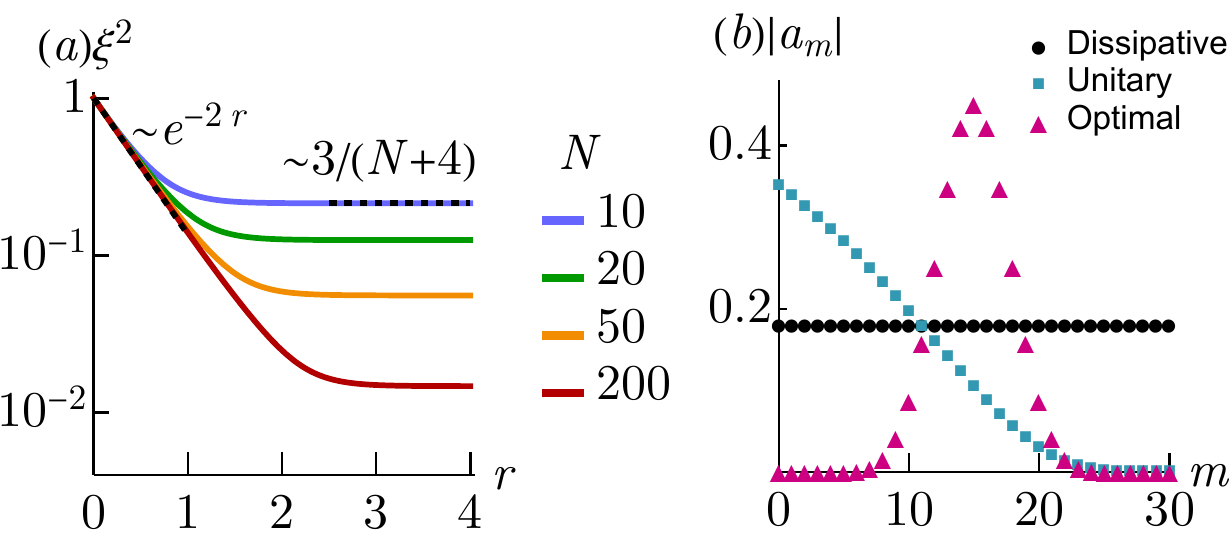}
\caption{(a) Wineland squeezing parameter of the squeezed operator $\hat{X}_{+}$ for the steady-state $\ket{\psi_G(r)}$ from Eq.~\eqref{eq_squeezing}, assuming both ensembles are spins of size $S=N/4$. (b) Coefficients $a_m$ of the symmetric state $\sum_{m=0}^{m_{\mathrm{max}}} a_{m} \ket{m,m}$ with fixed $m_{\mathrm{max}}=30$ for the state $\ket{\psi_G(r)}$ with $r \to \infty$, a squeezed state $\ket{\psi_{\mathrm{2M2A}}}$ generated via a unitary two-mode two-axis twisting Hamiltonian (see \cite{Supplement}), and an optimal state $\ket{\psi_{\mathrm{opt}}}$ found by maximizing the QFI over all possible $a_m$ (see \cite{Supplement}). The latter state coefficients exactly match binomial coefficients $a_m = (-1)^{m} m_{\mathrm{max}}! / [m! (m_{\mathrm{max}}-m)!]$.
}
\label{fig_Metrology}
\end{figure}

The Wineland parameter is plotted in Fig.~\ref{fig_Metrology}(a). For infinite spin size $S\to \infty$, we have $\xi^2 = e^{-2r} + \mathcal{O}(1/S)$, which matches the conventional bosonic result (as expected based on a linearized Holstein-Primakoff transformation). The opposite limit of infinite squeezing $r \to \infty$ yields,
\begin{equation}
\label{eq_winelandSScaling}
    \xi^2= \frac{3}{4(S+1)}+\frac{3+16S+16S^2}{20(S+1)}e^{-4r} + \mathcal{O}\left(e^{-6r}\right).
\end{equation}
If we take only the zeroth-order term, we again find a Heisenberg-scaling $\xi^2 \sim 3/N$. The absolute variance of measurements with a Ramsey-style experiment is $\xi^2/N$, which exactly coincides with the QFI, $\xi^2/N = 1/(\mathcal{N}_{Q}e^{2r})$ for any $N$ (not just asymptotically). Hence a generalized version of Ramsey-style metrology is an optimal measurement scheme to saturate the quantum Cramer-Rao bound for the entangled state~\eqref{eq_steadyState}.

\textit{Optimal paired states}
Before discussing how to prepare our GTMSS, we ask whether there exist paired states of the form $\sum_{m} a_m \ket{m,m}$ that (for the specific case of two spin ensembles) could be even more metrologically useful.  One candidate state of this form is produced by unitary evolution of a product, polarized state under a two-mode generalization of the two-axis twisting Hamiltonian.  The Hamiltonian here is  $\hat{H}_{\mathrm{2a2m}}= \hat{X}_{1} \hat{X}_{2}-\hat{Y}_{1}\hat{Y}_2$~\cite{kitzinger2020twoAxisTwoSpin}.  At an optimal evolution time, this dynamics generates squeezing of two collective spin variables that exhibit Heisenberg scaling, but with a slightly worse prefactor than our GTMSS: one finds a Wineland parameter $\xi^2 \sim 5.1/N$ (see \cite{Supplement}).
Figure~\ref{fig_Metrology}(b) compares the coefficients $a_m$ of the maximally squeezed state generated by this Hamiltonian to our steady-state. One can also employ a more conventional one-axis twisting Hamiltonian $\sim \hat{X}_{1} \hat{X}_{2}$ in the two-mode context, which also yields a state equally sensitive to two quadratures with a worse scaling $\sim 1/N^{2/3}$ (see \cite{Supplement}).

For a fixed $m_{\mathrm{max}}$ we can also optimize the coefficents $a_m$ to find the maximal QFI for two equally-sensitive operators (see \cite{Supplement}).
For spin-$S$ ensembles, the resulting optimal QFI is $N(\frac{N}{2}+1) \sim N^2/2$, which also happens to be the optimal bound for two-mode measurements of the two spin ensembles (see \cite{Supplement}). 
The optimal coefficients are also plotted in Fig.~\ref{fig_Metrology}(b). For spin ensembles these coefficients are staggered binomials $a_m = (-1)^{m} (2S)!/[m! (2S-m)!]$. However, the optimal measurement protocol to saturate the QCRB for such a state is less clear. Alternatively one may consider two-component superpositions such as GHZ states which also lead to a two-component simultaneous QFI scaling $\sim N^2/2$ but are rather difficult to generate (see \cite{Supplement}).

We stress that the TMSS in Eq.~\eqref{eq_steadyState} provides a significant advantage for simultaneous two-parameter estimation over entangled states prepared in each subsystem independently. Suppose one seeks to measure two correlated parameters coupling to $\hat{X}_{-}$ and $\hat{Y}_{+}$. There can be common-mode noise such as a random magnetic field $\sim \hat{X}_{+}$ affecting both ensembles equally during phase accumulation. If one makes independent measurements of each sub-ensemble without entangling the two, one would have to take a differential signal to suppress this noise, which inhibits measurement of the second sum parameter $\hat{Y}_{+}$. In contrast, our scheme still allows Heisenberg scaling for both parameters; we can simply choose the noisy quadrature to be anti-squeezed. This approach also works for sensing two differential fields $\hat{X}_{-}$ and $\hat{Y}_{-}$, as the choice of which two quadratures are squeezed is freely tuned via local rotations. In that case, our state is robust even in the presence of common-mode noise in both $\hat{X}_{+}$ and $\hat{Y}_{+}$. Even in the special case of noise that affects both correlated and single-ensemble measurements equally, our scheme still offers an improvement over independent sub-ensembles (see \cite{Supplement}).

\textit{Dissipative stabilization.}
We now discuss how to prepare and stabilize our GTMSS $\ket{\psi_G(r)}$ in 
Eq.~(\ref{eq_steadyState}) in the most general case of two arbitrary subsystems and arbitrary generalized lowering operator. This can be achieved by coupling the two subsystems to a common engineered dissipative reservoir, such that the resulting dynamics is described by the quantum master equation: 
\begin{equation}
\label{eq_masterEquation}
\begin{aligned}
\frac{d}{dt}\rho &= \gamma \sum_{j=1,2} \mathcal{D}[\cosh(r)\hat{O}_{j} + \sinh(r) \hat{O}_{\bar{j}}^{\dagger} ]\rho
\end{aligned}
\end{equation}
Here, $\mathcal{D}[\hat{O}]\rho = \hat{O}\rho \hat{O}^{\dagger} - \frac{1}{2}\hat{O}^{\dagger}\hat{O}\rho - \frac{1}{2}\rho \hat{O}^{\dagger}\hat{O}$, $\gamma$ is the dissipation rate, and $(\bar{1},\bar{2})=(2,1)$.  Remarkably, the pure state $\ket{\psi_G(r)}$ is always the unique steady state of this master equation, independent of additional details (i.e.~the specific choice of $m_{\rm max}$ and the coefficients $o(m)$ that define the generalized lowering operator); see \cite{Supplement}.
This follows from a special symmetry of the jump operators in this master equation under modular conjugation by the square root of the reduced steady-state density matrix of each subsystem (see \cite{Supplement})~\cite{pocklington2024solutionStructure}.

\begin{figure}[t]
	\centering
    \includegraphics[width=0.48\textwidth]{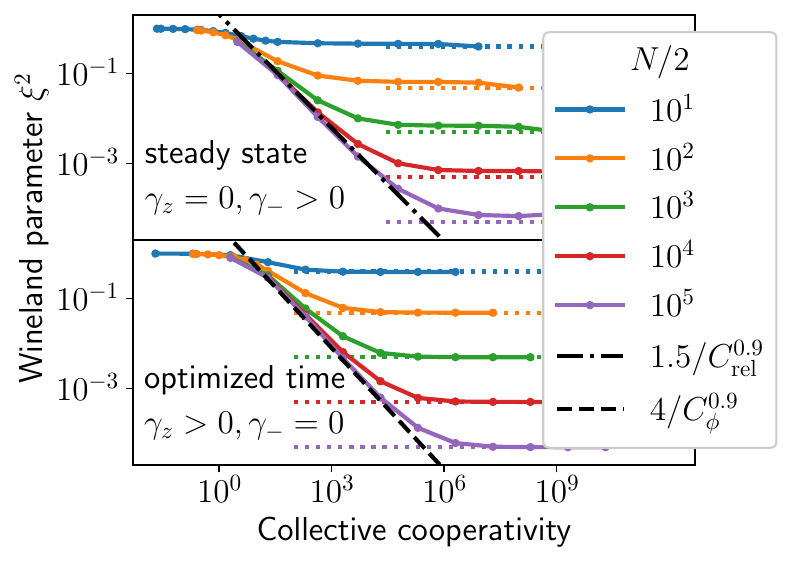}
	\caption{
		Impact of local dissipation competing with the collective decay in Eq.~\eqref{eq_masterEquation}.
        (a) Steady-state Wineland parameter $\lim_{t \to \infty} \xi^2$ in the presence of single-spin relaxation at a rate $\gammam$, optimized over the squeezing parameter $r$. 
        (b) Minimum Wineland parameter $\min_t \xi^2$ in the presence of single-spin dephasing at a rate $\gammaz$, optimized over the squeezing parameter $r$.
        In both cases, spin squeezing is observed if the collective cooperativities $C_{\mathrm{rel},\phi} = N \gamma/\gamma_{-,z}$ exceed unity. 
        Above this threshold, the Wineland parameter decreases $\propto C_{\mathrm{rel},\phi}^{-0.9}$ (dash-dotted and dashed black curves).
        The optimal Wineland parameter $\xi^2 \propto 1/(N+4)$ in the limit of infinite cooperativity is indicated by the dotted lines.
	}
	\label{fig_Local_Dissipation}
\end{figure}

Turning to the specific case where each subsystem is a spin ensemble, the above quantum master equation generalizes the dissipative preparation of single-ensemble spin-squeezed states to the two-ensemble case \cite{agarwal1989squeezedDissipator,dalla2013dissipatorImplementation,groszkowski2022reservoir}. 
Despite this superficial similarity, there are crucial differences.  The pure steady state in the single-ensemble case has in general no simple relation to a single-mode bosonic squeezed state, whereas in the two-ensemble case, our steady state in Eq.~(\ref{eq_steadyState}) has a form that directly mirrors a bosonic TMSS for arbitrary squeezing strength $r$. Note that Eq.~(\ref{eq_masterEquation}) for two spin ensembles was also explored in Ref.~\cite{parkins2006unconditionalTMS} in the context of entanglement generalization. Unlike our work, this previous work did not go beyond the weak squeezing, Gaussian limit, nor did it study quantum metrology properties. 
Ref~\cite{sundar2024squeezing} also studied an alternate dissipative approach for squeezing two ensembles, finding a Wineland parameter scaling of $\sim N^{-1/4}$.

One way to engineer Eq.~\eqref{eq_masterEquation} for a single ensemble is to use atoms with two spin states $\ket{\uparrow}$, $\ket{\downarrow}$ and two electronic states $\ket{g}$, $\ket{e}$ inside a lossy cavity, driven by a pair of light fields, as discussed in Ref.~\cite{dalla2013dissipatorImplementation}. The laser driving and cavity decay induce two indistinguishable effective spin-raising and lowering processes, leading to Lindblad-style dissipators of the form $\mathcal{D}[\cosh(r) \hat{O}^{+} + \sinh(r) \hat{O}^{-}]$. This idea can be extended to a two-mode scheme by considering two spatially separated ensembles with four effective raising and lowering processes~\cite{parkins2006unconditionalTMS, muschik2011experimentalImplementation, zheng2010experimentalImplementation}. A field gradient can generate a differential energy shift, 
that allows one to resolve the raising and lowering processes that contribute to the two dissipators in Eq.~\eqref{eq_masterEquation}.
A sample level scheme is discussed in \cite{Supplement}. 
One can also employ atoms with a larger internal structure, where different ensembles are encoded by different subsets of atomic levels~\cite{cooper2024multimodeExpt, sundar2024squeezing}. There are also studies of ensembles of larger-spin particles, such as spin-1 nematic condensates~\cite{hamley2012spinNematic, kunkel2019spinNematic, cao2023spinNematic} which enable quantum-enhanced multi-parameter encodings. Unlike these systems, where the two modes are encoded in two internal modes of the same spin ensemble, our dissipative approach can also generate two-mode squeezed states between spatially separated subensembles. Note that the timescale for dissipative stabilization is favorable, and remains almost independent of system size for a suitable choice of parameters (see \cite{Supplement} for details). Also note that, while the steady-state solution of Eq.~\eqref{eq_steadyState} assumes identical ensembles, properties such as squeezing persist for small deviations such as different atom numbers between the two ensembles; a numerical benchmark is provided in~\cite{Supplement}.

\textit{Robustness against local dissipation.}
The engineered dissipators in Eq.~\eqref{eq_masterEquation} typically compete with additional undesired dissipative processes.
We therefore analyze the robustness of the two-mode squeezing against their most prototypical forms -- single-spin relaxation at a rate $\gammam$ and single-spin dephasing at a rate $\gammaz$, modelled by adding dissipators $\gammam\sum_{i=1}^2 \sum_{j=1}^{N/2}\mathcal{D}[\hat{\sigma}^-_{i,j}] \rho$ and $\gammaz\sum_{i=1}^2 \sum_{j=1}^{N/2} \mathcal{D}[\hat{\sigma}^z_{i,j}]\rho$ to Eq.~\eqref{eq_masterEquation}.
Here, the index $i$ ($j$) denotes the two ensembles (different spins within an ensemble). 
To study large numbers of spins, we use a second-order mean-field-theory approach \cite{kubo1962generalizedcumulant,zens2019criticalphenomena,groszkowski2022reservoir}, where the full quantum dynamics is approximated by a set of coupled differential equations (see \cite{Supplement}). 
We compared the mean-field-theory solutions with the analytical results derived above and found excellent agreement. 
As shown in Fig.~\ref{fig_Local_Dissipation}, dissipative two-mode spin squeezing is present if the collective cooperativities $C_{\mathrm{rel},\phi} = N \gamma/\gamma_{-,z}$ exceed unity, and the Wineland parameter decreases $\propto 1/C_{\mathrm{rel},\phi}^{0.9}$ close to that threshold. 
Notably, this is a better scaling than what has been found for single-mode dissipative squeezing~\cite{groszkowski2022reservoir} and unitary single-mode spin-squeezing protocols~\cite{bennett2013,lewisswan2018}.

\textit{Conclusions and outlook.}
We have introduced a surprisingly simple and direct generalization of bosonic two-mode squeezed states to a general bipartite system, demonstrating that this state allows simultaneous Heisenberg-limited estimation of two independent parameters.  Further, we outlined a general dissipative dynamics that stabilizes such states.  For the specific case of two spin ensembles, our approach generalizes single-mode spin squeezing, and is compatible with several current experimental platforms.  Our structure is also applicable beyond spin ensembles; see \cite{Supplement} for a discussion of examples such as non-Gaussian two-photon two-mode squeezed states in bosonic systems~\cite{leghtas2015yaleTwoPhoton, reglade2024twoPhoton}.  
Our work opens up several new directions for research:
one could generalize the combination of entanglement with the dynamics of 
quantum-mechanics free subsystems \cite{TsangQMFS2012,Polzik2015,PolzikNature2017,Sillanpaa2021}
for enhanced sensing, as has been discussed for bosonic systems \cite{Didier2015}.  It would also be interesting to study whether our ideas could be usefully extended to the case for more than two ensembles.  
Finally, Heisenberg-limited spin-squeezing protocols require single-atom resolution to leverage their fully enhanced sensitivity. For single-mode squeezing, amplification protocols have been proposed to overcome this limitation~\cite{koppenhofer2023amplifier, davis2016twistUntwist, hosten2016magnification} and it would be interesting to generalize these concepts to multi-mode squeezing.

\begin{acknowledgments}
This work was primarily supported by the DOE Q-NEXT Center (Grant No. DOE 1F- 60579).  AC acknowledges support from the Simons Foundation (Grant No. 669487, A. C.). 
\end{acknowledgments}

\bibliography{DisBiblio}

\begin{thebibliography}{65}%
\makeatletter
\providecommand \@ifxundefined [1]{%
 \@ifx{#1\undefined}
}%
\providecommand \@ifnum [1]{%
 \ifnum #1\expandafter \@firstoftwo
 \else \expandafter \@secondoftwo
 \fi
}%
\providecommand \@ifx [1]{%
 \ifx #1\expandafter \@firstoftwo
 \else \expandafter \@secondoftwo
 \fi
}%
\providecommand \natexlab [1]{#1}%
\providecommand \enquote  [1]{``#1''}%
\providecommand \bibnamefont  [1]{#1}%
\providecommand \bibfnamefont [1]{#1}%
\providecommand \citenamefont [1]{#1}%
\providecommand \href@noop [0]{\@secondoftwo}%
\providecommand \href [0]{\begingroup \@sanitize@url \@href}%
\providecommand \@href[1]{\@@startlink{#1}\@@href}%
\providecommand \@@href[1]{\endgroup#1\@@endlink}%
\providecommand \@sanitize@url [0]{\catcode `\\12\catcode `\$12\catcode
  `\&12\catcode `\#12\catcode `\^12\catcode `\_12\catcode `\%12\relax}%
\providecommand \@@startlink[1]{}%
\providecommand \@@endlink[0]{}%
\providecommand \url  [0]{\begingroup\@sanitize@url \@url }%
\providecommand \@url [1]{\endgroup\@href {#1}{\urlprefix }}%
\providecommand \urlprefix  [0]{URL }%
\providecommand \Eprint [0]{\href }%
\providecommand \doibase [0]{https://doi.org/}%
\providecommand \selectlanguage [0]{\@gobble}%
\providecommand \bibinfo  [0]{\@secondoftwo}%
\providecommand \bibfield  [0]{\@secondoftwo}%
\providecommand \translation [1]{[#1]}%
\providecommand \BibitemOpen [0]{}%
\providecommand \bibitemStop [0]{}%
\providecommand \bibitemNoStop [0]{.\EOS\space}%
\providecommand \EOS [0]{\spacefactor3000\relax}%
\providecommand \BibitemShut  [1]{\csname bibitem#1\endcsname}%
\let\auto@bib@innerbib\@empty
\bibitem [{\citenamefont {Giovannetti}\ \emph {et~al.}(2006)\citenamefont
  {Giovannetti}, \citenamefont {Lloyd},\ and\ \citenamefont
  {Maccone}}]{giovannetti2006}%
  \BibitemOpen
  \bibfield  {author} {\bibinfo {author} {\bibfnamefont {V.}~\bibnamefont
  {Giovannetti}}, \bibinfo {author} {\bibfnamefont {S.}~\bibnamefont {Lloyd}},\
  and\ \bibinfo {author} {\bibfnamefont {L.}~\bibnamefont {Maccone}},\
  }\bibfield  {title} {\bibinfo {title} {Quantum metrology},\ }\href
  {https://doi.org/10.1103/PhysRevLett.96.010401} {\bibfield  {journal}
  {\bibinfo  {journal} {Phys. Rev. Lett.}\ }\textbf {\bibinfo {volume} {96}},\
  \bibinfo {pages} {010401} (\bibinfo {year} {2006})}\BibitemShut {NoStop}%
\bibitem [{\citenamefont {Degen}\ \emph {et~al.}(2017)\citenamefont {Degen},
  \citenamefont {Reinhard},\ and\ \citenamefont {Cappellaro}}]{degen2017}%
  \BibitemOpen
  \bibfield  {author} {\bibinfo {author} {\bibfnamefont {C.~L.}\ \bibnamefont
  {Degen}}, \bibinfo {author} {\bibfnamefont {F.}~\bibnamefont {Reinhard}},\
  and\ \bibinfo {author} {\bibfnamefont {P.}~\bibnamefont {Cappellaro}},\
  }\bibfield  {title} {\bibinfo {title} {Quantum sensing},\ }\href
  {https://doi.org/10.1103/RevModPhys.89.035002} {\bibfield  {journal}
  {\bibinfo  {journal} {Rev. Mod. Phys.}\ }\textbf {\bibinfo {volume} {89}},\
  \bibinfo {pages} {035002} (\bibinfo {year} {2017})}\BibitemShut {NoStop}%
\bibitem [{\citenamefont {Pezze}\ \emph {et~al.}(2018)\citenamefont {Pezze},
  \citenamefont {Smerzi}, \citenamefont {Oberthaler}, \citenamefont {Schmied},\
  and\ \citenamefont {Treutlein}}]{pezze2018squeezingReview}%
  \BibitemOpen
  \bibfield  {author} {\bibinfo {author} {\bibfnamefont {L.}~\bibnamefont
  {Pezze}}, \bibinfo {author} {\bibfnamefont {A.}~\bibnamefont {Smerzi}},
  \bibinfo {author} {\bibfnamefont {M.~K.}\ \bibnamefont {Oberthaler}},
  \bibinfo {author} {\bibfnamefont {R.}~\bibnamefont {Schmied}},\ and\ \bibinfo
  {author} {\bibfnamefont {P.}~\bibnamefont {Treutlein}},\ }\bibfield  {title}
  {\bibinfo {title} {Quantum metrology with nonclassical states of atomic
  ensembles},\ }\href {https://doi.org/10.1103/RevModPhys.90.035005} {\bibfield
   {journal} {\bibinfo  {journal} {Rev. Mod. Phys.}\ }\textbf {\bibinfo
  {volume} {90}},\ \bibinfo {pages} {035005} (\bibinfo {year}
  {2018})}\BibitemShut {NoStop}%
\bibitem [{\citenamefont {Ma}\ \emph {et~al.}(2011)\citenamefont {Ma},
  \citenamefont {Wang}, \citenamefont {Sun},\ and\ \citenamefont
  {Nori}}]{ma2011squeezingReview}%
  \BibitemOpen
  \bibfield  {author} {\bibinfo {author} {\bibfnamefont {J.}~\bibnamefont
  {Ma}}, \bibinfo {author} {\bibfnamefont {X.}~\bibnamefont {Wang}}, \bibinfo
  {author} {\bibfnamefont {C.-P.}\ \bibnamefont {Sun}},\ and\ \bibinfo {author}
  {\bibfnamefont {F.}~\bibnamefont {Nori}},\ }\bibfield  {title} {\bibinfo
  {title} {Quantum spin squeezing},\ }\href
  {https://doi.org/10.1016/j.physrep.2011.08.003} {\bibfield  {journal}
  {\bibinfo  {journal} {Physics Reports}\ }\textbf {\bibinfo {volume} {509}},\
  \bibinfo {pages} {89} (\bibinfo {year} {2011})}\BibitemShut {NoStop}%
\bibitem [{\citenamefont {Kitagawa}\ and\ \citenamefont
  {Ueda}(1993)}]{kitagawa1993squeezing}%
  \BibitemOpen
  \bibfield  {author} {\bibinfo {author} {\bibfnamefont {M.}~\bibnamefont
  {Kitagawa}}\ and\ \bibinfo {author} {\bibfnamefont {M.}~\bibnamefont
  {Ueda}},\ }\bibfield  {title} {\bibinfo {title} {Squeezed spin states},\
  }\href {https://doi.org/10.1103/PhysRevA.47.5138} {\bibfield  {journal}
  {\bibinfo  {journal} {Phys. Rev. A}\ }\textbf {\bibinfo {volume} {47}},\
  \bibinfo {pages} {5138} (\bibinfo {year} {1993})}\BibitemShut {NoStop}%
\bibitem [{\citenamefont {Leroux}\ \emph {et~al.}(2010)\citenamefont {Leroux},
  \citenamefont {Schleier-Smith},\ and\ \citenamefont
  {Vuleti\'{c}}}]{leroux2010}%
  \BibitemOpen
  \bibfield  {author} {\bibinfo {author} {\bibfnamefont {I.~D.}\ \bibnamefont
  {Leroux}}, \bibinfo {author} {\bibfnamefont {M.~H.}\ \bibnamefont
  {Schleier-Smith}},\ and\ \bibinfo {author} {\bibfnamefont {V.}~\bibnamefont
  {Vuleti\'{c}}},\ }\bibfield  {title} {\bibinfo {title} {Implementation of
  cavity squeezing of a collective atomic spin},\ }\href
  {https://doi.org/10.1103/PhysRevLett.104.073602} {\bibfield  {journal}
  {\bibinfo  {journal} {Phys. Rev. Lett.}\ }\textbf {\bibinfo {volume} {104}},\
  \bibinfo {pages} {073602} (\bibinfo {year} {2010})}\BibitemShut {NoStop}%
\bibitem [{\citenamefont {Gross}\ \emph {et~al.}(2010)\citenamefont {Gross},
  \citenamefont {Zibold}, \citenamefont {Nicklas}, \citenamefont {Est{\`e}ve},\
  and\ \citenamefont {Oberthaler}}]{gross2010}%
  \BibitemOpen
  \bibfield  {author} {\bibinfo {author} {\bibfnamefont {C.}~\bibnamefont
  {Gross}}, \bibinfo {author} {\bibfnamefont {T.}~\bibnamefont {Zibold}},
  \bibinfo {author} {\bibfnamefont {E.}~\bibnamefont {Nicklas}}, \bibinfo
  {author} {\bibfnamefont {J.}~\bibnamefont {Est{\`e}ve}},\ and\ \bibinfo
  {author} {\bibfnamefont {M.~K.}\ \bibnamefont {Oberthaler}},\ }\bibfield
  {title} {\bibinfo {title} {Nonlinear atom interferometer surpasses classical
  precision limit},\ }\href {https://doi.org/10.1038/nature08919} {\bibfield
  {journal} {\bibinfo  {journal} {Nature}\ }\textbf {\bibinfo {volume} {464}},\
  \bibinfo {pages} {1165} (\bibinfo {year} {2010})}\BibitemShut {NoStop}%
\bibitem [{\citenamefont {Riedel}\ \emph {et~al.}(2010)\citenamefont {Riedel},
  \citenamefont {B{\"o}hi}, \citenamefont {Li}, \citenamefont {H{\"a}nsch},
  \citenamefont {Sinatra},\ and\ \citenamefont {Treutlein}}]{Riedel2010}%
  \BibitemOpen
  \bibfield  {author} {\bibinfo {author} {\bibfnamefont {M.~F.}\ \bibnamefont
  {Riedel}}, \bibinfo {author} {\bibfnamefont {P.}~\bibnamefont {B{\"o}hi}},
  \bibinfo {author} {\bibfnamefont {Y.}~\bibnamefont {Li}}, \bibinfo {author}
  {\bibfnamefont {T.~W.}\ \bibnamefont {H{\"a}nsch}}, \bibinfo {author}
  {\bibfnamefont {A.}~\bibnamefont {Sinatra}},\ and\ \bibinfo {author}
  {\bibfnamefont {P.}~\bibnamefont {Treutlein}},\ }\bibfield  {title} {\bibinfo
  {title} {Atom-chip-based generation of entanglement for quantum metrology},\
  }\href {https://doi.org/10.1038/nature08988} {\bibfield  {journal} {\bibinfo
  {journal} {Nature}\ }\textbf {\bibinfo {volume} {464}},\ \bibinfo {pages}
  {1170} (\bibinfo {year} {2010})}\BibitemShut {NoStop}%
\bibitem [{\citenamefont {Strobel}\ \emph {et~al.}(2014)\citenamefont
  {Strobel}, \citenamefont {Muessel}, \citenamefont {Linnemann}, \citenamefont
  {Zibold}, \citenamefont {Hume}, \citenamefont {Pezz{\`e}}, \citenamefont
  {Smerzi},\ and\ \citenamefont {Oberthaler}}]{Strobel2014}%
  \BibitemOpen
  \bibfield  {author} {\bibinfo {author} {\bibfnamefont {H.}~\bibnamefont
  {Strobel}}, \bibinfo {author} {\bibfnamefont {W.}~\bibnamefont {Muessel}},
  \bibinfo {author} {\bibfnamefont {D.}~\bibnamefont {Linnemann}}, \bibinfo
  {author} {\bibfnamefont {T.}~\bibnamefont {Zibold}}, \bibinfo {author}
  {\bibfnamefont {D.~B.}\ \bibnamefont {Hume}}, \bibinfo {author}
  {\bibfnamefont {L.}~\bibnamefont {Pezz{\`e}}}, \bibinfo {author}
  {\bibfnamefont {A.}~\bibnamefont {Smerzi}},\ and\ \bibinfo {author}
  {\bibfnamefont {M.~K.}\ \bibnamefont {Oberthaler}},\ }\bibfield  {title}
  {\bibinfo {title} {Fisher information and entanglement of non-gaussian spin
  states},\ }\href {https://doi.org/10.1126/science.1250147} {\bibfield
  {journal} {\bibinfo  {journal} {Science}\ }\textbf {\bibinfo {volume}
  {345}},\ \bibinfo {pages} {424} (\bibinfo {year} {2014})}\BibitemShut
  {NoStop}%
\bibitem [{\citenamefont {Muessel}\ \emph {et~al.}(2015)\citenamefont
  {Muessel}, \citenamefont {Strobel}, \citenamefont {Linnemann}, \citenamefont
  {Zibold}, \citenamefont {Juli\'a-D\'{\i}az},\ and\ \citenamefont
  {Oberthaler}}]{Muessel2015}%
  \BibitemOpen
  \bibfield  {author} {\bibinfo {author} {\bibfnamefont {W.}~\bibnamefont
  {Muessel}}, \bibinfo {author} {\bibfnamefont {H.}~\bibnamefont {Strobel}},
  \bibinfo {author} {\bibfnamefont {D.}~\bibnamefont {Linnemann}}, \bibinfo
  {author} {\bibfnamefont {T.}~\bibnamefont {Zibold}}, \bibinfo {author}
  {\bibfnamefont {B.}~\bibnamefont {Juli\'a-D\'{\i}az}},\ and\ \bibinfo
  {author} {\bibfnamefont {M.~K.}\ \bibnamefont {Oberthaler}},\ }\bibfield
  {title} {\bibinfo {title} {Twist-and-turn spin squeezing in bose-einstein
  condensates},\ }\href {https://doi.org/10.1103/PhysRevA.92.023603} {\bibfield
   {journal} {\bibinfo  {journal} {Phys. Rev. A}\ }\textbf {\bibinfo {volume}
  {92}},\ \bibinfo {pages} {023603} (\bibinfo {year} {2015})}\BibitemShut
  {NoStop}%
\bibitem [{\citenamefont {Hosten}\ \emph
  {et~al.}(2016{\natexlab{a}})\citenamefont {Hosten}, \citenamefont {Engelsen},
  \citenamefont {Krishnakumar},\ and\ \citenamefont
  {Kasevich}}]{hosten2016squeezing}%
  \BibitemOpen
  \bibfield  {author} {\bibinfo {author} {\bibfnamefont {O.}~\bibnamefont
  {Hosten}}, \bibinfo {author} {\bibfnamefont {N.~J.}\ \bibnamefont
  {Engelsen}}, \bibinfo {author} {\bibfnamefont {R.}~\bibnamefont
  {Krishnakumar}},\ and\ \bibinfo {author} {\bibfnamefont {M.~A.}\ \bibnamefont
  {Kasevich}},\ }\bibfield  {title} {\bibinfo {title} {Measurement noise 100
  times lower than the quantum-projection limit using entangled atoms},\ }\href
  {https://doi.org/10.1038/nature16176} {\bibfield  {journal} {\bibinfo
  {journal} {Nature}\ }\textbf {\bibinfo {volume} {529}},\ \bibinfo {pages}
  {505} (\bibinfo {year} {2016}{\natexlab{a}})}\BibitemShut {NoStop}%
\bibitem [{\citenamefont {Bohnet}\ \emph {et~al.}(2016)\citenamefont {Bohnet},
  \citenamefont {Sawyer}, \citenamefont {Britton}, \citenamefont {Wall},
  \citenamefont {Rey}, \citenamefont {Foss-Feig},\ and\ \citenamefont
  {Bollinger}}]{bohnet2016squeezing}%
  \BibitemOpen
  \bibfield  {author} {\bibinfo {author} {\bibfnamefont {J.~G.}\ \bibnamefont
  {Bohnet}}, \bibinfo {author} {\bibfnamefont {B.~C.}\ \bibnamefont {Sawyer}},
  \bibinfo {author} {\bibfnamefont {J.~W.}\ \bibnamefont {Britton}}, \bibinfo
  {author} {\bibfnamefont {M.~L.}\ \bibnamefont {Wall}}, \bibinfo {author}
  {\bibfnamefont {A.~M.}\ \bibnamefont {Rey}}, \bibinfo {author} {\bibfnamefont
  {M.}~\bibnamefont {Foss-Feig}},\ and\ \bibinfo {author} {\bibfnamefont
  {J.~J.}\ \bibnamefont {Bollinger}},\ }\bibfield  {title} {\bibinfo {title}
  {Quantum spin dynamics and entanglement generation with hundreds of trapped
  ions},\ }\href {https://doi.org/10.1126/science.aad9958} {\bibfield
  {journal} {\bibinfo  {journal} {Science}\ }\textbf {\bibinfo {volume}
  {352}},\ \bibinfo {pages} {1297} (\bibinfo {year} {2016})}\BibitemShut
  {NoStop}%
\bibitem [{\citenamefont {Braverman}\ \emph {et~al.}(2019)\citenamefont
  {Braverman}, \citenamefont {Kawasaki}, \citenamefont {Pedrozo-Pe\~nafiel},
  \citenamefont {Colombo}, \citenamefont {Shu}, \citenamefont {Li},
  \citenamefont {Mendez}, \citenamefont {Yamoah}, \citenamefont {Salvi},
  \citenamefont {Akamatsu}, \citenamefont {Xiao},\ and\ \citenamefont
  {Vuleti\ifmmode~\acute{c}\else \'{c}\fi{}}}]{braverman2019}%
  \BibitemOpen
  \bibfield  {author} {\bibinfo {author} {\bibfnamefont {B.}~\bibnamefont
  {Braverman}}, \bibinfo {author} {\bibfnamefont {A.}~\bibnamefont {Kawasaki}},
  \bibinfo {author} {\bibfnamefont {E.}~\bibnamefont {Pedrozo-Pe\~nafiel}},
  \bibinfo {author} {\bibfnamefont {S.}~\bibnamefont {Colombo}}, \bibinfo
  {author} {\bibfnamefont {C.}~\bibnamefont {Shu}}, \bibinfo {author}
  {\bibfnamefont {Z.}~\bibnamefont {Li}}, \bibinfo {author} {\bibfnamefont
  {E.}~\bibnamefont {Mendez}}, \bibinfo {author} {\bibfnamefont
  {M.}~\bibnamefont {Yamoah}}, \bibinfo {author} {\bibfnamefont
  {L.}~\bibnamefont {Salvi}}, \bibinfo {author} {\bibfnamefont
  {D.}~\bibnamefont {Akamatsu}}, \bibinfo {author} {\bibfnamefont
  {Y.}~\bibnamefont {Xiao}},\ and\ \bibinfo {author} {\bibfnamefont
  {V.}~\bibnamefont {Vuleti\ifmmode~\acute{c}\else \'{c}\fi{}}},\ }\bibfield
  {title} {\bibinfo {title} {Near-unitary spin squeezing in
  $^{171}\mathrm{Yb}$},\ }\href
  {https://doi.org/10.1103/PhysRevLett.122.223203} {\bibfield  {journal}
  {\bibinfo  {journal} {Phys. Rev. Lett.}\ }\textbf {\bibinfo {volume} {122}},\
  \bibinfo {pages} {223203} (\bibinfo {year} {2019})}\BibitemShut {NoStop}%
\bibitem [{\citenamefont {Colombo}\ \emph {et~al.}(2022)\citenamefont
  {Colombo}, \citenamefont {Pedrozo-Pe{\~n}afiel}, \citenamefont {Adiyatullin},
  \citenamefont {Li}, \citenamefont {Mendez}, \citenamefont {Shu},\ and\
  \citenamefont {Vuleti{\'c}}}]{colombo2022heisenbergScaledSqz}%
  \BibitemOpen
  \bibfield  {author} {\bibinfo {author} {\bibfnamefont {S.}~\bibnamefont
  {Colombo}}, \bibinfo {author} {\bibfnamefont {E.}~\bibnamefont
  {Pedrozo-Pe{\~n}afiel}}, \bibinfo {author} {\bibfnamefont {A.~F.}\
  \bibnamefont {Adiyatullin}}, \bibinfo {author} {\bibfnamefont
  {Z.}~\bibnamefont {Li}}, \bibinfo {author} {\bibfnamefont {E.}~\bibnamefont
  {Mendez}}, \bibinfo {author} {\bibfnamefont {C.}~\bibnamefont {Shu}},\ and\
  \bibinfo {author} {\bibfnamefont {V.}~\bibnamefont {Vuleti{\'c}}},\
  }\bibfield  {title} {\bibinfo {title} {Time-reversal-based quantum metrology
  with many-body entangled states},\ }\href
  {https://doi.org/10.1038/s41567-022-01653-5} {\bibfield  {journal} {\bibinfo
  {journal} {Nature Physics}\ }\textbf {\bibinfo {volume} {18}},\ \bibinfo
  {pages} {925} (\bibinfo {year} {2022})}\BibitemShut {NoStop}%
\bibitem [{\citenamefont {Hines}\ \emph {et~al.}(2023)\citenamefont {Hines},
  \citenamefont {Rajagopal}, \citenamefont {Moreau}, \citenamefont {Wahrman},
  \citenamefont {Lewis}, \citenamefont {Markovi{\'c}},\ and\ \citenamefont
  {Schleier-Smith}}]{hines2023squeezing}%
  \BibitemOpen
  \bibfield  {author} {\bibinfo {author} {\bibfnamefont {J.~A.}\ \bibnamefont
  {Hines}}, \bibinfo {author} {\bibfnamefont {S.~V.}\ \bibnamefont
  {Rajagopal}}, \bibinfo {author} {\bibfnamefont {G.~L.}\ \bibnamefont
  {Moreau}}, \bibinfo {author} {\bibfnamefont {M.~D.}\ \bibnamefont {Wahrman}},
  \bibinfo {author} {\bibfnamefont {N.~A.}\ \bibnamefont {Lewis}}, \bibinfo
  {author} {\bibfnamefont {O.}~\bibnamefont {Markovi{\'c}}},\ and\ \bibinfo
  {author} {\bibfnamefont {M.}~\bibnamefont {Schleier-Smith}},\ }\bibfield
  {title} {\bibinfo {title} {Spin squeezing by rydberg dressing in an array of
  atomic ensembles},\ }\href {https://doi.org/10.1103/PhysRevLett.131.063401}
  {\bibfield  {journal} {\bibinfo  {journal} {Phys. Rev. Lett.}\ }\textbf
  {\bibinfo {volume} {131}},\ \bibinfo {pages} {063401} (\bibinfo {year}
  {2023})}\BibitemShut {NoStop}%
\bibitem [{\citenamefont {Franke}\ \emph {et~al.}(2023)\citenamefont {Franke},
  \citenamefont {Muleady}, \citenamefont {Kaubruegger}, \citenamefont {Kranzl},
  \citenamefont {Blatt}, \citenamefont {Rey}, \citenamefont {Joshi},\ and\
  \citenamefont {Roos}}]{franke2023squeezing}%
  \BibitemOpen
  \bibfield  {author} {\bibinfo {author} {\bibfnamefont {J.}~\bibnamefont
  {Franke}}, \bibinfo {author} {\bibfnamefont {S.~R.}\ \bibnamefont {Muleady}},
  \bibinfo {author} {\bibfnamefont {R.}~\bibnamefont {Kaubruegger}}, \bibinfo
  {author} {\bibfnamefont {F.}~\bibnamefont {Kranzl}}, \bibinfo {author}
  {\bibfnamefont {R.}~\bibnamefont {Blatt}}, \bibinfo {author} {\bibfnamefont
  {A.~M.}\ \bibnamefont {Rey}}, \bibinfo {author} {\bibfnamefont {M.~K.}\
  \bibnamefont {Joshi}},\ and\ \bibinfo {author} {\bibfnamefont {C.~F.}\
  \bibnamefont {Roos}},\ }\bibfield  {title} {\bibinfo {title}
  {Quantum-enhanced sensing on optical transitions through finite-range
  interactions},\ }\href {https://doi.org/10.1038/s41586-023-06472-z}
  {\bibfield  {journal} {\bibinfo  {journal} {Nature}\ }\textbf {\bibinfo
  {volume} {621}},\ \bibinfo {pages} {740} (\bibinfo {year}
  {2023})}\BibitemShut {NoStop}%
\bibitem [{\citenamefont {Eckner}\ \emph {et~al.}(2023)\citenamefont {Eckner},
  \citenamefont {Darkwah~Oppong}, \citenamefont {Cao}, \citenamefont {Young},
  \citenamefont {Milner}, \citenamefont {Robinson}, \citenamefont {Ye},\ and\
  \citenamefont {Kaufman}}]{eckner2023squeezing}%
  \BibitemOpen
  \bibfield  {author} {\bibinfo {author} {\bibfnamefont {W.~J.}\ \bibnamefont
  {Eckner}}, \bibinfo {author} {\bibfnamefont {N.}~\bibnamefont
  {Darkwah~Oppong}}, \bibinfo {author} {\bibfnamefont {A.}~\bibnamefont {Cao}},
  \bibinfo {author} {\bibfnamefont {A.~W.}\ \bibnamefont {Young}}, \bibinfo
  {author} {\bibfnamefont {W.~R.}\ \bibnamefont {Milner}}, \bibinfo {author}
  {\bibfnamefont {J.~M.}\ \bibnamefont {Robinson}}, \bibinfo {author}
  {\bibfnamefont {J.}~\bibnamefont {Ye}},\ and\ \bibinfo {author}
  {\bibfnamefont {A.~M.}\ \bibnamefont {Kaufman}},\ }\bibfield  {title}
  {\bibinfo {title} {Realizing spin squeezing with rydberg interactions in an
  optical clock},\ }\href {https://doi.org/10.1038/s41586-023-06360-6}
  {\bibfield  {journal} {\bibinfo  {journal} {Nature}\ }\textbf {\bibinfo
  {volume} {621}},\ \bibinfo {pages} {734} (\bibinfo {year}
  {2023})}\BibitemShut {NoStop}%
\bibitem [{\citenamefont {Bornet}\ \emph {et~al.}(2023)\citenamefont {Bornet},
  \citenamefont {Emperauger}, \citenamefont {Chen}, \citenamefont {Ye},
  \citenamefont {Block}, \citenamefont {Bintz}, \citenamefont {Boyd},
  \citenamefont {Barredo}, \citenamefont {Comparin}, \citenamefont {Mezzacapo}
  \emph {et~al.}}]{bornet2023squeezing}%
  \BibitemOpen
  \bibfield  {author} {\bibinfo {author} {\bibfnamefont {G.}~\bibnamefont
  {Bornet}}, \bibinfo {author} {\bibfnamefont {G.}~\bibnamefont {Emperauger}},
  \bibinfo {author} {\bibfnamefont {C.}~\bibnamefont {Chen}}, \bibinfo {author}
  {\bibfnamefont {B.}~\bibnamefont {Ye}}, \bibinfo {author} {\bibfnamefont
  {M.}~\bibnamefont {Block}}, \bibinfo {author} {\bibfnamefont
  {M.}~\bibnamefont {Bintz}}, \bibinfo {author} {\bibfnamefont {J.~A.}\
  \bibnamefont {Boyd}}, \bibinfo {author} {\bibfnamefont {D.}~\bibnamefont
  {Barredo}}, \bibinfo {author} {\bibfnamefont {T.}~\bibnamefont {Comparin}},
  \bibinfo {author} {\bibfnamefont {F.}~\bibnamefont {Mezzacapo}}, \emph
  {et~al.},\ }\bibfield  {title} {\bibinfo {title} {Scalable spin squeezing in
  a dipolar rydberg atom array},\ }\href
  {https://doi.org/10.1038/s41586-023-06414-9} {\bibfield  {journal} {\bibinfo
  {journal} {Nature}\ }\textbf {\bibinfo {volume} {621}},\ \bibinfo {pages}
  {728} (\bibinfo {year} {2023})}\BibitemShut {NoStop}%
\bibitem [{\citenamefont {Robinson}\ \emph {et~al.}(2024)\citenamefont
  {Robinson}, \citenamefont {Miklos}, \citenamefont {Tso}, \citenamefont
  {Kennedy}, \citenamefont {Bothwell}, \citenamefont {Kedar}, \citenamefont
  {Thompson},\ and\ \citenamefont {Ye}}]{robinson2024squeezing}%
  \BibitemOpen
  \bibfield  {author} {\bibinfo {author} {\bibfnamefont {J.~M.}\ \bibnamefont
  {Robinson}}, \bibinfo {author} {\bibfnamefont {M.}~\bibnamefont {Miklos}},
  \bibinfo {author} {\bibfnamefont {Y.~M.}\ \bibnamefont {Tso}}, \bibinfo
  {author} {\bibfnamefont {C.~J.}\ \bibnamefont {Kennedy}}, \bibinfo {author}
  {\bibfnamefont {T.}~\bibnamefont {Bothwell}}, \bibinfo {author}
  {\bibfnamefont {D.}~\bibnamefont {Kedar}}, \bibinfo {author} {\bibfnamefont
  {J.~K.}\ \bibnamefont {Thompson}},\ and\ \bibinfo {author} {\bibfnamefont
  {J.}~\bibnamefont {Ye}},\ }\bibfield  {title} {\bibinfo {title} {Direct
  comparison of two spin-squeezed optical clock ensembles at the 10-17 level},\
  }\href {https://doi.org/10.1038/s41567-023-02310-1} {\bibfield  {journal}
  {\bibinfo  {journal} {Nature Physics}\ }\textbf {\bibinfo {volume} {20}},\
  \bibinfo {pages} {208} (\bibinfo {year} {2024})}\BibitemShut {NoStop}%
\bibitem [{\citenamefont {Genoni}\ \emph {et~al.}(2013)\citenamefont {Genoni},
  \citenamefont {Paris}, \citenamefont {Adesso}, \citenamefont {Nha},
  \citenamefont {Knight},\ and\ \citenamefont
  {Kim}}]{genoni2013multiParameter}%
  \BibitemOpen
  \bibfield  {author} {\bibinfo {author} {\bibfnamefont {M.~G.}\ \bibnamefont
  {Genoni}}, \bibinfo {author} {\bibfnamefont {M.~G.}\ \bibnamefont {Paris}},
  \bibinfo {author} {\bibfnamefont {G.}~\bibnamefont {Adesso}}, \bibinfo
  {author} {\bibfnamefont {H.}~\bibnamefont {Nha}}, \bibinfo {author}
  {\bibfnamefont {P.~L.}\ \bibnamefont {Knight}},\ and\ \bibinfo {author}
  {\bibfnamefont {M.}~\bibnamefont {Kim}},\ }\bibfield  {title} {\bibinfo
  {title} {Optimal estimation of joint parameters in phase space},\ }\href
  {https://doi.org/10.1103/physreva.87.012107} {\bibfield  {journal} {\bibinfo
  {journal} {Phys. Rev. A}\ }\textbf {\bibinfo {volume} {87}},\ \bibinfo
  {pages} {012107} (\bibinfo {year} {2013})}\BibitemShut {NoStop}%
\bibitem [{\citenamefont {Humphreys}\ \emph {et~al.}(2013)\citenamefont
  {Humphreys}, \citenamefont {Barbieri}, \citenamefont {Datta},\ and\
  \citenamefont {Walmsley}}]{humphreys2013multiParameter}%
  \BibitemOpen
  \bibfield  {author} {\bibinfo {author} {\bibfnamefont {P.~C.}\ \bibnamefont
  {Humphreys}}, \bibinfo {author} {\bibfnamefont {M.}~\bibnamefont {Barbieri}},
  \bibinfo {author} {\bibfnamefont {A.}~\bibnamefont {Datta}},\ and\ \bibinfo
  {author} {\bibfnamefont {I.~A.}\ \bibnamefont {Walmsley}},\ }\bibfield
  {title} {\bibinfo {title} {Quantum enhanced multiple phase estimation},\
  }\href {https://doi.org/10.1103/physrevlett.111.070403} {\bibfield  {journal}
  {\bibinfo  {journal} {Phys. Rev. Lett.}\ }\textbf {\bibinfo {volume} {111}},\
  \bibinfo {pages} {070403} (\bibinfo {year} {2013})}\BibitemShut {NoStop}%
\bibitem [{\citenamefont {Zhang}\ and\ \citenamefont
  {Fan}(2014)}]{zhang2014multiParameter}%
  \BibitemOpen
  \bibfield  {author} {\bibinfo {author} {\bibfnamefont {Y.-R.}\ \bibnamefont
  {Zhang}}\ and\ \bibinfo {author} {\bibfnamefont {H.}~\bibnamefont {Fan}},\
  }\bibfield  {title} {\bibinfo {title} {Quantum metrological bounds for vector
  parameters},\ }\href {https://doi.org/10.1103/PhysRevA.90.043818} {\bibfield
  {journal} {\bibinfo  {journal} {Phys. Rev. A}\ }\textbf {\bibinfo {volume}
  {90}},\ \bibinfo {pages} {043818} (\bibinfo {year} {2014})}\BibitemShut
  {NoStop}%
\bibitem [{\citenamefont {Gao}\ and\ \citenamefont
  {Lee}(2014)}]{gao2014multiParameter}%
  \BibitemOpen
  \bibfield  {author} {\bibinfo {author} {\bibfnamefont {Y.}~\bibnamefont
  {Gao}}\ and\ \bibinfo {author} {\bibfnamefont {H.}~\bibnamefont {Lee}},\
  }\bibfield  {title} {\bibinfo {title} {Bounds on quantum multiple-parameter
  estimation with gaussian state},\ }\href
  {https://doi.org/10.1140/epjd/e2014-50560-1} {\bibfield  {journal} {\bibinfo
  {journal} {The European Physical Journal D}\ }\textbf {\bibinfo {volume}
  {68}},\ \bibinfo {pages} {1} (\bibinfo {year} {2014})}\BibitemShut {NoStop}%
\bibitem [{\citenamefont {Baumgratz}\ and\ \citenamefont
  {Datta}(2016)}]{baumgratz2016multiParameter}%
  \BibitemOpen
  \bibfield  {author} {\bibinfo {author} {\bibfnamefont {T.}~\bibnamefont
  {Baumgratz}}\ and\ \bibinfo {author} {\bibfnamefont {A.}~\bibnamefont
  {Datta}},\ }\bibfield  {title} {\bibinfo {title} {Quantum enhanced estimation
  of a multidimensional field},\ }\href
  {https://doi.org/10.1103/physrevlett.116.030801} {\bibfield  {journal}
  {\bibinfo  {journal} {Phys. Rev. Lett.}\ }\textbf {\bibinfo {volume} {116}},\
  \bibinfo {pages} {030801} (\bibinfo {year} {2016})}\BibitemShut {NoStop}%
\bibitem [{\citenamefont {Gessner}\ \emph {et~al.}(2020)\citenamefont
  {Gessner}, \citenamefont {Smerzi},\ and\ \citenamefont
  {Pezz{\`e}}}]{gessner2020multiparameter}%
  \BibitemOpen
  \bibfield  {author} {\bibinfo {author} {\bibfnamefont {M.}~\bibnamefont
  {Gessner}}, \bibinfo {author} {\bibfnamefont {A.}~\bibnamefont {Smerzi}},\
  and\ \bibinfo {author} {\bibfnamefont {L.}~\bibnamefont {Pezz{\`e}}},\
  }\bibfield  {title} {\bibinfo {title} {Multiparameter squeezing for optimal
  quantum enhancements in sensor networks},\ }\href
  {https://doi.org/10.1038/s41467-020-17471-3} {\bibfield  {journal} {\bibinfo
  {journal} {Nature communications}\ }\textbf {\bibinfo {volume} {11}},\
  \bibinfo {pages} {3817} (\bibinfo {year} {2020})}\BibitemShut {NoStop}%
\bibitem [{\citenamefont {Baamara}\ \emph {et~al.}(2023)\citenamefont
  {Baamara}, \citenamefont {Gessner},\ and\ \citenamefont
  {Sinatra}}]{baamara2023quantum}%
  \BibitemOpen
  \bibfield  {author} {\bibinfo {author} {\bibfnamefont {Y.}~\bibnamefont
  {Baamara}}, \bibinfo {author} {\bibfnamefont {M.}~\bibnamefont {Gessner}},\
  and\ \bibinfo {author} {\bibfnamefont {A.}~\bibnamefont {Sinatra}},\
  }\bibfield  {title} {\bibinfo {title} {Quantum-enhanced multiparameter
  estimation and compressed sensing of a field},\ }\href
  {https://doi.org/10.21468/SciPostPhys.14.3.050} {\bibfield  {journal}
  {\bibinfo  {journal} {SciPost Physics}\ }\textbf {\bibinfo {volume} {14}},\
  \bibinfo {pages} {050} (\bibinfo {year} {2023})}\BibitemShut {NoStop}%
\bibitem [{\citenamefont {Fabre}\ and\ \citenamefont
  {Treps}(2020)}]{fabre2020bosonicReview}%
  \BibitemOpen
  \bibfield  {author} {\bibinfo {author} {\bibfnamefont {C.}~\bibnamefont
  {Fabre}}\ and\ \bibinfo {author} {\bibfnamefont {N.}~\bibnamefont {Treps}},\
  }\bibfield  {title} {\bibinfo {title} {Modes and states in quantum optics},\
  }\href {https://doi.org/10.1103/RevModPhys.92.035005} {\bibfield  {journal}
  {\bibinfo  {journal} {Rev. Mod. Phys.}\ }\textbf {\bibinfo {volume} {92}},\
  \bibinfo {pages} {035005} (\bibinfo {year} {2020})}\BibitemShut {NoStop}%
\bibitem [{\citenamefont {Agarwal}\ and\ \citenamefont
  {Puri}(1989)}]{agarwal1989squeezedDissipator}%
  \BibitemOpen
  \bibfield  {author} {\bibinfo {author} {\bibfnamefont {G.}~\bibnamefont
  {Agarwal}}\ and\ \bibinfo {author} {\bibfnamefont {R.}~\bibnamefont {Puri}},\
  }\bibfield  {title} {\bibinfo {title} {Nonequilibrium phase transitions in a
  squeezed cavity and the generation of spin states satisfying uncertainty
  equality},\ }\href {https://doi.org/10.1016/0030-4018(89)90113-2} {\bibfield
  {journal} {\bibinfo  {journal} {Optics communications}\ }\textbf {\bibinfo
  {volume} {69}},\ \bibinfo {pages} {267} (\bibinfo {year} {1989})}\BibitemShut
  {NoStop}%
\bibitem [{\citenamefont {Dalla~Torre}\ \emph {et~al.}(2013)\citenamefont
  {Dalla~Torre}, \citenamefont {Otterbach}, \citenamefont {Demler},
  \citenamefont {Vuletic},\ and\ \citenamefont
  {Lukin}}]{dalla2013dissipatorImplementation}%
  \BibitemOpen
  \bibfield  {author} {\bibinfo {author} {\bibfnamefont {E.~G.}\ \bibnamefont
  {Dalla~Torre}}, \bibinfo {author} {\bibfnamefont {J.}~\bibnamefont
  {Otterbach}}, \bibinfo {author} {\bibfnamefont {E.}~\bibnamefont {Demler}},
  \bibinfo {author} {\bibfnamefont {V.}~\bibnamefont {Vuletic}},\ and\ \bibinfo
  {author} {\bibfnamefont {M.~D.}\ \bibnamefont {Lukin}},\ }\bibfield  {title}
  {\bibinfo {title} {Dissipative preparation of spin squeezed atomic ensembles
  in a steady state},\ }\href {https://doi.org/10.1103/PhysRevLett.110.120402}
  {\bibfield  {journal} {\bibinfo  {journal} {Phys. Rev. Lett.}\ }\textbf
  {\bibinfo {volume} {110}},\ \bibinfo {pages} {120402} (\bibinfo {year}
  {2013})}\BibitemShut {NoStop}%
\bibitem [{\citenamefont {Groszkowski}\ \emph {et~al.}(2022)\citenamefont
  {Groszkowski}, \citenamefont {Koppenh{\"o}fer}, \citenamefont {Lau},\ and\
  \citenamefont {Clerk}}]{groszkowski2022reservoir}%
  \BibitemOpen
  \bibfield  {author} {\bibinfo {author} {\bibfnamefont {P.}~\bibnamefont
  {Groszkowski}}, \bibinfo {author} {\bibfnamefont {M.}~\bibnamefont
  {Koppenh{\"o}fer}}, \bibinfo {author} {\bibfnamefont {H.-K.}\ \bibnamefont
  {Lau}},\ and\ \bibinfo {author} {\bibfnamefont {A.}~\bibnamefont {Clerk}},\
  }\bibfield  {title} {\bibinfo {title} {Reservoir-engineered spin squeezing:
  Macroscopic even-odd effects and hybrid-systems implementations},\ }\href
  {https://doi.org/10.1103/PhysRevX.12.011015} {\bibfield  {journal} {\bibinfo
  {journal} {Phys. Rev. X}\ }\textbf {\bibinfo {volume} {12}},\ \bibinfo
  {pages} {011015} (\bibinfo {year} {2022})}\BibitemShut {NoStop}%
\bibitem [{\citenamefont {Gessner}\ \emph {et~al.}(2018)\citenamefont
  {Gessner}, \citenamefont {Pezz{\`e}},\ and\ \citenamefont
  {Smerzi}}]{gessner2018sensitivity}%
  \BibitemOpen
  \bibfield  {author} {\bibinfo {author} {\bibfnamefont {M.}~\bibnamefont
  {Gessner}}, \bibinfo {author} {\bibfnamefont {L.}~\bibnamefont {Pezz{\`e}}},\
  and\ \bibinfo {author} {\bibfnamefont {A.}~\bibnamefont {Smerzi}},\
  }\bibfield  {title} {\bibinfo {title} {Sensitivity bounds for multiparameter
  quantum metrology},\ }\href {https://doi.org/10.1103/PhysRevLett.121.130503}
  {\bibfield  {journal} {\bibinfo  {journal} {Phys. Rev. Lett.}\ }\textbf
  {\bibinfo {volume} {121}},\ \bibinfo {pages} {130503} (\bibinfo {year}
  {2018})}\BibitemShut {NoStop}%
\bibitem [{\citenamefont {Fadel}\ \emph {et~al.}(2023)\citenamefont {Fadel},
  \citenamefont {Yadin}, \citenamefont {Mao}, \citenamefont {Byrnes},\ and\
  \citenamefont {Gessner}}]{Fadel2023}%
  \BibitemOpen
  \bibfield  {author} {\bibinfo {author} {\bibfnamefont {M.}~\bibnamefont
  {Fadel}}, \bibinfo {author} {\bibfnamefont {B.}~\bibnamefont {Yadin}},
  \bibinfo {author} {\bibfnamefont {Y.}~\bibnamefont {Mao}}, \bibinfo {author}
  {\bibfnamefont {T.}~\bibnamefont {Byrnes}},\ and\ \bibinfo {author}
  {\bibfnamefont {M.}~\bibnamefont {Gessner}},\ }\bibfield  {title} {\bibinfo
  {title} {Multiparameter quantum metrology and mode entanglement with
  spatially split nonclassical spin ensembles},\ }\href
  {https://doi.org/10.1088/1367-2630/ace1a0} {\bibfield  {journal} {\bibinfo
  {journal} {New Journal of Physics}\ }\textbf {\bibinfo {volume} {25}},\
  \bibinfo {pages} {073006} (\bibinfo {year} {2023})}\BibitemShut {NoStop}%
\bibitem [{\citenamefont {Liu}\ \emph {et~al.}(2020)\citenamefont {Liu},
  \citenamefont {Yuan}, \citenamefont {Lu},\ and\ \citenamefont
  {Wang}}]{liu2020multiparReview}%
  \BibitemOpen
  \bibfield  {author} {\bibinfo {author} {\bibfnamefont {J.}~\bibnamefont
  {Liu}}, \bibinfo {author} {\bibfnamefont {H.}~\bibnamefont {Yuan}}, \bibinfo
  {author} {\bibfnamefont {X.-M.}\ \bibnamefont {Lu}},\ and\ \bibinfo {author}
  {\bibfnamefont {X.}~\bibnamefont {Wang}},\ }\bibfield  {title} {\bibinfo
  {title} {Quantum fisher information matrix and multiparameter estimation},\
  }\href {https://doi.org/10.1088/1751-8121/ab5d4d} {\bibfield  {journal}
  {\bibinfo  {journal} {Journal of Physics A: Mathematical and Theoretical}\
  }\textbf {\bibinfo {volume} {53}},\ \bibinfo {pages} {023001} (\bibinfo
  {year} {2020})}\BibitemShut {NoStop}%
\bibitem [{\citenamefont {Szczykulska}\ \emph {et~al.}(2016)\citenamefont
  {Szczykulska}, \citenamefont {Baumgratz},\ and\ \citenamefont
  {Datta}}]{szczykulska2016multiparReview}%
  \BibitemOpen
  \bibfield  {author} {\bibinfo {author} {\bibfnamefont {M.}~\bibnamefont
  {Szczykulska}}, \bibinfo {author} {\bibfnamefont {T.}~\bibnamefont
  {Baumgratz}},\ and\ \bibinfo {author} {\bibfnamefont {A.}~\bibnamefont
  {Datta}},\ }\bibfield  {title} {\bibinfo {title} {Multi-parameter quantum
  metrology},\ }\href {https://doi.org/10.1080/23746149.2016.1230476}
  {\bibfield  {journal} {\bibinfo  {journal} {Advances in Physics: X}\ }\textbf
  {\bibinfo {volume} {1}},\ \bibinfo {pages} {621} (\bibinfo {year}
  {2016})}\BibitemShut {NoStop}%
\bibitem [{\citenamefont {Giovannetti}\ \emph {et~al.}(2011)\citenamefont
  {Giovannetti}, \citenamefont {Lloyd},\ and\ \citenamefont
  {Maccone}}]{giovannetti2011multiparReview}%
  \BibitemOpen
  \bibfield  {author} {\bibinfo {author} {\bibfnamefont {V.}~\bibnamefont
  {Giovannetti}}, \bibinfo {author} {\bibfnamefont {S.}~\bibnamefont {Lloyd}},\
  and\ \bibinfo {author} {\bibfnamefont {L.}~\bibnamefont {Maccone}},\
  }\bibfield  {title} {\bibinfo {title} {Advances in quantum metrology},\
  }\href {https://doi.org/10.1038/nphoton.2011.35} {\bibfield  {journal}
  {\bibinfo  {journal} {Nature photonics}\ }\textbf {\bibinfo {volume} {5}},\
  \bibinfo {pages} {222} (\bibinfo {year} {2011})}\BibitemShut {NoStop}%
\bibitem [{Sup()}]{Supplement}%
  \BibitemOpen
  \href@noop {} {}\bibinfo {note} {See Supplemental Material for derivations,
  details on measurement protocols and numerical benchmarking, which also cites
  References~\cite{liu2020multiparReview,matsumoto2002,Ragy2016,Pezze2017,
  groszkowski2022reservoir, pocklington2024solutionStructure,
  braunstein1987generalizedSqueezing, leghtas2015yaleTwoPhoton,
  reglade2024twoPhoton}.}\BibitemShut {Stop}%
\bibitem [{\citenamefont {Matsumoto}(2002)}]{matsumoto2002}%
  \BibitemOpen
  \bibfield  {author} {\bibinfo {author} {\bibfnamefont {K.}~\bibnamefont
  {Matsumoto}},\ }\bibfield  {title} {\bibinfo {title} {A new approach to the
  cramér-rao-type bound of the pure-state model},\ }\href
  {https://doi.org/10.1088/0305-4470/35/13/307} {\bibfield  {journal} {\bibinfo
   {journal} {Journal of Physics A: Mathematical and General}\ }\textbf
  {\bibinfo {volume} {35}},\ \bibinfo {pages} {3111} (\bibinfo {year}
  {2002})}\BibitemShut {NoStop}%
\bibitem [{\citenamefont {Ragy}\ \emph {et~al.}(2016)\citenamefont {Ragy},
  \citenamefont {Jarzyna},\ and\ \citenamefont
  {Demkowicz-Dobrzański}}]{Ragy2016}%
  \BibitemOpen
  \bibfield  {author} {\bibinfo {author} {\bibfnamefont {S.}~\bibnamefont
  {Ragy}}, \bibinfo {author} {\bibfnamefont {M.}~\bibnamefont {Jarzyna}},\ and\
  \bibinfo {author} {\bibfnamefont {R.}~\bibnamefont {Demkowicz-Dobrzański}},\
  }\bibfield  {title} {\bibinfo {title} {Compatibility in multiparameter
  quantum metrology},\ }\href {https://doi.org/10.1103/physreva.94.052108}
  {\bibfield  {journal} {\bibinfo  {journal} {Phys. Rev. A}\ }\textbf {\bibinfo
  {volume} {94}},\ \bibinfo {pages} {052108} (\bibinfo {year}
  {2016})}\BibitemShut {NoStop}%
\bibitem [{\citenamefont {Pezzè}\ \emph {et~al.}(2017)\citenamefont {Pezzè},
  \citenamefont {Ciampini}, \citenamefont {Spagnolo}, \citenamefont
  {Humphreys}, \citenamefont {Datta}, \citenamefont {Walmsley}, \citenamefont
  {Barbieri}, \citenamefont {Sciarrino},\ and\ \citenamefont
  {Smerzi}}]{Pezze2017}%
  \BibitemOpen
  \bibfield  {author} {\bibinfo {author} {\bibfnamefont {L.}~\bibnamefont
  {Pezzè}}, \bibinfo {author} {\bibfnamefont {M.~A.}\ \bibnamefont
  {Ciampini}}, \bibinfo {author} {\bibfnamefont {N.}~\bibnamefont {Spagnolo}},
  \bibinfo {author} {\bibfnamefont {P.~C.}\ \bibnamefont {Humphreys}}, \bibinfo
  {author} {\bibfnamefont {A.}~\bibnamefont {Datta}}, \bibinfo {author}
  {\bibfnamefont {I.~A.}\ \bibnamefont {Walmsley}}, \bibinfo {author}
  {\bibfnamefont {M.}~\bibnamefont {Barbieri}}, \bibinfo {author}
  {\bibfnamefont {F.}~\bibnamefont {Sciarrino}},\ and\ \bibinfo {author}
  {\bibfnamefont {A.}~\bibnamefont {Smerzi}},\ }\bibfield  {title} {\bibinfo
  {title} {Optimal measurements for simultaneous quantum estimation of multiple
  phases},\ }\href {https://doi.org/10.1103/physrevlett.119.130504} {\bibfield
  {journal} {\bibinfo  {journal} {Physical Review Letters}\ }\textbf {\bibinfo
  {volume} {119}},\ \bibinfo {pages} {130504} (\bibinfo {year}
  {2017})}\BibitemShut {NoStop}%
\bibitem [{Note1()}]{Note1}%
  \BibitemOpen
  \bibinfo {note} {Note that one can construct examples where the QFI does not
  monotonically increase with $r$. However, for most cases of interest,
  including the crucial case of two spin ensembles, we do have a monotonic
  increase.}\BibitemShut {Stop}%
\bibitem [{\citenamefont {Kitzinger}\ \emph {et~al.}(2020)\citenamefont
  {Kitzinger}, \citenamefont {Chaudhary}, \citenamefont {Kondappan},
  \citenamefont {Ivannikov},\ and\ \citenamefont
  {Byrnes}}]{kitzinger2020twoAxisTwoSpin}%
  \BibitemOpen
  \bibfield  {author} {\bibinfo {author} {\bibfnamefont {J.}~\bibnamefont
  {Kitzinger}}, \bibinfo {author} {\bibfnamefont {M.}~\bibnamefont
  {Chaudhary}}, \bibinfo {author} {\bibfnamefont {M.}~\bibnamefont
  {Kondappan}}, \bibinfo {author} {\bibfnamefont {V.}~\bibnamefont
  {Ivannikov}},\ and\ \bibinfo {author} {\bibfnamefont {T.}~\bibnamefont
  {Byrnes}},\ }\bibfield  {title} {\bibinfo {title} {Two-axis two-spin squeezed
  states},\ }\href {https://doi.org/10.1103/PhysRevResearch.2.033504}
  {\bibfield  {journal} {\bibinfo  {journal} {Phys. Rev. Res.}\ }\textbf
  {\bibinfo {volume} {2}},\ \bibinfo {pages} {033504} (\bibinfo {year}
  {2020})}\BibitemShut {NoStop}%
\bibitem [{\citenamefont {Pocklington}\ and\ \citenamefont
  {Clerk}(2024)}]{pocklington2024solutionStructure}%
  \BibitemOpen
  \bibfield  {author} {\bibinfo {author} {\bibfnamefont {A.}~\bibnamefont
  {Pocklington}}\ and\ \bibinfo {author} {\bibfnamefont {A.~A.}\ \bibnamefont
  {Clerk}},\ }\bibfield  {title} {\bibinfo {title} {Universal time-entanglement
  trade-off in open quantum systems},\ }\href
  {https://dx.doi.org/10.48550/arXiv.2404.03625} {\bibfield  {journal}
  {\bibinfo  {journal} {arXiv preprint arXiv:2404.03625}\ } (\bibinfo {year}
  {2024})}\BibitemShut {NoStop}%
\bibitem [{\citenamefont {Parkins}\ \emph {et~al.}(2006)\citenamefont
  {Parkins}, \citenamefont {Solano},\ and\ \citenamefont
  {Cirac}}]{parkins2006unconditionalTMS}%
  \BibitemOpen
  \bibfield  {author} {\bibinfo {author} {\bibfnamefont {A.~S.}\ \bibnamefont
  {Parkins}}, \bibinfo {author} {\bibfnamefont {E.}~\bibnamefont {Solano}},\
  and\ \bibinfo {author} {\bibfnamefont {J.~I.}\ \bibnamefont {Cirac}},\
  }\bibfield  {title} {\bibinfo {title} {Unconditional two-mode squeezing of
  separated atomic ensembles},\ }\href
  {https://doi.org/10.1103/physrevlett.96.053602} {\bibfield  {journal}
  {\bibinfo  {journal} {Phys. Rev. Lett.}\ }\textbf {\bibinfo {volume} {96}},\
  \bibinfo {pages} {053602} (\bibinfo {year} {2006})}\BibitemShut {NoStop}%
\bibitem [{\citenamefont {Sundar}\ \emph {et~al.}(2024)\citenamefont {Sundar},
  \citenamefont {Barberena}, \citenamefont {Rey},\ and\ \citenamefont
  {Orioli}}]{sundar2024squeezing}%
  \BibitemOpen
  \bibfield  {author} {\bibinfo {author} {\bibfnamefont {B.}~\bibnamefont
  {Sundar}}, \bibinfo {author} {\bibfnamefont {D.}~\bibnamefont {Barberena}},
  \bibinfo {author} {\bibfnamefont {A.~M.}\ \bibnamefont {Rey}},\ and\ \bibinfo
  {author} {\bibfnamefont {A.~P.}\ \bibnamefont {Orioli}},\ }\bibfield  {title}
  {\bibinfo {title} {Squeezing multilevel atoms in dark states via cavity
  superradiance},\ }\href {https://doi.org/10.1103/PhysRevLett.132.033601}
  {\bibfield  {journal} {\bibinfo  {journal} {Phys. Rev. Lett.}\ }\textbf
  {\bibinfo {volume} {132}},\ \bibinfo {pages} {033601} (\bibinfo {year}
  {2024})}\BibitemShut {NoStop}%
\bibitem [{\citenamefont {Muschik}\ \emph {et~al.}(2011)\citenamefont
  {Muschik}, \citenamefont {Polzik},\ and\ \citenamefont
  {Cirac}}]{muschik2011experimentalImplementation}%
  \BibitemOpen
  \bibfield  {author} {\bibinfo {author} {\bibfnamefont {C.~A.}\ \bibnamefont
  {Muschik}}, \bibinfo {author} {\bibfnamefont {E.~S.}\ \bibnamefont
  {Polzik}},\ and\ \bibinfo {author} {\bibfnamefont {J.~I.}\ \bibnamefont
  {Cirac}},\ }\bibfield  {title} {\bibinfo {title} {Dissipatively driven
  entanglement of two macroscopic atomic ensembles},\ }\href
  {https://doi.org//10.1103/PhysRevA.83.052312} {\bibfield  {journal} {\bibinfo
   {journal} {Phys. Rev. A}\ }\textbf {\bibinfo {volume} {83}},\ \bibinfo
  {pages} {052312} (\bibinfo {year} {2011})}\BibitemShut {NoStop}%
\bibitem [{\citenamefont {Zheng}\ \emph {et~al.}(2010)\citenamefont {Zheng},
  \citenamefont {Yang},\ and\ \citenamefont
  {Xia}}]{zheng2010experimentalImplementation}%
  \BibitemOpen
  \bibfield  {author} {\bibinfo {author} {\bibfnamefont {S.-B.}\ \bibnamefont
  {Zheng}}, \bibinfo {author} {\bibfnamefont {Z.-B.}\ \bibnamefont {Yang}},\
  and\ \bibinfo {author} {\bibfnamefont {Y.}~\bibnamefont {Xia}},\ }\bibfield
  {title} {\bibinfo {title} {Generation of two-mode squeezed states for two
  separated atomic ensembles via coupled cavities},\ }\href
  {https://doi.org/10.1103/PhysRevA.81.015804} {\bibfield  {journal} {\bibinfo
  {journal} {Phys. Rev. A}\ }\textbf {\bibinfo {volume} {81}},\ \bibinfo
  {pages} {015804} (\bibinfo {year} {2010})}\BibitemShut {NoStop}%
\bibitem [{\citenamefont {Cooper}\ \emph {et~al.}(2024)\citenamefont {Cooper},
  \citenamefont {Kunkel}, \citenamefont {Periwal},\ and\ \citenamefont
  {Schleier-Smith}}]{cooper2024multimodeExpt}%
  \BibitemOpen
  \bibfield  {author} {\bibinfo {author} {\bibfnamefont {E.~S.}\ \bibnamefont
  {Cooper}}, \bibinfo {author} {\bibfnamefont {P.}~\bibnamefont {Kunkel}},
  \bibinfo {author} {\bibfnamefont {A.}~\bibnamefont {Periwal}},\ and\ \bibinfo
  {author} {\bibfnamefont {M.}~\bibnamefont {Schleier-Smith}},\ }\bibfield
  {title} {\bibinfo {title} {Graph states of atomic ensembles engineered by
  photon-mediated entanglement},\ }\href
  {https://doi.org/10.1038/s41567-024-02407-1} {\bibfield  {journal} {\bibinfo
  {journal} {Nature Physics}\ ,\ \bibinfo {pages} {1}} (\bibinfo {year}
  {2024})}\BibitemShut {NoStop}%
\bibitem [{\citenamefont {Hamley}\ \emph {et~al.}(2012)\citenamefont {Hamley},
  \citenamefont {Gerving}, \citenamefont {Hoang}, \citenamefont {Bookjans},\
  and\ \citenamefont {Chapman}}]{hamley2012spinNematic}%
  \BibitemOpen
  \bibfield  {author} {\bibinfo {author} {\bibfnamefont {C.~D.}\ \bibnamefont
  {Hamley}}, \bibinfo {author} {\bibfnamefont {C.}~\bibnamefont {Gerving}},
  \bibinfo {author} {\bibfnamefont {T.~M.}\ \bibnamefont {Hoang}}, \bibinfo
  {author} {\bibfnamefont {E.~M.}\ \bibnamefont {Bookjans}},\ and\ \bibinfo
  {author} {\bibfnamefont {M.~S.}\ \bibnamefont {Chapman}},\ }\bibfield
  {title} {\bibinfo {title} {Spin-nematic squeezed vacuum in a quantum gas},\
  }\href@noop {} {\bibfield  {journal} {\bibinfo  {journal} {Nature Physics}\
  }\textbf {\bibinfo {volume} {8}},\ \bibinfo {pages} {305} (\bibinfo {year}
  {2012})}\BibitemShut {NoStop}%
\bibitem [{\citenamefont {Kunkel}\ \emph {et~al.}(2019)\citenamefont {Kunkel},
  \citenamefont {Pr{\"u}fer}, \citenamefont {Lannig}, \citenamefont
  {Rosa-Medina}, \citenamefont {Bonnin}, \citenamefont {G{\"a}rttner},
  \citenamefont {Strobel},\ and\ \citenamefont
  {Oberthaler}}]{kunkel2019spinNematic}%
  \BibitemOpen
  \bibfield  {author} {\bibinfo {author} {\bibfnamefont {P.}~\bibnamefont
  {Kunkel}}, \bibinfo {author} {\bibfnamefont {M.}~\bibnamefont {Pr{\"u}fer}},
  \bibinfo {author} {\bibfnamefont {S.}~\bibnamefont {Lannig}}, \bibinfo
  {author} {\bibfnamefont {R.}~\bibnamefont {Rosa-Medina}}, \bibinfo {author}
  {\bibfnamefont {A.}~\bibnamefont {Bonnin}}, \bibinfo {author} {\bibfnamefont
  {M.}~\bibnamefont {G{\"a}rttner}}, \bibinfo {author} {\bibfnamefont
  {H.}~\bibnamefont {Strobel}},\ and\ \bibinfo {author} {\bibfnamefont {M.~K.}\
  \bibnamefont {Oberthaler}},\ }\bibfield  {title} {\bibinfo {title}
  {Simultaneous readout of noncommuting collective spin observables beyond the
  standard quantum limit},\ }\href@noop {} {\bibfield  {journal} {\bibinfo
  {journal} {Physical review letters}\ }\textbf {\bibinfo {volume} {123}},\
  \bibinfo {pages} {063603} (\bibinfo {year} {2019})}\BibitemShut {NoStop}%
\bibitem [{\citenamefont {Cao}\ \emph {et~al.}(2023)\citenamefont {Cao},
  \citenamefont {Li}, \citenamefont {Mao}, \citenamefont {Xu},\ and\
  \citenamefont {You}}]{cao2023spinNematic}%
  \BibitemOpen
  \bibfield  {author} {\bibinfo {author} {\bibfnamefont {J.}~\bibnamefont
  {Cao}}, \bibinfo {author} {\bibfnamefont {X.}~\bibnamefont {Li}}, \bibinfo
  {author} {\bibfnamefont {T.}~\bibnamefont {Mao}}, \bibinfo {author}
  {\bibfnamefont {W.}~\bibnamefont {Xu}},\ and\ \bibinfo {author}
  {\bibfnamefont {L.}~\bibnamefont {You}},\ }\bibfield  {title} {\bibinfo
  {title} {Joint estimation of a two-phase spin rotation beyond classical
  limit},\ }\href@noop {} {\bibfield  {journal} {\bibinfo  {journal} {arXiv
  preprint arXiv:2312.10480}\ } (\bibinfo {year} {2023})}\BibitemShut {NoStop}%
\bibitem [{\citenamefont {Kubo}(1962)}]{kubo1962generalizedcumulant}%
  \BibitemOpen
  \bibfield  {author} {\bibinfo {author} {\bibfnamefont {R.}~\bibnamefont
  {Kubo}},\ }\bibfield  {title} {\bibinfo {title} {Generalized cumulant
  expansion method},\ }\href {https://doi.org/10.1143/JPSJ.17.1100} {\bibfield
  {journal} {\bibinfo  {journal} {Journal of the Physical Society of Japan}\
  }\textbf {\bibinfo {volume} {17}},\ \bibinfo {pages} {1100} (\bibinfo {year}
  {1962})}\BibitemShut {NoStop}%
\bibitem [{\citenamefont {Zens}\ \emph {et~al.}(2019)\citenamefont {Zens},
  \citenamefont {Krimer},\ and\ \citenamefont
  {Rotter}}]{zens2019criticalphenomena}%
  \BibitemOpen
  \bibfield  {author} {\bibinfo {author} {\bibfnamefont {M.}~\bibnamefont
  {Zens}}, \bibinfo {author} {\bibfnamefont {D.~O.}\ \bibnamefont {Krimer}},\
  and\ \bibinfo {author} {\bibfnamefont {S.}~\bibnamefont {Rotter}},\
  }\bibfield  {title} {\bibinfo {title} {Critical phenomena and nonlinear
  dynamics in a spin ensemble strongly coupled to a cavity. ii.
  semiclassical-to-quantum boundary},\ }\href
  {https://doi.org/10.1103/PhysRevA.100.013856} {\bibfield  {journal} {\bibinfo
   {journal} {Phys. Rev. A}\ }\textbf {\bibinfo {volume} {100}},\ \bibinfo
  {pages} {013856} (\bibinfo {year} {2019})}\BibitemShut {NoStop}%
\bibitem [{\citenamefont {Bennett}\ \emph {et~al.}(2013)\citenamefont
  {Bennett}, \citenamefont {Yao}, \citenamefont {Otterbach}, \citenamefont
  {Zoller}, \citenamefont {Rabl},\ and\ \citenamefont {Lukin}}]{bennett2013}%
  \BibitemOpen
  \bibfield  {author} {\bibinfo {author} {\bibfnamefont {S.~D.}\ \bibnamefont
  {Bennett}}, \bibinfo {author} {\bibfnamefont {N.~Y.}\ \bibnamefont {Yao}},
  \bibinfo {author} {\bibfnamefont {J.}~\bibnamefont {Otterbach}}, \bibinfo
  {author} {\bibfnamefont {P.}~\bibnamefont {Zoller}}, \bibinfo {author}
  {\bibfnamefont {P.}~\bibnamefont {Rabl}},\ and\ \bibinfo {author}
  {\bibfnamefont {M.~D.}\ \bibnamefont {Lukin}},\ }\bibfield  {title} {\bibinfo
  {title} {Phonon-induced spin-spin interactions in diamond nanostructures:
  Application to spin squeezing},\ }\href
  {https://doi.org/10.1103/PhysRevLett.110.156402} {\bibfield  {journal}
  {\bibinfo  {journal} {Phys. Rev. Lett.}\ }\textbf {\bibinfo {volume} {110}},\
  \bibinfo {pages} {156402} (\bibinfo {year} {2013})}\BibitemShut {NoStop}%
\bibitem [{\citenamefont {Lewis-Swan}\ \emph {et~al.}(2018)\citenamefont
  {Lewis-Swan}, \citenamefont {Norcia}, \citenamefont {Cline}, \citenamefont
  {Thompson},\ and\ \citenamefont {Rey}}]{lewisswan2018}%
  \BibitemOpen
  \bibfield  {author} {\bibinfo {author} {\bibfnamefont {R.~J.}\ \bibnamefont
  {Lewis-Swan}}, \bibinfo {author} {\bibfnamefont {M.~A.}\ \bibnamefont
  {Norcia}}, \bibinfo {author} {\bibfnamefont {J.~R.~K.}\ \bibnamefont
  {Cline}}, \bibinfo {author} {\bibfnamefont {J.~K.}\ \bibnamefont
  {Thompson}},\ and\ \bibinfo {author} {\bibfnamefont {A.~M.}\ \bibnamefont
  {Rey}},\ }\bibfield  {title} {\bibinfo {title} {Robust spin squeezing via
  photon-mediated interactions on an optical clock transition},\ }\href
  {https://doi.org/10.1103/PhysRevLett.121.070403} {\bibfield  {journal}
  {\bibinfo  {journal} {Phys. Rev. Lett.}\ }\textbf {\bibinfo {volume} {121}},\
  \bibinfo {pages} {070403} (\bibinfo {year} {2018})}\BibitemShut {NoStop}%
\bibitem [{\citenamefont {Leghtas}\ \emph {et~al.}(2015)\citenamefont
  {Leghtas}, \citenamefont {Touzard}, \citenamefont {Pop}, \citenamefont {Kou},
  \citenamefont {Vlastakis}, \citenamefont {Petrenko}, \citenamefont {Sliwa},
  \citenamefont {Narla}, \citenamefont {Shankar}, \citenamefont {Hatridge}
  \emph {et~al.}}]{leghtas2015yaleTwoPhoton}%
  \BibitemOpen
  \bibfield  {author} {\bibinfo {author} {\bibfnamefont {Z.}~\bibnamefont
  {Leghtas}}, \bibinfo {author} {\bibfnamefont {S.}~\bibnamefont {Touzard}},
  \bibinfo {author} {\bibfnamefont {I.~M.}\ \bibnamefont {Pop}}, \bibinfo
  {author} {\bibfnamefont {A.}~\bibnamefont {Kou}}, \bibinfo {author}
  {\bibfnamefont {B.}~\bibnamefont {Vlastakis}}, \bibinfo {author}
  {\bibfnamefont {A.}~\bibnamefont {Petrenko}}, \bibinfo {author}
  {\bibfnamefont {K.~M.}\ \bibnamefont {Sliwa}}, \bibinfo {author}
  {\bibfnamefont {A.}~\bibnamefont {Narla}}, \bibinfo {author} {\bibfnamefont
  {S.}~\bibnamefont {Shankar}}, \bibinfo {author} {\bibfnamefont {M.~J.}\
  \bibnamefont {Hatridge}}, \emph {et~al.},\ }\bibfield  {title} {\bibinfo
  {title} {Confining the state of light to a quantum manifold by engineered
  two-photon loss},\ }\href
  {http://science.sciencemag.org/content/347/6224/853} {\bibfield  {journal}
  {\bibinfo  {journal} {Science}\ }\textbf {\bibinfo {volume} {347}},\ \bibinfo
  {pages} {853} (\bibinfo {year} {2015})}\BibitemShut {NoStop}%
\bibitem [{\citenamefont {R{\'e}glade}\ \emph {et~al.}(2024)\citenamefont
  {R{\'e}glade}, \citenamefont {Bocquet}, \citenamefont {Gautier},
  \citenamefont {Cohen}, \citenamefont {Marquet}, \citenamefont {Albertinale},
  \citenamefont {Pankratova}, \citenamefont {Hall{\'e}n}, \citenamefont
  {Rautschke}, \citenamefont {Sellem} \emph {et~al.}}]{reglade2024twoPhoton}%
  \BibitemOpen
  \bibfield  {author} {\bibinfo {author} {\bibfnamefont {U.}~\bibnamefont
  {R{\'e}glade}}, \bibinfo {author} {\bibfnamefont {A.}~\bibnamefont
  {Bocquet}}, \bibinfo {author} {\bibfnamefont {R.}~\bibnamefont {Gautier}},
  \bibinfo {author} {\bibfnamefont {J.}~\bibnamefont {Cohen}}, \bibinfo
  {author} {\bibfnamefont {A.}~\bibnamefont {Marquet}}, \bibinfo {author}
  {\bibfnamefont {E.}~\bibnamefont {Albertinale}}, \bibinfo {author}
  {\bibfnamefont {N.}~\bibnamefont {Pankratova}}, \bibinfo {author}
  {\bibfnamefont {M.}~\bibnamefont {Hall{\'e}n}}, \bibinfo {author}
  {\bibfnamefont {F.}~\bibnamefont {Rautschke}}, \bibinfo {author}
  {\bibfnamefont {L.-A.}\ \bibnamefont {Sellem}}, \emph {et~al.},\ }\bibfield
  {title} {\bibinfo {title} {Quantum control of a cat qubit with bit-flip times
  exceeding ten seconds},\ }\href@noop {} {\bibfield  {journal} {\bibinfo
  {journal} {Nature}\ ,\ \bibinfo {pages} {1}} (\bibinfo {year}
  {2024})}\BibitemShut {NoStop}%
\bibitem [{\citenamefont {Tsang}\ and\ \citenamefont
  {Caves}(2012)}]{TsangQMFS2012}%
  \BibitemOpen
  \bibfield  {author} {\bibinfo {author} {\bibfnamefont {M.}~\bibnamefont
  {Tsang}}\ and\ \bibinfo {author} {\bibfnamefont {C.~M.}\ \bibnamefont
  {Caves}},\ }\bibfield  {title} {\bibinfo {title} {Evading quantum mechanics:
  Engineering a classical subsystem within a quantum environment},\ }\href
  {https://doi.org/10.1103/PhysRevX.2.031016} {\bibfield  {journal} {\bibinfo
  {journal} {Phys. Rev. X}\ }\textbf {\bibinfo {volume} {2}},\ \bibinfo {pages}
  {031016} (\bibinfo {year} {2012})}\BibitemShut {NoStop}%
\bibitem [{\citenamefont {Polzik}\ and\ \citenamefont
  {Hammerer}(2015)}]{Polzik2015}%
  \BibitemOpen
  \bibfield  {author} {\bibinfo {author} {\bibfnamefont {E.~S.}\ \bibnamefont
  {Polzik}}\ and\ \bibinfo {author} {\bibfnamefont {K.}~\bibnamefont
  {Hammerer}},\ }\bibfield  {title} {\bibinfo {title} {Trajectories without
  quantum uncertainties},\ }\href
  {https://doi.org/https://doi.org/10.1002/andp.201400099} {\bibfield
  {journal} {\bibinfo  {journal} {Ann. der Physik}\ }\textbf {\bibinfo {volume}
  {527}},\ \bibinfo {pages} {A15} (\bibinfo {year} {2015})}\BibitemShut
  {NoStop}%
\bibitem [{\citenamefont {M{\o}ller}\ \emph {et~al.}(2017)\citenamefont
  {M{\o}ller}, \citenamefont {Thomas}, \citenamefont {Vasilakis}, \citenamefont
  {Zeuthen}, \citenamefont {Tsaturyan}, \citenamefont {Balabas}, \citenamefont
  {Jensen}, \citenamefont {Schliesser}, \citenamefont {Hammerer},\ and\
  \citenamefont {Polzik}}]{PolzikNature2017}%
  \BibitemOpen
  \bibfield  {author} {\bibinfo {author} {\bibfnamefont {C.~B.}\ \bibnamefont
  {M{\o}ller}}, \bibinfo {author} {\bibfnamefont {R.~A.}\ \bibnamefont
  {Thomas}}, \bibinfo {author} {\bibfnamefont {G.}~\bibnamefont {Vasilakis}},
  \bibinfo {author} {\bibfnamefont {E.}~\bibnamefont {Zeuthen}}, \bibinfo
  {author} {\bibfnamefont {Y.}~\bibnamefont {Tsaturyan}}, \bibinfo {author}
  {\bibfnamefont {M.}~\bibnamefont {Balabas}}, \bibinfo {author} {\bibfnamefont
  {K.}~\bibnamefont {Jensen}}, \bibinfo {author} {\bibfnamefont
  {A.}~\bibnamefont {Schliesser}}, \bibinfo {author} {\bibfnamefont
  {K.}~\bibnamefont {Hammerer}},\ and\ \bibinfo {author} {\bibfnamefont
  {E.~S.}\ \bibnamefont {Polzik}},\ }\bibfield  {title} {\bibinfo {title}
  {{Quantum back-action-evading measurement of motion in a negative mass
  reference frame}},\ }\href {https://doi.org/10.1038/nature22980} {\bibfield
  {journal} {\bibinfo  {journal} {Nature}\ }\textbf {\bibinfo {volume} {547}},\
  \bibinfo {pages} {191 } (\bibinfo {year} {2017})}\BibitemShut {NoStop}%
\bibitem [{\citenamefont {de~Lépinay}\ \emph {et~al.}(2021)\citenamefont
  {de~Lépinay}, \citenamefont {Ockeloen-Korppi}, \citenamefont {Woolley},\
  and\ \citenamefont {Sillanpää}}]{Sillanpaa2021}%
  \BibitemOpen
  \bibfield  {author} {\bibinfo {author} {\bibfnamefont {L.~M.}\ \bibnamefont
  {de~Lépinay}}, \bibinfo {author} {\bibfnamefont {C.~F.}\ \bibnamefont
  {Ockeloen-Korppi}}, \bibinfo {author} {\bibfnamefont {M.~J.}\ \bibnamefont
  {Woolley}},\ and\ \bibinfo {author} {\bibfnamefont {M.~A.}\ \bibnamefont
  {Sillanpää}},\ }\bibfield  {title} {\bibinfo {title} {Quantum
  mechanics–free subsystem with mechanical oscillators},\ }\href
  {https://doi.org/10.1126/science.abf5389} {\bibfield  {journal} {\bibinfo
  {journal} {Science}\ }\textbf {\bibinfo {volume} {372}},\ \bibinfo {pages}
  {625} (\bibinfo {year} {2021})}\BibitemShut {NoStop}%
\bibitem [{\citenamefont {Didier}\ \emph {et~al.}(2015)\citenamefont {Didier},
  \citenamefont {Kamal}, \citenamefont {Oliver}, \citenamefont {Blais},\ and\
  \citenamefont {Clerk}}]{Didier2015}%
  \BibitemOpen
  \bibfield  {author} {\bibinfo {author} {\bibfnamefont {N.}~\bibnamefont
  {Didier}}, \bibinfo {author} {\bibfnamefont {A.}~\bibnamefont {Kamal}},
  \bibinfo {author} {\bibfnamefont {W.~D.}\ \bibnamefont {Oliver}}, \bibinfo
  {author} {\bibfnamefont {A.}~\bibnamefont {Blais}},\ and\ \bibinfo {author}
  {\bibfnamefont {A.~A.}\ \bibnamefont {Clerk}},\ }\bibfield  {title} {\bibinfo
  {title} {Heisenberg-limited qubit read-out with two-mode squeezed light},\
  }\href {https://doi.org/10.1103/PhysRevLett.115.093604} {\bibfield  {journal}
  {\bibinfo  {journal} {Phys. Rev. Lett.}\ }\textbf {\bibinfo {volume} {115}},\
  \bibinfo {pages} {093604} (\bibinfo {year} {2015})}\BibitemShut {NoStop}%
\bibitem [{\citenamefont {Koppenh{\"o}fer}\ \emph {et~al.}(2023)\citenamefont
  {Koppenh{\"o}fer}, \citenamefont {Groszkowski},\ and\ \citenamefont
  {Clerk}}]{koppenhofer2023amplifier}%
  \BibitemOpen
  \bibfield  {author} {\bibinfo {author} {\bibfnamefont {M.}~\bibnamefont
  {Koppenh{\"o}fer}}, \bibinfo {author} {\bibfnamefont {P.}~\bibnamefont
  {Groszkowski}},\ and\ \bibinfo {author} {\bibfnamefont {A.}~\bibnamefont
  {Clerk}},\ }\bibfield  {title} {\bibinfo {title} {Squeezed superradiance
  enables robust entanglement-enhanced metrology even with highly imperfect
  readout},\ }\href@noop {} {\bibfield  {journal} {\bibinfo  {journal}
  {Physical Review Letters}\ }\textbf {\bibinfo {volume} {131}},\ \bibinfo
  {pages} {060802} (\bibinfo {year} {2023})}\BibitemShut {NoStop}%
\bibitem [{\citenamefont {Davis}\ \emph {et~al.}(2016)\citenamefont {Davis},
  \citenamefont {Bentsen},\ and\ \citenamefont
  {Schleier-Smith}}]{davis2016twistUntwist}%
  \BibitemOpen
  \bibfield  {author} {\bibinfo {author} {\bibfnamefont {E.}~\bibnamefont
  {Davis}}, \bibinfo {author} {\bibfnamefont {G.}~\bibnamefont {Bentsen}},\
  and\ \bibinfo {author} {\bibfnamefont {M.}~\bibnamefont {Schleier-Smith}},\
  }\bibfield  {title} {\bibinfo {title} {Approaching the heisenberg limit
  without single-particle detection},\ }\href@noop {} {\bibfield  {journal}
  {\bibinfo  {journal} {Physical review letters}\ }\textbf {\bibinfo {volume}
  {116}},\ \bibinfo {pages} {053601} (\bibinfo {year} {2016})}\BibitemShut
  {NoStop}%
\bibitem [{\citenamefont {Hosten}\ \emph
  {et~al.}(2016{\natexlab{b}})\citenamefont {Hosten}, \citenamefont
  {Krishnakumar}, \citenamefont {Engelsen},\ and\ \citenamefont
  {Kasevich}}]{hosten2016magnification}%
  \BibitemOpen
  \bibfield  {author} {\bibinfo {author} {\bibfnamefont {O.}~\bibnamefont
  {Hosten}}, \bibinfo {author} {\bibfnamefont {R.}~\bibnamefont
  {Krishnakumar}}, \bibinfo {author} {\bibfnamefont {N.~J.}\ \bibnamefont
  {Engelsen}},\ and\ \bibinfo {author} {\bibfnamefont {M.~A.}\ \bibnamefont
  {Kasevich}},\ }\bibfield  {title} {\bibinfo {title} {Quantum phase
  magnification},\ }\href@noop {} {\bibfield  {journal} {\bibinfo  {journal}
  {Science}\ }\textbf {\bibinfo {volume} {352}},\ \bibinfo {pages} {1552}
  (\bibinfo {year} {2016}{\natexlab{b}})}\BibitemShut {NoStop}%
\bibitem [{\citenamefont {Braunstein}\ and\ \citenamefont
  {McLachlan}(1987)}]{braunstein1987generalizedSqueezing}%
  \BibitemOpen
  \bibfield  {author} {\bibinfo {author} {\bibfnamefont {S.~L.}\ \bibnamefont
  {Braunstein}}\ and\ \bibinfo {author} {\bibfnamefont {R.~I.}\ \bibnamefont
  {McLachlan}},\ }\bibfield  {title} {\bibinfo {title} {Generalized
  squeezing},\ }\href {https://doi.org/10.1103/PhysRevA.35.1659} {\bibfield
  {journal} {\bibinfo  {journal} {Phys. Rev. A}\ }\textbf {\bibinfo {volume}
  {35}},\ \bibinfo {pages} {1659} (\bibinfo {year} {1987})}\BibitemShut
  {NoStop}%
\end{thebibliography}%

\clearpage

\onecolumngrid

\begin{center}
\textbf{\large Supplemental Material for\\Non-Gaussian generalized two-mode squeezing: applications to two-ensemble spin squeezing and beyond}
\end{center}

\begin{center}
Mikhail Mamaev,${}^1$ Martin Koppenh\"{o}fer,${}^2$ Andrew Pocklington,${}^{3,1}$ and Aashish A. Clerk${}^1$\\
\emph{${}^1$Pritzker School of Molecular Engineering, University of Chicago, Chicago, Illinois, USA\\
${}^2$Fraunhofer Institute for Applied Solid State Physics IAF, Tullastr.~72, 79108 Freiburg, Germany\\
${}^3$Department of Physics, University of Chicago, Chicago, Illinois, USA}\\
(Dated: \today)
\end{center}

\setcounter{equation}{0}
\setcounter{figure}{0}
\setcounter{table}{0}
\setcounter{page}{1}
\makeatletter
\renewcommand{\theequation}{S\arabic{equation}}
\renewcommand{\thefigure}{S\arabic{figure}}
\renewcommand{\bibnumfmt}[1]{[S#1]}
\renewcommand{\citenumfont}[1]{#1}


\tableofcontents


\section{Calculation of the symmetric logarithmic derivatives}

In this section, we derive the explicit form of the symmetric logarithmic derivative (SLD) operators stated in Eq.~(5) of the main text. The QFIM may alternatively be written as
\begin{equation}
\mathcal{Q}_{i,j} = \frac{1}{2} \text{tr} \left[\rho_{G} \left(\hat{L}_{\hat{W}_i} \hat{L}_{\hat{W}_j} + \hat{L}_{\hat{W}_j} \hat{L}_{\hat{W}_i}\right)\right],
\end{equation}
where $\rho_{G} = \ket{\psi_G(r)}\bra{\psi_G (r)}$, and $\hat{L}_{\hat{W}}$ are the SLD operators for $\hat{W} \in \{\hat{X}_{+}, \hat{X}_{-}, \hat{Y}_{+}, \hat{Y}_{-}\}$.
The SLD operators satisfy
\begin{equation}
\hat{L}_{\hat{W}} \rho_{\hat{W}} + \rho_{\hat{W}} \hat{L}_{\hat{W}} = 2 \frac{\partial \rho_{\hat{W}}}{\partial \theta},
\end{equation}
where $\rho_{\hat{W}} = \ket{\psi_{\hat{W}}}\bra{\psi_{\hat{W}}}$ is the general squeezed state subject to a unitary rotation $\ket{\psi_{\hat{W}}} = \exp (-i \theta \hat{W})\ket{\psi_G(r)}$ in the limit $\theta \to 0$. In this limit, the density matrix is unperturbed, $\rho_{\hat{W}} = \rho_{G}$, but has a non-zero derivative $\frac{\partial \rho_{\hat{W}}}{\partial \theta} =- i \hat{W} \rho_{G}+ i \rho_{G} \hat{W}$. We must thus solve the equation,
\begin{equation}
\label{eq_SLDSupplementDef}
\hat{L}_{\hat{W}} \rho_{G} + \rho_{G} \hat{L}_{\hat{W}} = 2\left(- i \hat{W} \rho_{G} + i \rho_{G}\hat{W}\right).
\end{equation}
While the SLD operators are not unique, we show that one simple parametrization is given by the quadratures themselves up to constant prefactors,
\begin{align}
\hat{L}_{\hat{X}_{\pm}} &= 2 e^{\mp 2r} \hat{Y}_{\pm}, &
\hat{L}_{\hat{Y}_{\pm}} &= -2 e^{\pm 2r} \hat{X}_{\pm}.
\label{eqn:SM:SLDs}
\end{align}
As an example, for $\hat{L}_{\hat{X}_{+}} = 2 e^{-2r}\hat{Y}_{+}$, the first term on the left hand side of Eq.~\eqref{eq_SLDSupplementDef} is,
\begin{equation}
\begin{aligned}
\hat{L}_{\hat{X}_+} \rho_{G} &= 2 e^{-2r} \hat{Y}_{+}\rho_{G}\\
&=-i e^{-2r} \left(\hat{O}_1^{\dagger} - \hat{O}_{1} + \hat{O}_{2}^{\dagger} -\hat{O}_{2}\right)\rho_{G}\\
&=-i \mathcal{N}^2 e^{-2r} \left(\hat{O}_1^{\dagger} - \hat{O}_{2} - \hat{O}_{1} + \hat{O}_{2}^{\dagger}\right)\sum_{m,m'=0}^{m_{\mathrm{max}}}[-\tanh (r)]^{m+m'}\ket{m,m}\bra{m',m'}\\
&=-i \mathcal{N}^2 e^{-2r} \sum_{m,m'}[-\tanh(r)]^{m+m'} \bigg[o(m+1)\ket{m+1,m}- o(m)\ket{m,m-1}\\
&\quad\quad\quad\quad\quad\quad\quad\quad\quad\quad\quad\quad\quad\quad- o(m) \ket{m-1,m} + o(m+1)\ket{m,m+1}\bigg]\bra{m',m'}\\
&=-i \mathcal{N}^2 e^{-2r} \sum_{m,m'}[-\tanh(r)]^{m+m'} \bigg[\frac{o(m)}{-\tanh(r)}-o(m)\bigg]\big(\ket{m,m-1}+\ket{m-1,m}\big)\bra{m',m'}\\
&=-i \mathcal{N}^2 \sum_{m,m'}[-\tanh(r)]^{m+m'} \bigg[\frac{-e^{-2r}}{\tanh(r)}-e^{-2r}\bigg]o(m)\big(\ket{m,m-1}+\ket{m-1,m}\big)\bra{m',m'}\\
&=-i \mathcal{N}^2 \sum_{m,m'}[-\tanh(r)]^{m+m'} \bigg[\frac{-1}{\tanh(r)}+1\bigg]o(m)\big(\ket{m,m-1}+\ket{m-1,m}\big)\bra{m',m'}.
\end{aligned}
\end{equation}
We compare this to the first term on the right hand side of Eq.~\eqref{eq_SLDSupplementDef},
\begin{equation}
\begin{aligned}
-2i \hat{X}_{+}\rho_{G} &= -i \left(\hat{O_1}^{\dagger} + \hat{O}_{1} + \hat{O}_{2}^{\dagger} + \hat{O}_{2}\right)\rho_{G}\\
&= -i \mathcal{N}^2 \left(\hat{O_1}^{\dagger} + \hat{O}_{2} + \hat{O}_{1} + \hat{O}_{2}^{\dagger} \right)\sum_{m,m} [-\tanh(r)]^{m+m'}\ket{m,m}\bra{m',m'}\\
&= - i \mathcal{N}^2 \sum_{m,m'} [-\tanh(r)]^{m+m'}\bigg[o(m+1) \ket{m+1,m} + o(m) \ket{m,m-1}\\
&\quad\quad\quad\quad\quad\quad\quad\quad\quad\quad\quad\quad\quad + o(m)\ket{m-1,m} + o(m+1)\ket{m,m+1}\bigg]\bra{m',m'}\\
&=-i \mathcal{N}^2 \sum_{m,m'}[-\tanh(r)]^{m+m'} \left[\frac{-1}{\tanh(r)} + 1\right] o(m)(\ket{m,m-1} + \ket{m-1,m})\bra{m',m'},
\end{aligned}
\end{equation}
which matches the previous result, showing that $\hat{L}_{\hat{X}_{+}}\rho_{G} = -2 i \hat{X}_{+}\rho_{G}$. An analogous calculation shows that the second terms on the left- and right- hand sides of Eq.~\eqref{eq_SLDSupplementDef} also match. This procedure can be repeated for all four SLDs in the same way.

\clearpage
\section{Sample commuting SLD operators}
The previous section showed that the SLD operators are equivalent to the quadratures themselves up to constant prefactors. These SLD operators do not commute in general, therefore, it is not possible to measure the corresponding observables simultaneously.
However, in principle one can construct more complex operators that do commute and are therefore simultaneously measurable, since the commutator of the SLDs vanishes on average~\cite{liu2020multiparReview,matsumoto2002,Ragy2016,Pezze2017}. Here we give a simple demonstrative example of such a construction.

Consider each subsystem to be a single spin-1/2 ($S=1/2$). The full system Hilbert space can be written as $\{\ket{\uparrow,\uparrow},\ket{\uparrow,\downarrow},\ket{\downarrow,\uparrow},\ket{\downarrow,\downarrow}\}$. The quadrature operators (proportional to the SLD's) expressed in this basis read,
\begin{equation}
\hat{L}_{\hat{Y}_{+}} \sim \hat{X}_{+} = \frac{1}{2}\left(\begin{array}{cccc}0&1&1&0\\1&0&0&1\\1&0&0&1\\0&1&1&0 \end{array}\right),\>\>\>\>\hat{L}_{\hat{X}_{-}} \sim \hat{Y}_{-} = \frac{1}{2}\left(\begin{array}{cccc}0&i&-i&0\\-i&0&0&-i\\i&0&0&i\\0&i&-i&0 \end{array}\right).
\end{equation}
While these do not commute, we note that the two-mode squeezed state here takes the form of $\ket{\psi_G(r)}= \left(-\tanh(r),0,0,1\right)/\sqrt{1+\tanh^2(r)}$ (with weight only on the symmetric components $\ket{\uparrow,\uparrow}$, $\ket{\downarrow,\downarrow}$). Hence any additional matrix elements acting only on the subspace $\ket{\uparrow,\downarrow}$, $\ket{\downarrow,\uparrow}$ can be added to make new SLD's:
\begin{equation}
\hat{L}^{'}_{\hat{Y}_{+}} \sim \frac{1}{2}\left(\begin{array}{cccc}0&1&1&0\\1&a_{11}&a_{12}&1\\1&a_{21}&a_{22}&1\\0&1&1&0 \end{array}\right), \>\>\>\hat{L}^{'}_{\hat{X}_{-}} \sim \frac{1}{2}\left(\begin{array}{cccc}0&i&-i&0\\-i&b_{11}&b_{12}&-i\\i&b_{21}&b_{22}&i\\0&i&-i&0 \end{array}\right),
\end{equation}
where the coefficients $a_{ij}$, $b_{ij}$ are arbitrary (aside from the operators needing to remain Hermitian).

For this small-dimensional system, it is straightforward to find a set of coefficients $a_{11}=a_{22}=b_{11}=b_{22}=0$, $a_{12}=a_{21}=\sqrt{2}$, $b_{12}=-b_{21}=\sqrt{2}i$ for which the new SLD operators commute, $[\hat{L}_{\hat{Y}_+}^{'},\hat{L}_{\hat{X}_-}^{'}] = 0$, and can thus be measured independently without disturbing each other. These new operators correspond to,
\begin{equation}
\hat{L}_{\hat{Y}_{+}}^{'} \sim \hat{X}_{+} + \frac{1}{\sqrt{2}}\left(\hat{O}_{1}^{\dagger}\hat{O}_{2}+h.c.\right),\>\>\>\>\hat{L}_{\hat{X}_{-}}^{'} \sim \hat{Y}_{-} + \frac{i}{\sqrt{2}}\left(\hat{O}_{1}^{\dagger}\hat{O}_{2}- h.c.\right).
\end{equation}
However, these new SLDs are now non-local. Extending to larger-sized spins or more arbitrary system structures can lead to SLDs that are challenging to even construct, let alone measure. We discuss an alternate approach using the original SLDs proportional to the quadratures in the next section.

\clearpage
\section{Simultaneous two-parameter estimation saturating QCRB}

As mentioned in the main text, the simplest case enabling simultaneous estimation of two parameters with an error saturating the QCRB is when the corresponding SLDs commute.   In this case, the joint estimation just involves simultaneous measurement of the two SLDs.  Joint estimation of both parameters at the QCRB is also possible when a weaker condition is met:  the corresponding SLDs can fail to commute, but the expectation value of the commutator should vanish in the sensing state \cite{liu2020multiparReview,matsumoto2002,Ragy2016,Pezze2017}.  
In this case however, the optimal measurements are less straightforward.  One needs to find two new, equivalent SLDs that commute with one another, and then measure these operators.  While in principle this is always possible, in practice the new commuting SLD operators may correspond to operators that are far more complex than the original SLDs, and hence correspond to measurements that are difficult to implement (e.g., while the original SLD operators are sums of single-spin operators, the new SLDs could involve high-weight operators).  

Here, we show that for our generalized TMSS (realized using two spin ensembles), one can do extremely well by simply measuring the original SLD operators (which here are simple collective spin variables).  One can simultaneously extract the two parameters encoded in $\hat{X}_+$ and $\hat{Y}_-$ with an estimation error that exhibits Heisenberg scaling, 
and which misses the fundamental QCRB bound by just a prefactor of the order of unity.  The origin of this  remarkable result is that, in our case, the commutator of the SLDs is a non-zero operator, but the sensing state is in its null space.  This implies that all powers of the commutator have zero expectation value in the sensing state.  Hence, while there is non-zero measurement backaction, its impact is minimal.

We consider an initial state of our two spin ensembles (with $N$ spins total) to be a generalized two-mode squeezed state (GTMSS)    which has been imprinted by the two infinitesimal parameters $\theta_A, \theta_B$ of interest:
\begin{align}
    \ket{\psi_{\rm spins}(0)} = e^{-i (\theta_{A} \hat{Y}_{+} + \theta_{B}\hat{X}_{-})}\ket{\psi_{G}(r)}.
\end{align}
We imagine first making a variable strength measurement of $X_+$ to extract the parameter $\theta_A$.  The goal is to have this measurement be strong enough to still have Heisenberg scaling of the estimation error, without being so strong that the measurement backaction degrades a subsequent measurement of $Y_{-}$.  
To model the finite-strength measurement of $\hat{X}_+$, we consider a simple model of an ideal detector:  an infinitely-heavy free test mass with position (momentum) operator $\hat{q}$ ($\hat{p}$), which is coupled to the spin ensembles via
\begin{align}
    \hat{H}_\mathrm{int} = - \lambda\hat{X}_+ \hat{q}.
\end{align}
We imagine that the detector mass starts in a state described by a Gaussian wavefunction $\phi(q)$ with zero mean position and momentum, and with position variance $\sigma_q$, $\phi(q) = (2 \pi \sigma_q)^{-1/4} \exp \left(-q^2 / 4 \sigma_q \right)$.
The corresponding momentum variance is then (setting $\hbar=1$) $\sigma_p = 1 / (2 \sigma_q)$.  There is no initial entanglement between the detector mass and the spins.      

 The basic idea of the measurement is that the momentum of the detector will be displaced by an amount proportional to the collective spin variables of interest, i.e.
\begin{align}
    \frac{\mathrm{d}}{\mathrm{d}t} \hat{p} = \lambda \hat{X}_+.
\end{align}
A final measurement of $p$ then allows one to infer the value of $\hat{X}_+$.  
Assuming an evolution time $T$, 
and letting $\langle \langle \hat{A}^2 \rangle \rangle$ denote the variance of $\hat{A}$, 
the signal-to-noise ratio associated with our measurement is
\begin{align}
    \left( \mathrm{SNR} \right)_{X_+} & \equiv 
        \frac{ \langle \hat{p}(T) \rangle^2 } { \langle \langle \hat{p}(T)^2 \rangle \rangle} 
        =
        \frac{ \left(  \lambda T \langle \hat{X}_+(0) \rangle \right)^2 }
        {\sigma_p + 
         (\lambda T)^2 \langle \langle \hat{X}_+^2(0) \rangle \rangle}
         \equiv
       \frac{ \left(  \langle \hat{X}_+(0) \rangle \right)^2 }
        {\langle \langle \hat{X}_+^2 \rangle \rangle_{\rm imp} + 
         \langle \langle \hat{X}_+^2(0) \rangle \rangle}.         
        \label{eq:SNR}
\end{align}
Here, we have defined the added noise of the measurement (the imprecision noise) to be:
\begin{equation}
    \langle \langle X_+^2 \rangle \rangle_{\rm imp} = 
        \frac{\sigma_p}{\lambda^2 T^2}
    = \frac{1}{2 \lambda^2 T^2 \sigma_q} \equiv \frac{1}{2 \Lambda}.
    \label{eq:XImprecision}
\end{equation}
where we introduce the parameter $\Lambda$ to denote the effective measurement strength. A strong measurement is one where the measurement strength $\Lambda \gg 1 / \langle \langle \hat{X}_+^2 \rangle \rangle$, implying that the estimation error will be only limited by the intrinsic fluctuations in the state $|\psi_G(r)\rangle$.  

We next characterize how this measurement of $X_+$ degrades the information on 
$\theta_B$ encoded in $Y_{-}$.  The relevant quantity here is the signal to noise ratio associated with $Y_{-}$ (which is simply proportional to the inverse of the Wineland parameter $\xi^2$):
\begin{equation}
    ( \mathrm{SNR} )_{Y_{-}}(t) \equiv 
        \frac{ \left ( \langle \hat{Y}_{-}(t) \rangle \right)^2}
        {\langle \langle \hat{Y}_{-}^2(t) \rangle \rangle}.
        \label{eq:SNRY}
\end{equation}
We would like to see how this SNR is reduced at time $T$ (after the first measurement of $X_+$) compared to its initial value at time $t=0$.  Note that as we are interested in infinitesimal (local) parameter sensing, we only need this quantity to order $\theta^2$, and the numerator is already necessarily $\theta^2$.  

To that end, note that, at the end of the measurement, the total state of the detector plus spin ensemble can be written simply as:
\begin{equation}
    |\psi_{\rm tot}(T)\rangle = 
        \int_{-\infty}^{\infty} dq \, \phi(q) |q \rangle \otimes
        \exp\left(i \lambda q \hat{X}_+ T \right) |\psi_{\rm spins} \rangle  .
\end{equation}
The backation of the measurement on the spins is thus easy to describe: both spin ensembles experience a random rotation about the $X$ axis by an angle $\lambda q T$, where $q$ is a zero mean Gaussian random variable with variance $\sigma_q$.  We are interested in how this random rotation impacts $\hat{Y}_{-}$.  This rotation simply rotates $\hat Y_{-}$ in the $\hat Y_{-}$-$\hat Z_{-}$ plane, i.e.:
\begin{equation}
    \exp\left(-i \theta \hat{X}_+ \right) \hat{Y}_{-} \exp\left(i \theta \hat{X}_+ \right) = 
        \cos \theta \hat{Y}_{-} + \sin \theta \hat{Z}_{-}.
\end{equation}
Using this and the fact that $\langle \hat{Z}_{-} \rangle$ vanishes to order $\theta_A, \theta_B$ in the state $|\psi_{\rm spins}(0) \rangle$, we find that to order $\theta_B$:
\begin{equation}
    \langle \hat{Y}_{-}(T) \rangle = 
        \langle \hat{Y}_{-}(0) \rangle \int dq |\phi(q)|^2 
            \cos(\lambda q T) =  
            \langle \hat{Y}_{-}(0) \rangle e^{-\Lambda/2},
\end{equation}
where $\Lambda$ is the previously introduced measurement strength.  
We can calculate the variance of $\hat{Y}_{-}$ in a similar manner; we only require this to zeroth order in $\theta_A, \theta_B$.  
\begin{equation}
    \langle \hat{Y}_{-}(T)^2 \rangle = 
        \langle \hat{Y}_{-}^2(0) \rangle \int dq |\phi(q)|^2 
            \cos^2(\lambda q T) =  
            \langle \hat{Y}_{-}^2(0) \rangle \cosh \Lambda e^{-\Lambda}.
\end{equation}

Combining these results, we have:
\begin{equation}
    \left( \mathrm{SNR} \right)_{Y_{-}}(T) = 
    \left( \mathrm{SNR} \right)_{Y_{-}}(0) \frac{1}{\cosh \Lambda} .
    \label{eqn:SNRYT}
\end{equation}
Hence, there is indeed a backaction effect: as we increase the strength $\Lambda$ of the first $X_+$ measurement, the SNR associated with a subsequent measurement of $Y_{-}$ is exponentially suppressed.  

Despite this exponential suppression, the backaction effect here is in fact mild enough to allow estimation of both $\theta_A$ and $\theta_B$ at the Heisenberg limit.  To see this, we first constrain the measurement strength $\Lambda$ to be a constant $\Lambda_0$ of the order of unity that is independent of the the number of spins $N$.  It follows from Eq.~(\ref{eqn:SNRYT}) that the measurement is weak enough that the Wineland parameter associated with a $Y_{-}$ measurement will only be enhanced by
an $N$-independent constant prefactor $\cosh \Lambda_0$.  With this choice, the backaction of our $X_{+}$ measurement is weak enough that the scaling of the $Y_{-}$ Wineland parameter with $N$ is not changed.  

While a $\Lambda = \Lambda_0$ measurement strength is sufficient to have a minimal backaction effect on our $Y_{-}$ measurement, the question remains whether it is too weak to enable a good estimation of $X_{+}$ and hence $\theta_A$.  Turning to 
Eq.~(\ref{eq:XImprecision}), we find that the imprecision noise of the $X_+$ measurement is a constant of the order of unity,
$\langle \langle \hat{X}_+^2 \rangle \rangle_{\rm imp} = 1/2 \Lambda_0$.  As such, from Eq.~(\ref{eq:SNR}), there is a limit to how much squeezing we can usefully employ:  there is no point in squeezing $X_+$ below $\langle \langle \hat{X}_+^2 \rangle \rangle \sim 1/2 \Lambda_0$, i.e., the level of the imprecision noise.  

For concreteness, consider a squeezing parameter $r$ chosen to scale with $N$ as
\begin{equation}
    \exp(2r) = \frac{N}{4}.
    \label{eq:rSpecial}
\end{equation}
For large $N$, and using Eq.~(4) of the main text, this yields a GTMSS state whose squeezed quadrature variances are $N$-independent constants:
\begin{equation}
    \langle \langle \hat{X}_+^2 \rangle \rangle = 
    \langle \langle \hat{Y}_-^2 \rangle \rangle = \frac{1}{2 C_0}, \quad
    C_0 \simeq 0.667.
    \label{eq:SqzVarianceChoice}
\end{equation}
For this choice of $r$, and for large $N$, the Wineland parameter [c.f. Eq.~(8) of the main text] is found to be 
\begin{equation}
    \xi^2_X = \xi^2_Y = \frac{C_1}{N}.
\end{equation}
where the constant $C_1 \simeq 5.33$.  
Crucially, the value of $r$ given in Eq.~(\ref{eq:rSpecial}) is enough to permit Heisenberg scaling of the Wineland parameters.  

We now also set the measurement strength so that the imprecision noise of the $X_+$ measurement is equal to the value of the squeezed variances in Eq.~(\ref{eq:SqzVarianceChoice}), i.e., we take
$\Lambda_0 = C_0 \simeq 0.667$.  
We stress that this choice of measurement strength is independent of $N$.  
With this choice of measurement strength, the scaling of the estimation error (i.e., Wineland parameters) including the imprecision noise and backaction of the first measurement are only modified by a prefactor.  For the estimation of $\theta_A$ via measurement of $X_+$, our choice of a finite measurement strength $\Lambda_0$ causes the measurement imprecision to double the effective variance of $X_+$, and hence doubles the Wineland parameter:
\begin{equation}
    \xi^2_{X_+} \rightarrow \frac{2 C_1}{N}.
\end{equation}
Similarly, our choice of the measurement strength $\Lambda$ has a backaction that degrades the SNR associated with $Y_{-}$ (c.f.~Eq.~(\ref{eq:SNRY})), implying that the corresponding Wineland parameter is also enhanced by a constant:
\begin{equation}
      \xi^2_{Y_{-}} \rightarrow \frac{(\cosh \Lambda_0) C_1}{N} \simeq \frac{1.23 C_1}{N}.
\end{equation}

We thus have our final result:  by picking a finite strength first measurement of $X_+$ and a finite squeezing strength $r$, we can achieve Heisenberg-limited, $1/N$ scaling simultaneously for the estimation errors of the parameters $\theta_A$ and $\theta_B$.  There is a non-zero backaction effect due to the commutator of $X_+$ and $Y_-$ being non-zero, but in our system this is not strong enough to preclude Heisenberg limited scaling.  

\clearpage
\section{Comparison to unitary spin-squeezing protocols}

\begin{figure}[h]
\center
\includegraphics[width=0.5\columnwidth]{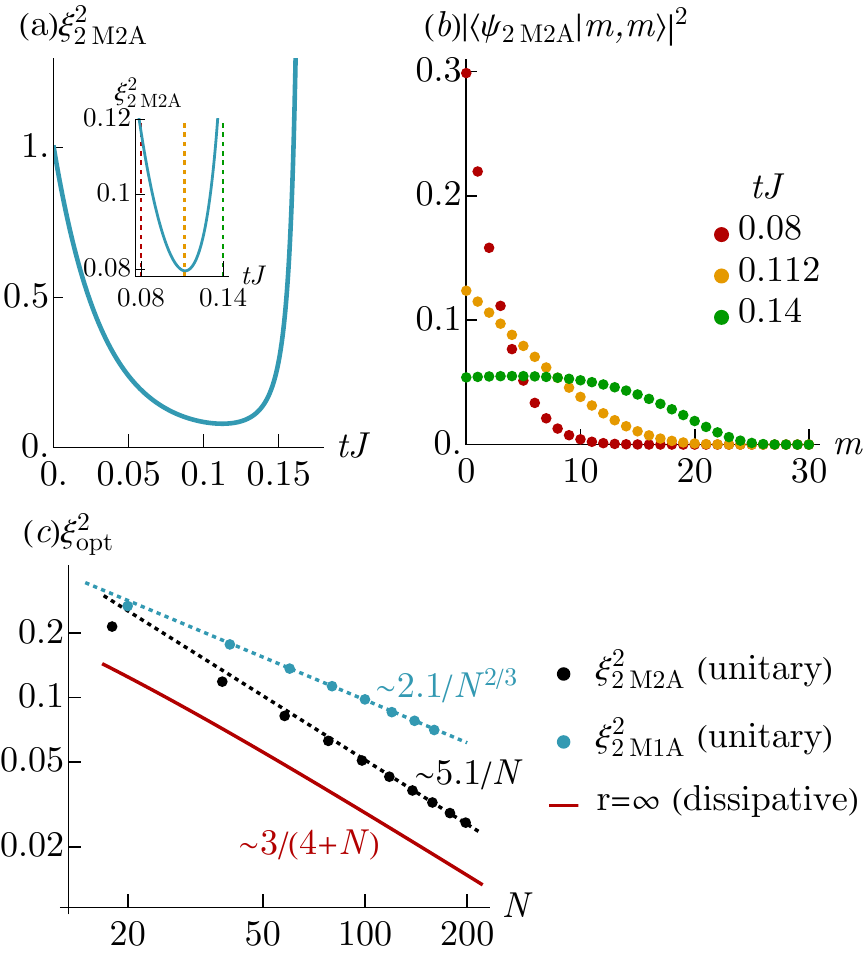}
\caption{(a) Time-evolution of squeezing generated by the unitary two-mode two-axis (2M2A) protocol in the equally-squeezed operators $(\hat{X}_{+} + \hat{Y}_{+})/\sqrt{2}$ and $(\hat{X}_{-} - \hat{Y}_{-})/\sqrt{2}$ using spin operators with fixed size $S=15$, hence $m_{\mathrm{max}}=30$. (b) Wavefunction coefficients of the pure state for times before, at, and after the optimal time $t_{\mathrm{opt}}$ at which squeezing is highest. (c) Comparison of optimal squeezing for the unitary protocol (dots) and general squeezed state $\ket{\psi_{G}(r)}$ for $r \to \infty$, realizeable via dissipative stabilization (solid line).}
\label{fig_Unitary}
\end{figure}

Here, we consider two-mode spin-squeezed states generated by a typical unitary protocol using time-evolution under an entangling Hamiltonian.  The simplest candidate interaction for comparison is a two-mode one-axis twisting Hamiltonian (2M1A),
\begin{equation}
\hat{H}_{\mathrm{2M1A}} = J \hat{X}_{1}\hat{X}_{2}.
\end{equation}
which can generate squeezing when applied to an initial state $\ket{\psi_{0}}=\ket{0,0}$. Evolving under $\hat{H}_{\mathrm{2M1A}}$ generates (equal) squeezing in two combinations of quadratures $\sin(\theta) \hat{X}_{+} + \cos(\theta)\hat{Y}_{+}$ and $\sin(\theta)\hat{X}_{-}-\cos(\theta)\hat{Y}_{-}$, where one must optimize over all $\theta$ to find the lowest variance. This yields a (two-mode) squeezing parameter $\xi^2_{\mathrm{2M1A}} = N \text{min}_{\theta} \langle[\sin(\theta) \hat{X}_{+} + \cos(\theta)\hat{Y}_{+}]^2\rangle/ (\langle\hat{Z}_1\rangle + \langle\hat{Z}_2\rangle)^2$. Another candidate interaction is a two-mode two-axis twisting Hamiltonian (2M2A),
\begin{equation}
\hat{H}_{\mathrm{2M2A}} = J \left(\hat{X}_{1}\hat{X}_{2} - \hat{Y}_{1}\hat{Y}_{2}\right),
\end{equation}
which can also generate squeezing when applied to the initial state $\ket{\psi_{0}}$; in this case the squeezed quadratures are the combined operators $(\hat{X}_{+} + \hat{Y}_{+})/\sqrt{2}$ and $(\hat{X}_{-} - \hat{Y}_{-})/\sqrt{2}$, while the squeezing parameter is $\xi^2_{\mathrm{2M2A}} = \frac{N}{2} \langle(\hat{X}_{+} + \hat{Y}_{+})^2\rangle / (\langle\hat{Z}_1\rangle + \langle\hat{Z}_2\rangle)^2$. In Fig.~\ref{fig_Unitary}(a) we show a sample time-evolution profile of the squeezing for this latter Hamiltonian.

Any (pure) squeezed state $\ket{\psi_{\mathrm{2M2A}}} = e^{-i t \hat{H}_{\mathrm{2M2A}}}\ket{\psi_0}$ generated by the 2M2A unitary protocol has non-zero matrix elements only for wavefunction components of equal excitation number $\ket{m,m}$ by symmetry. Figure~\ref{fig_Unitary}(b) shows these wavefunction components before, at, and after the optimal squeezing time. Main text Fig.2(b) shows these coefficients at the optimal time.

In Fig.~\ref{fig_Unitary}(c) we compare the optimal squeezing generated by the unitary evolution under both 2M1A and 2M2A unitary protocols to that of the general squeezed state $\ket{\psi_{G}(r)}$ obtainable via dissipative stabilization. The 2M1A scales as $(\xi^2_{\mathrm{2M2A}})_{\mathrm{opt}}\sim 2.1 /N^{2/3}$, analogous to the single-mode one-axis twisting result. The 2M2A scales as $(\xi^2_{\mathrm{2M2A}})_{\mathrm{opt}}\sim 5.1 /N$ (numerically fitted). The dissipative protocol exhibits the analytically-computed optimal scaling $\xi^2 \sim 3/N$ in the limit $r \to \infty$.

\clearpage
\section{Quantum Fisher information optimization for symmetric states}

Here, we discuss the computation of the optimal QFI of states optimized for measurements of two equally-sensitive operators. We consider pure states $\ket{\psi}$, for which the diagonal QFI matrix elements are directly proportional to the variance of the state:
\begin{equation}
\mathcal{Q}_{i,i} = 4 \bra{\psi} \hat{W}_{i}^2 \ket{\psi} - 4 (\bra{\psi}\hat{W}_{i}\ket{\psi})^2.
\end{equation}
We are interested in symmetric states of the form
\begin{equation}
\ket{\psi} = \sum_{m=0}^{m_{\mathrm{max}}} a_m \ket{m,m},    
\end{equation}
with arbitrary normalized complex coefficients $a_m$ for a fixed $m_{\mathrm{max}}$. For such states, off-diagonal matrix elements of the QFIM $\mathcal{Q}_{i,j}$ for $i \neq j$ are guaranteed to vanish due to the symmetry between the two ensembles. We can evaluate the diagonal matrix elements explicitly. For example, for $\hat{W}_{i} = \hat{X}_{-}$ we have:
\begin{equation}
\begin{aligned}
\mathcal{Q}_{i,i} = &\sum_{m,m'}a_{m}^{*}a_{m'} \bra{m,m} \left(\hat{O}_{1}^{\dagger} + \hat{O}_{1} - \hat{O}_{2}^{\dagger}-\hat{O}_{2}\right)^2\ket{m',m'} \\
- &\sum_{m,m'}a_{m}^{*}a_{m'} \left[\bra{m,m} \left(\hat{O}_{1}^{\dagger} + \hat{O}_{1} - \hat{O}_{2}^{\dagger}-\hat{O}_{2}\right)\ket{m',m'}\right]^2.
\end{aligned}
\end{equation}
The second line vanishes, while the first simplifies to,
\begin{equation}
\begin{aligned}
\mathcal{Q}_{i,i} &= \sum_{m,m'}a_{m}^{*}a_{m'} \bra{m,m} \left[\hat{O}_{1}^{\dagger}\hat{O}_{1}+\hat{O}_{1}\hat{O}_{1}^{\dagger} + \hat{O}_{2}^{\dagger}\hat{O}_{2} + \hat{O}_{2}\hat{O}_{2}^{\dagger}-2\hat{O}_1 \hat{O}_2 -2 \hat{O}_1^{\dagger}\hat{O}_2^{\dagger}\right]\ket{m',m'}\\
&=2\sum_{m}\big[|a_m|^2 o^2(m) + |a_{m+1}|^2 o^2(m+1) - a_{m}^{*}a_{m+1}o^2(m+1) - a_{m}^{*}a_{m-1}o^2(m)\big]\\
&=2\sum_{m} |a_{m}-a_{m-1}|^2 o^2(m).
\end{aligned}
\end{equation}
The other operators are computed analogously.

We can optimize the above expression over all possible $a_m$. For spin-$S$ ensembles with $m_{\mathrm{max}}=2S$ and $o(m)=\sqrt{S(S+1)-(m-S)(m-S-1)}$, the coefficients $a_m$ of the state with maximal QFI exactly match binomial coefficients with a staggered minus sign:
\begin{equation}
\label{eq_optimalCoefficients}
a_m  = (-1)^{m} \left(\begin{array}{c} 2S \\ m \end{array}\right)= (-1)^{m} \frac{(2S)!}{m! (2S-m)!}.
\end{equation}
The corresponding maximal QFI is exactly $\mathcal{Q}_{i,i}=N(\frac{N}{2}+1)$. Note to avoid confusion that the states here are labeled $m = 0 \dots 2S$.

To show the above is true, we show that the QFI is at a global optimum for these coefficients. We write the QFI out explicitly for spin ensembles, including a normalization constant:
\begin{equation}
\mathcal{Q}_{i,i} = 2 \frac{\sum_{m} \left(a_m - a_{m-1}\right)^2 [S(S+1)-(m-S)(m-S-1)]}{\sum_{m}(a_m)^2}.
\end{equation}
Note that we have assumed the coefficients $a_m$ to be real-valued. It is straightforward to see that this must be true, as the magnitude of each contributing term in the sum is maximized for $a_m$, $a_{m-1}$ of opposite sign. If we choose the first coefficient to be real without loss of generality, the remaining ones must be as well.

To maximize, we now take the partial derivative of the QFI with respect to an arbitrary coefficient $a_m$ and set it to zero:
\begin{equation}    
\begin{aligned}
\frac{\partial \mathcal{Q}_{i,i}}{\partial a_m} = 0 = &-4 \frac{(a_{m+1}-a_{m})[S(S+1)-(m-S)(m-S+1)]}{\sum_{m'}(a_{m'})^2} + 4 \frac{(a_{m}-a_{m-1})[S(S+1)-(m-S)(m-S-1)]}{\sum_{m'}(a_{m'})^2}\\
&-4 a_m \frac{\sum_{m'}(a_{m'}-a_{m'-1})^2[S(S+1)-(m'-S)(m'-S-1)]}{\left(\sum_{m'}(a_{m'})^2\right)^2}.
\end{aligned}
\end{equation}
Simplifying, the requisite expression is:
\begin{equation}
\begin{aligned}
&a_m  \sum_{m'}(a_{m'}-a_{m'-1})^2[S(S+1)-(m'-S)(m'-S-1)]\\
=& -\sum_{m'}(a_{m'})^2 \bigg[(a_{m+1}-a_{m})[S(S+1)-(m-S)(m-S+1)] \\
&\quad\quad\quad\quad\quad+ (a_{m-1}-a_{m})[S(S+1)-(m-S)(m-S-1)]\bigg].
\end{aligned}
\end{equation}
We insert the coefficients from Eq.~\eqref{eq_optimalCoefficients} into the above expression, and employ the following helpful identities:
\begin{equation}
\begin{aligned}
&\sum_{m=0}^{2S} \left(\begin{array}{c} 2S \\ m \end{array}\right)^2 = \left(\begin{array}{c} 4S \\ 2S\end{array}\right),\\
&\sum_{m=1}^{2S} \left[\left(\begin{array}{c} 2S \\ m \end{array}\right) + \left(\begin{array}{c} 2S \\ m-1 \end{array}\right)\right]^2 [S(S+1)-(m-S)(m-S-1)] = 2S(2S+1)\left(\begin{array}{c} 4S \\ 2S\end{array}\right).\\
\end{aligned}
\end{equation}
Using these the expression simplifies to:
\begin{equation}
\begin{aligned}
\left(\begin{array}{c}2S \\ m \end{array}\right) 2S(2S+1) =  \bigg[&\left[\left(\begin{array}{c}2S \\ m+1 \end{array}\right)+\left(\begin{array}{c}2S \\ m \end{array}\right)\right][S(S+1)-(m-S)(m-S+1)] \\
+ &\left[\left(\begin{array}{c}2S \\ m-1 \end{array}\right)+\left(\begin{array}{c}2S \\ m \end{array}\right)\right][S(S+1)-(m-S)(m-S-1)]\bigg]
\end{aligned}
\end{equation}
It is straightforward to verify that the above expression holds true for any integer or half-integer $S$, and any $m \in \{0, \dots 2S\}$. Hence $\frac{\partial \mathcal{Q}_{i,i}}{\partial a_m} = 0$ for all $m$, and the staggered binomial coefficients are the optimal state coefficients for two-mode measurements with two equal spin ensembles. Note that, while this optimization is performed over a single operator $\hat{X}_{-}$, the symmetry of the ansatz state and the operator under exchange of the ensembles ensures that the complementary operator $\hat{Y}_{+}$ has equal maximized QFI for the same state.

We note that the maximal QFI $N(\frac{N}{2}+1)$ obtained by this optimization is the optimal bound for two-mode measurements, even if we allow non-symmetric states $\sum_{m,m'}a_{m,m'}\ket{m,m'}$. This can be seen by observing that the total angular momentum $\langle\hat{X}_{+}^2\rangle + \langle\hat{Y}_{+}^2\rangle + \langle\hat{Z}_{+}^2\rangle = \frac{N}{2}(\frac{N}{2}+1)$, where $\hat{Z}_{+} = \hat{Z}_1 + \hat{Z}_2$. If we assume the $x$ and $y$ components are equally sensitive, and bound $\langle\hat{Z}_{+}^2\rangle\geq 0$, we find that $\langle\hat{Y}_{+}^2\rangle \geq \frac{N}{4}(\frac{N}{2}+1)$. For pure states the QFI is four times the variance, and is hence bounded by $\geq N(\frac{N}{2}+1)$, matching the QFI obtained by optimization over the symmetric states. Note that while the sensitive quadratures used in that optimization are $\hat{Y}_{+}$ and $\hat{X}_{-}$, the sensitive quadratures can be changed via local rotations on one ensemble only, which should not affect the QFI bound.

\clearpage
\section{Comparison to GHZ states and single-ensemble approaches}

Here we compute the QFI of a GHZ-style state for two spin-$S$ ensembles. Before two-operator correlated measurements, we consider a GHZ state sensitive to just one operator $\hat{W}_{i}=\hat{X}_{-}$, which can be written as,
\begin{equation}
\ket{\mathrm{GHZ}_{1}} = \frac{1}{\sqrt{2}}e^{-i \frac{\pi}{2}\hat{Y}_{-}} \left(\ket{0,0} + \ket{m_{\mathrm{max}},m_{\mathrm{max}}}\right).
\end{equation}
This is a pure state, so we can again use the variance:
\begin{equation}
\begin{aligned}
\mathcal{Q}_{i,i} &= 4 \bra{\mathrm{GHZ}_1} \hat{X}_{-}^2\ket{\mathrm{GHZ}_1} - 4 \left(\bra{\mathrm{GHZ}_1} \hat{X}_{-}\ket{\mathrm{GHZ}_1}\right)^2\\
&=2\left(\bra{0,0} + \bra{m_{\mathrm{max}},m_{\mathrm{max}}}\right) e^{i \frac{\pi}{2}\hat{Y}_{-}}\hat{X}_{-}^2 e^{-i \frac{\pi}{2}\hat{Y}_{-}}\left(\ket{0,0} + \ket{m_{\mathrm{max}},m_{\mathrm{max}}}\right)\\
&=2\left(\bra{0,0} + \bra{m_{\mathrm{max}},m_{\mathrm{max}}}\right) \left(\hat{Z}_1 + \hat{Z}_2\right)^2\left(\ket{0,0} + \ket{m_{\mathrm{max}},m_{\mathrm{max}}}\right)\\
&=2 \left(\bra{0,0} + \bra{m_{\mathrm{max}},m_{\mathrm{max}}}\right)\left(\left[2\left(-\frac{m_{\mathrm{max}}}{2}\right)\right]^2\ket{0,0} + \left[2\left(m_{\mathrm{max}}-\frac{m_{\mathrm{max}}}{2}\right)\right]^2\ket{m_{\mathrm{max}},m_{\mathrm{max}}}\right)\\
&=4 m_{\mathrm{max}}^2\\
\end{aligned}
\end{equation}
Note that going from the second to the third line requires commutation relations which hold for spin operators (not in general). Furthermore, going from the third to the fourth line we used the matrix elements
\begin{equation}
\hat{Z}_i\ket{m}_i = (m-S)\ket{m}_i = \left(m-\frac{m_{\mathrm{max}}}{2}\right)\ket{m}_i,
\end{equation}
which are adjusted from the usual longitudinal spin projection operator since we count our states from $m=0$ rather than $m= - S$.
Since $m_{\mathrm{max}}=2S$ and the total atom number is $N = 4S$ we have,
\begin{equation}
\mathcal{Q}_{i,i} = 16S^2 = N^2.
\end{equation}
A similar calculation shows that the sensitivity of the other operators $\hat{W}_{i} = \{\hat{X}_{+}, \hat{Y}_{+}, \hat{Y}_{-}\}$ for this state is $\{0,N,N\}$ respectively.

Next we consider a two-mode analogue, which is maximally sensitive to two correlated operators. Inspired by the single-mode sensitive GHZ state, we write an analogous state with a different rotation,
\begin{equation}
\begin{aligned}
\rho_{\mathrm{GHZ},2} &=\ket{\mathrm{GHZ}_{2}} \bra{\mathrm{GHZ}_{2}},\\
\ket{\mathrm{GHZ}_{2}} &= \frac{1}{\sqrt{2}}e^{-i \frac{\pi}{2\sqrt{2}}(\hat{X}_{+}+\hat{Y}_{-})} \left(\ket{0,0} + \ket{m_{\mathrm{max}},m_{\mathrm{max}}}\right).
\end{aligned}
\end{equation}
Much of the preceding calculation remains the same, except there are now two (equally) sensitive operators $\hat{W}_{i} = \hat{X}_{-}, \hat{Y}_{+}$. The QFI for them reads,
\begin{equation}
\begin{aligned}
\mathcal{Q}_{i,i} &= 2(\bra{0,0} + \bra{m_{\mathrm{max}},m_{\mathrm{max}}})e^{i \frac{\pi}{2\sqrt{2}}(\hat{X}_{+} + \hat{Y}_{-})}\hat{X}_{-}^2 e^{-i \frac{\pi}{2\sqrt{2}}(\hat{X}_{+} + \hat{Y}_{-})}(\ket{0,0} - \ket{m_{\mathrm{max}},m_{\mathrm{max}}})\\
&=2 m_{\mathrm{max}}(2 m_{\mathrm{max}}+1) =\frac{N(N+1)}{2}.
\end{aligned}
\end{equation}
Notably, this is almost the optimal scaling found in the prior section, albeit smaller by $N/2$ [the prior section found $N(N+2)/2$].

We note that the Heisenberg-limited scaling we find with our GTMSS $\ket{\psi_G (r)}$ is much stronger than what can be accomplished with independent entangled states prepared separately in each ensemble (such as a pair of GHZ states $\ket{\mathrm{GHZ}_1} \otimes \ket{\mathrm{GHZ}_1}$). As argued in the main text, if there is common-mode or differential noise affecting both sub-ensembles, single-mode protocols will have no way of suppressing it if one seeks to measure two correlated parameters simultaneously. Furthermore, even in the very special case of noise that affects measurement outcomes for both sub-ensembles equally, the scaling we find is better than what can be accomplished with single-mode approaches. The best single-mode strategy using single-mode GHZ states would measure one parameter on each subensemble with a maximal QFI of $(N/2)^2 = N^2/4$, which is worse than the GTMSS result $\simeq N^2/3$. Instead considering independently squeezed states for each ensemble (for a more direct comparison, as GHZ states are more fragile) following Ref.~\cite{groszkowski2022reservoir}, the optimal QFI for a single-ensemble approach would instead scale as $\simeq N^2/8$. Finally, if one seeks to measure specific 2D combinations of fields that are not perfectly correlated across the ensembles a priori, one would need entangled states along each direction for each ensemble, requiring one to further split each subensemble in half and yielding a QFI scaling of $\simeq N^2/16$ for GHZ states and $\simeq N^2/32$ for squeezed states at best.

\clearpage

\section{Verification of the steady-state solution}
Here, we show that $\ket{\psi_{G}(r)}$ is the unique dark state of the dissipation. The action of one of the dissipators reads
\begin{equation}
\begin{aligned}
\left(\cosh(r)\hat{O}_{1} + \sinh(r) \hat{O}_{2}^{\dagger}\right)\ket{\psi_{G}(r)} =&\mathcal{N} \cosh(r) \sum_{m=1}^{m_{\mathrm{max}}}\left[-\tanh(r)\right]^{m}o(m)\ket{m-1,m} \\
+ &\mathcal{N}\sinh(r) \sum_{m=0}^{m_{\mathrm{max}}-1}\left[-\tanh(r)\right]^{m}o(m+1)\ket{m,m+1}\\
=&\mathcal{N} \cosh(r) \sum_{m=0}^{m_{\mathrm{max}}-1}\left[-\tanh(r)\right]^{m+1}o(m+1)\ket{m,m+1} \\
+ &\mathcal{N}\sinh(r) \sum_{m=0}^{m_{\mathrm{max}}-1}\left(-\tanh(r)\right)^{m}o(m+1)\ket{m,m+1}\\
=&\mathcal{N} \sum_{m=0}^{m_{\mathrm{max}}-1} \left[-\tanh(r)\right)^{-m} \big[-\cosh(r)\tanh(r) o(m+1) \\
&\quad\quad\quad\quad\quad\quad\quad\quad\quad\quad\quad\quad\quad\quad+ \sinh(r)o(m+1)\big]\ket{m,m+1}\\
=& 0.
\end{aligned}
\end{equation}
The other dissipator has exactly the same dark state structure, except exchanging $m$ and $m+1$ in the Dirac ket.

\clearpage
\section{General structure of the steady-state solution}
We can understand the steady state solution $\ket{\psi_G (r)}$ more generally by considering the construction presented in Ref.~\cite{pocklington2024solutionStructure}. There, it was observed that for any bipartite system with a jump operator of the form
\begin{align}
    \hat \Gamma &= \hat O_1 \otimes 1 + 1  \otimes \hat O_2, \label{eqn:sum_of_locals}
\end{align}
a pure steady state $|\psi \rangle$ must satisfy
\begin{align}
    \hat O_1 &= -\hat \Psi \hat K \hat O_2^\dagger \hat K^{-1}\hat \Psi^{-1}, \label{eqn:jump_consturction}
\end{align}
where $\hat \Psi = \sqrt{\mathrm{tr}_2 |\psi \rangle \langle \psi |}$ and $\hat K$ is complex conjugation in the Schmidt basis. Now, any state $|\psi \rangle = \sum_m \psi_m |m,m\rangle$ is already diagonal in the Schmidt basis, and assuming all of the matrix elements are real, we find that this implies:
\begin{align}
    (\hat O_1)_{mn} &= -(\hat O_2)_{nm} \frac{\psi_m}{\psi_n}.
\end{align}
Finally, observing that $(O_1)_{mn} = o(m) \delta_{m,n + 1}$ and $\psi_m = \tanh^m(r)$, we see that this implies
\begin{align}
    \hat O_1 &= -\tanh^{-1}(r) \hat O_2^\dagger,
\end{align}
which can be trivially satisfied by rescaling $\hat L = \cosh(r) \hat O_1 \otimes 1 + \sinh(r) 1 \otimes \hat O_1^\dagger$, and we can see that this will always be a steady state.

This constructive example, though, now allows us to create even more complicated steady states. For example, the ideal state for maximizing the QFI is given by a binomial distribution
\begin{align}
    |\psi \rangle &= \mathcal{N} \sum_{m = 0}^{2S} \frac{(-1)^m (2S)!}{(2S - m!)m!} |m,m \rangle. \label{eqn:binomial_dist_state}
\end{align}
\textit{A priori} it is not obvious what jump operators would be able to stabilize such a state; however, using Eq.~\ref{eqn:jump_consturction} we can directly calculate the matrix coefficients that would be required. For example, let's assume that there is a jump operator as in Eq.~\ref{eqn:sum_of_locals}, and take $\hat O_1 = \hat S_1^-$ is just a simple spin lowering operator. Then we can use Eq.~\ref{eqn:jump_consturction} to find that $\hat O_2$ must be a kind of generalized spin raising operator of the form 
\begin{align}
    \hat O_2 &= \sum_{m = 0}^{2S - 1} \tilde o(m) |m + 1 \rangle \langle m |, \\
    \tilde o(m) &= -\sqrt{S (S + 1) - (m-S)(m-S+1)} \frac{(2S - m)!m!}{(2S - m - 1)!(m + 1)!} \nonumber  \\
    &= \sqrt{S (S + 1) - (m-S)(m-S+1)} \frac{2S - m}{m + 1}.
\end{align}
Using this special form, we can define a jump operator (and its partner under exchanging subsystems)
\begin{align}
    \hat{\Gamma}_a &= \hat S^- \otimes 1 + 1 \otimes \hat O_2, \\
    \hat{\Gamma}_b &= 1 \otimes \hat S^- + \hat O_2 \otimes 1,
\end{align}
such that the quantum master equation $\partial_t \hat \rho = \mathcal{D}[\hat{\Gamma}_a] \hat \rho + \mathcal{D}[\hat{\Gamma}_b] \hat \rho$ uniquely stabilizes the ideal steady state $|\psi \rangle
$ given in Eq.~\ref{eqn:binomial_dist_state}.

\clearpage
\section{Unequal ensemble sizes}

\begin{figure}[h!]
\center
\includegraphics[width=.9\columnwidth]{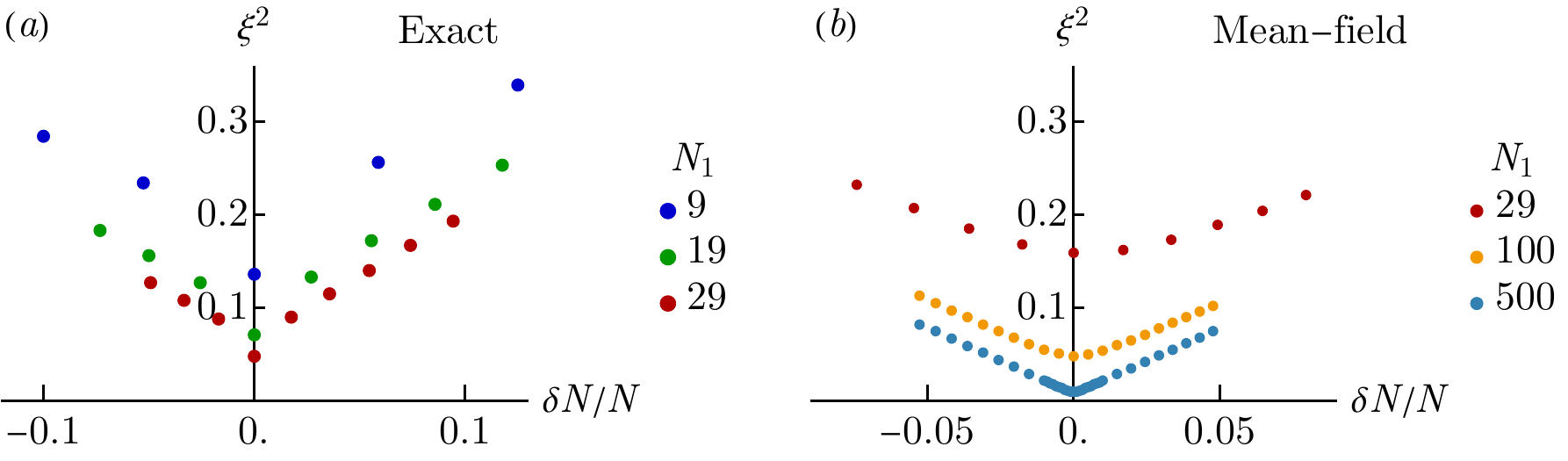}
\caption{(a) Steady-state two-mode squeezing for spin ensembles with unequal atom numbers $N_1$, $N_2$ (hence corresponding spin sizes $S_1 = N_1/2$, $S_2 = N_2/2$). We vary the difference in atom number $\delta N = N_2 - N_1$ normalized by the total atom number $N = N_1 + N_2$. For $N_1 = N_2$, we use the analytical result predicting maximum squeezing at $r \to \infty$. For $N_1 \neq N_2$ the steady-state is computed via exact numerical time-evolution of the master equation; in this case the squeezing is non-monotonic as a function of $r$, and the plotted result is optimized over $r$ (although we do not find rapid variations with changing $r$). (b) Same steady-state two-mode squeezing for larger system sizes, using a mean-field theory approach (see Section~\ref{sec_MFT} for details).}
\label{fig_MismatchedSize}
\end{figure}

The GTMSS $\ket{\psi_G(r)}$ is the steady-state of the engineered dissipation considered in the main text under the assumption of identical ensembles. However, properties such as squeezing and relevant observable expectation values remain robust for slight discrepancies between the subsystems. Here we benchmark the two-mode squeezing for spin ensembles with slightly different sizes. Fig.~\ref{fig_MismatchedSize}(a) shows the steady-state two-mode squeezing, found by computing the steady state numerically exactly, for subsystems consisting of spin ensembles composed of unequal numbers of atoms $N_1$ and $N_2$. We observe that the squeezing remains robust provided the discrepancy is small. Fig.~\ref{fig_MismatchedSize}(b) extends these calculations to larger system sizes with a mean-field theory computation (see Section~\ref{sec_MFT} for details), finding analogous behavior.

\clearpage
\section{Experimental implementation of the engineered dissipation}

\begin{figure}[h!]
\center
\includegraphics[width=.6\columnwidth]{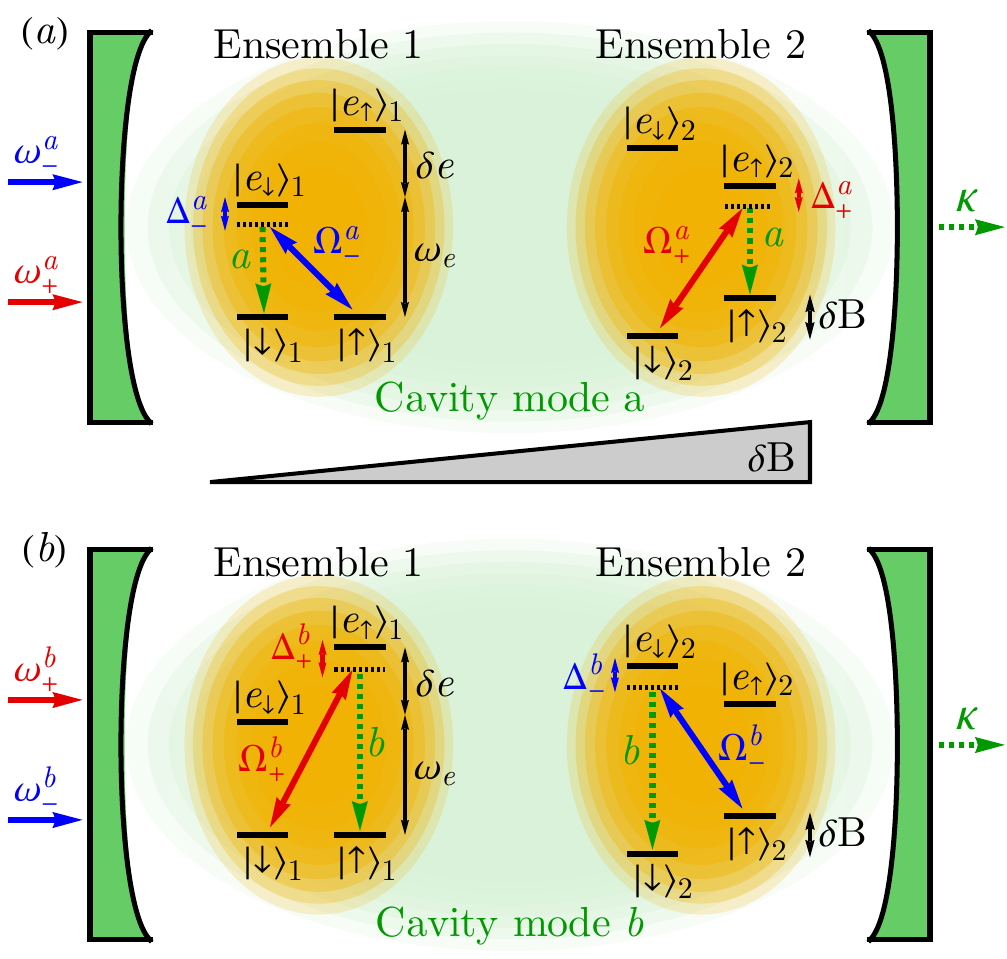}
\caption{Schematic of cavity-QED implementation of the engineered dissipators (a) $\hat{\Gamma}_{a}$ and (b) $\hat{\Gamma}_b$.}
\label{fig_Experimental}
\end{figure}

Here, we discuss a potential experimental implementation of the scheme in a cavity-QED architecture. The master equation from the main text that we seek to realize contains two engineered dissipators, which we denote $\hat{\Gamma}_{a}$, $\hat{\Gamma}_{b}$:
\begin{equation}
\label{eq_dissipators}
\begin{aligned}
\hat{\Gamma}_{a} &= \cosh(r) \hat{O}_1 + \sinh(r) \hat{O}_{2}^{\dagger},\\
\hat{\Gamma}_{b} &= \cosh(r) \hat{O}_2 + \sinh(r) \hat{O}_{1}^{\dagger}.
\end{aligned}
\end{equation}
We consider two spatially separated ensembles $i = 1,2$ of trapped atoms inside a lossy cavity. The scheme is depicted in Fig.~\ref{fig_Experimental}. Atoms in each ensemble have two ground states $\ket{\downarrow}_{i}$, $\ket{\uparrow}_{i}$ (encoding the spin degree of freedom), and two excited states $\ket{e_{\downarrow}}_{i}$, $\ket{e_{\uparrow}}_{i}$. For the first ensemble, the $\ket{e_{\downarrow}}_{1}$ and $\ket{e_{\uparrow}}_{1}$ states have an energy splitting $\delta e$. For the second manifold, the excited states $\ket{e_{\downarrow}}_{2}$ and $\ket{e_{\uparrow}}_{2}$ instead have a splitting $- \delta e$ (reversed from the first ensemble). While this difference in energies could be realized with electromagnetic fields, it is easier to just swap the labeling by taking the atomic states associated with $\downarrow$, $\uparrow$ for ensemble 1 to be $\uparrow$, $\downarrow$ for ensemble 2. In addition, the system is subject to a field gradient that introduces an additional energy shift $-\delta B/2$, $+\delta B/2$ for $\downarrow$, $\uparrow$ states of ensemble 2 relative to ensemble 1 respectively. The Hamiltonian for these levels reads,
\begin{equation}
\hat{H}_{0} = \omega_{e} \sum_{i=1,2}\left(\ket{e_{\downarrow}}_i \bra{e_{\downarrow}}_i + \ket{e_{\uparrow}}_i \bra{e_{\uparrow}}_i\right) + \delta e \left(\ket{e_{\uparrow}}_1 \bra{e_{\uparrow}}_1 - \ket{e_{\downarrow}}_2 \bra{e_{\downarrow}}_2\right) 
+ \frac{\delta B}{2}\left(\ket{e_{\uparrow}}_2 \bra{e_{\uparrow}}_2 - \ket{e_{\downarrow}}_2 \bra{e_{\downarrow}}_2 + \ket{\uparrow}_2 \bra{\uparrow}_2 - \ket{\downarrow}_2 \bra{\downarrow}_2\right).
\end{equation}
Each engineered dissipator can be realized with a pair of coherent laser fields driving the cavity. The dissipator $\hat{\Gamma}_{a}$ is shown in Fig.~\ref{fig_Experimental}(a). One laser with frequency, Rabi frequency and detuning ($\omega_{-}^{a}$, $\Omega_{-}^{a}$, $\Delta_{-}^{a}$) excites atoms from $\ket{\uparrow}_{1}$ to $\ket{e_{\downarrow}}_{1}$, while a second laser ($\omega_{+}^{a}$, $\Omega_{+}^{a}$, $\Delta_{+}^{a}$) excites from $\ket{\downarrow}_{2}$ to $\ket{e_{\uparrow}}_{2}$. The field gradient and splitting of the excited states ensures that only these transitions are resonant. The cavity is assumed to have a mode with frequency $\omega_{a} = \omega_e$ that matches the energy difference from $\ket{e_{\downarrow}}_{1}$ to $\ket{\downarrow}_{1}$ (equal to the difference from  $\ket{e_{\uparrow}}_{2}$ to $\ket{\uparrow}_{2}$ by construction) up to the detunings $\Delta_\pm^a$.

Any atom excited by the laser drives will exchange its excitation into a cavity photon at a rate $g$ set by the spin-cavity coupling strength. This photon will then leak out at a rate $\kappa$. Provided $\kappa$ is much larger than the Rabi frequencies and $g$, the excited states can be adiabatically eliminated. Since one cannot distinguish whether the photon was generated by the first laser (which would lead to an effective spin-lowering from $\ket{\uparrow}_{1}$ to $\ket{\downarrow}_{1}$) or the second laser (which would lead to an effective spin-raising from $\ket{\downarrow}_2$ to $\ket{\uparrow}_2$), the net effect of the loss manifests as correlated dissipation of the form $\hat{\Gamma}_{a}$:
\begin{equation}
\begin{aligned}
\hat{\Gamma}_a &\sim \sqrt{\frac{g^2 \tilde{\Omega}^2}{\kappa}} \left(\frac{1}{\tilde{\Omega}}\frac{\Omega_{-}^a}{\Delta_{-}^a}\hat{O}_{1}^{-} + \frac{1}{\tilde{\Omega}}\frac{\Omega_{+}^{a}}{\Delta_{+}^a} \hat{O}_{2}^{+}\right),\\
\tilde{\Omega} &= \sqrt{\left(\frac{\Omega_{-}^a}{\Delta_{-}^a}\right)^2 - \left(\frac{\Omega_{+}^a}{\Delta_{+}^a}\right)^2}.
\end{aligned}
\end{equation}
The rate prefactor is $\gamma \sim g^2 \tilde{\Omega}^2 / \kappa$. The squeezing parameter is set by $\cosh(r) = \frac{1}{\tilde{\Omega}}\frac{\Omega_{-}^a}{\Delta_{-}^a}$ and $\sinh(r) = \frac{1}{\tilde{\Omega}}\frac{\Omega_{+}^a}{\Delta_{+}^a}$, which yields $\tanh(r)= \frac{\Omega^{a}_{+}}{\Omega^{a}_{-}} \frac{\Delta_{-}^{a}}{\Delta_{+}^{a}}$. Strong squeezing is realized when the laser drives are almost equal in strength; note however that we cannot bring them exactly equal, as the dissipative stabilization timescales would become too long (see next Supplementary section for details).

The other dissipator is realized with the same level scheme, but a different pair of lasers $(\omega_{-}^{b}, \Omega_{-}^{b}, \Delta_{-}^{b})$ and $(\omega_{+}^{b}, \Omega_{+}^{b}, \Delta_{+}^{b})$ depicted in Fig.~\ref{fig_Experimental}(b). These again induce excitations that are resonant with a different cavity mode of frequency $\omega_{b} = \omega_e + \delta e$ (up to the detunings $\Delta_{\pm}^{b}$), yielding $\hat{\Gamma}_{b}$.

\clearpage
\section{Dissipative stabilization timescale}

\begin{figure}[h!]
\center
\includegraphics[width=0.65\columnwidth]{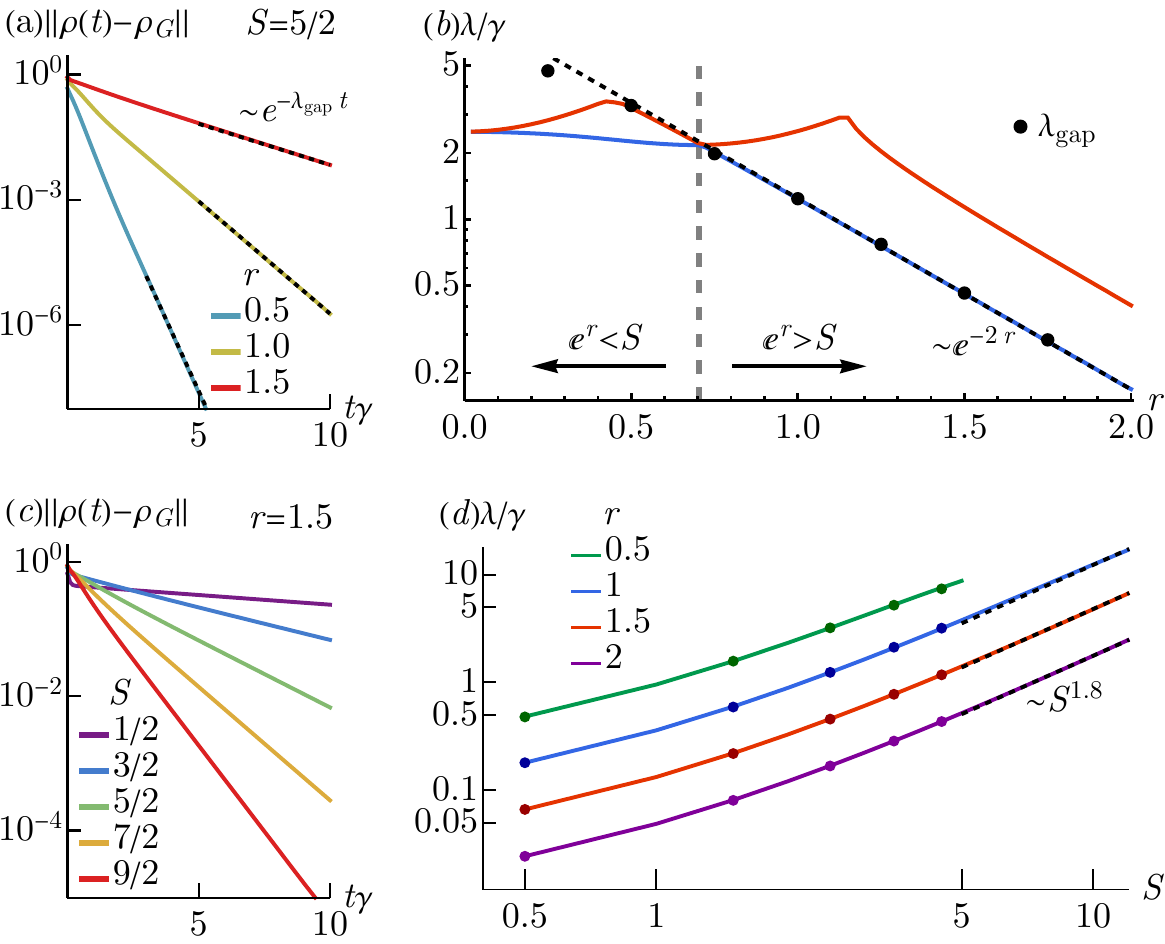}
\caption{(a) Time-evolution of infidelity with the steady-state, assuming the ensembles to be spins of fixed size $S=5/2$, for different values of $r$. The black dashed lines are a long-time exponential decay $e^{- \lambda_{\mathrm{gap}}t }$ with $\lambda_{\mathrm{gap}}$ the relevant dissipative gap. (b) Negative real part of the smallest and second-smallest non-zero Liouvillian eigenvalue (blue and red lines respectively). The black points are the numerically fitted rates from the infidelity dynamics in panel (a), which fall on one particular branch of the Liouvillian spectrum scaling as $\sim e^{-2r}$. (c) Time-evolution of infidelity for fixed $r = 1.5$ and varied $S$. (d) Negative real part of the smallest non-zero Liouvillian eigenvalue, shown as solid lines for different $r$. Solid points are corresponding fitted decay rates from the dynamics in panel (c). For large $r$, the relevant dissipative gap empirically scales as $\sim S^{1.8}$.}
\label{fig_Timescale}
\end{figure}

Here, we discuss the timescale needed to generate the entangled steady-state for two spin-$S$ ensembles. Figure~\ref{fig_Timescale}(a) plots the time evolution of the infidelity $||\rho(t) - \rho_{G}||$ under the engineered dissipation, starting from the initial state $\ket{\psi_0} = \ket{0,0}$. Aside from short-time transient dynamics, the infidelity scales exponentially as $\sim e^{- \lambda_{\mathrm{gap}} t}$, where the decay rate $\lambda_{\mathrm{gap}}$ is typically set by the Liouvillian dissipative gap. Formally, this gap is obtained by writing the dissipators as a Liouvillian superoperator,
\begin{equation}
\mathcal{L} = \gamma \sum_{\nu = a,b}\left[ \hat{\Gamma}_{\nu} \otimes \hat{\Gamma}_{\nu}^{*} - \frac{1}{2}\hat{\Gamma}_{\nu}^{\dagger}\hat{\Gamma}_{\nu} \otimes \mathbbm{1} - \frac{1}{2} \mathbbm{1}\otimes \left(\hat{\Gamma}_{\nu}^{\dagger}\hat{\Gamma}_{\nu}\right)^{*}\right],
\end{equation}
where $\hat{\Gamma}_{\nu}$ are the dissipators from Eq.~\eqref{eq_dissipators}.
We can diagonalize the Liouvillian and obtain the right eigenvectors,
\begin{equation}
\mathcal{L} \ket{\lambda_i} = \lambda_i \ket{\lambda_i}.
\end{equation}
There is one (unique) steady-state with eigenvalue $\lambda_0=0$ corresponding to our solution $\ket{\lambda_0} = \text{vec}(\ket{\psi_{G}(r)}\bra{\psi_{G}(r)})$, where $\text{vec}(\rho)$ indicates the vectorization (column-stacking) of a density matrix $\rho$.

Figure~\ref{fig_Timescale}(b) plots the next two smallest non-zero eigenvalues as a function of $r$; the corresponding decay rates $\lambda_{\mathrm{gap}}$ from the previous panel are also shown as dots. Normally, the dissipative gap $\lambda_{\mathrm{gap}}$ is the smallest eigenvalue; here this only holds true in the limit $e^{r} \gg S$. Outside that limit, the rate is determined by a higher eigenvalue of the Liouvillian (i.e. there is a specific relevant gap for our choice of initial state). Regardless, in our regime of interest $e^{r} \gtrsim S$ the relevant gap always scales as $\sim e^{-2r}$.

While this scaling is unfavorable with $r$, the gap increases for larger spin size $S$. Figure~\ref{fig_Timescale}(c) plots the infidelity for fixed $r$ and varying $S$, demonstrating an improvement in timescale for larger $S$. The decay rate and smallest two Liouvillian eigenvalues are plotted in Figure~\ref{fig_Timescale}(d). We find a scaling of $\sim S^{1.8}$, growing with system size, which can help compensate for the slowdown caused by increasing $r$. The overall effective dissipation rate is $\sim \gamma S^{1.8}e^{-2r}$. Note, however, that our dissipators' prefactors grow with $r$ due to the $\cosh(r)$, $\sinh(r)$ factors chosen for notational convenience. If we further adjust these dissipator prefactors to remain approximately constant with $r$, which would be the case for e.g. fixed laser power in an experiment, we must multiply the rate by another $e^{-2r}\sim S^{-1}$, yielding a rate $\sim \gamma S^{0.8}e^{-2r}$.

In general, we want to have an $r$ sufficiently large to generate close-to-optimal squeezing, and not any larger to avoid slowdown. This generally requires $e^{r} \gtrsim S$. For simplicity we can assume $e^{2r} \sim S$. This yields a rate $\sim \gamma S^{-0.2}$, which favorably remains almost constant with growing system size.

\clearpage
\section{Second-order mean-field theory analysis of local dissipation}
\label{sec_MFT}

Here, we discuss the mean-field-theory analysis of the two-mode spin-squeezing dynamics in the presence of local dissipation.
We start with the quantum master equation
\begin{align}
	\frac{\mathrm{d}}{\mathrm{d} t} \hat{\rho} = \gamma \left( \mathcal{D}[\hat{\Gamma}_a] + \mathcal{D}[\hat{\Gamma}_b] \right) \hat{\rho} + \sum_{i=1}^2 \sum_{j=1}^{N/2} \gammam \mathcal{D}[\hat{\sigma}_{i,j}^-] \hat{\rho} + \sum_{i=1}^2 \sum_{j=1}^{N/2} \gammaz \mathcal{D}[\hat{\sigma}_{i,j}^z] \hat{\rho} ~,
\end{align}
from which we derive equations of motion for the collective expectation values $S_i^\alpha = \cave{\hat{\alpha}_i}$ as well as their (co)variances $C_{ii'}^{\alpha\beta} = \cave{(\hat{\alpha}_i \hat{\beta}_{i'} + \hat{\beta}_{i'} \hat{\alpha}_i)}/2 - \cave{\hat{\alpha}_i} \cave{\hat{\beta}_{i'}}$, with $i,i'\in \{1,2\}$ and $\alpha,\beta \in \{X,Y,Z\}$. 
We consider an initial state where all spins are in their ground state, $S_1^z = S_2^z = - N/4$, $C_{ii}^{xx} = C_{ii}^{yy} = N/8$ and all other $S_i^\alpha$, $C_{ii'}^{\alpha\beta}$ are zero (note that $N$ denotes the \emph{total} number of spins, i.e., the sum of both ensembles).
For this initial state, the only nontrivial equations of motion are
\begin{align}
	\frac{\mathrm{d}}{\mathrm{d}t} S_i^z &= \gammam \left( S_i^z - \frac{N}{4} \right) - \gamma  \left[ C_{ii}^{xx} + C_{ii}^{yy} + \cosh(2r) S_i^z \right] ~, 
    \label{eq:SM:MFTSaz}\\
	\frac{\mathrm{d}}{\mathrm{d}t} C_{ii}^{xx} &= \left( \gammam + 4 \gammaz \right) \left( \frac{N}{8} - C_{ii}^{xx} \right) + \gamma \left[ \cosh(2 r) (C_{ii}^{zz} - C_{ii}^{xx} + (S_i^z)^2) - \frac{1}{2} S_i^z + 2 C_{ii}^{xx} S_i^z \right] ~, \\
	\frac{\mathrm{d}}{\mathrm{d}t} C_{ii}^{yy} &= \left( \gammam + 4 \gammaz \right) \left( \frac{N}{8} - C_{ii}^{yy} \right) + \gamma \left[ \cosh(2r) (C_{ii}^{zz} - C_{ii}^{yy} +  (S_i^z)^2) - \frac{1}{2} S_i^z + 2 C_{ii}^{yy} S_i^z \right] ~, \\
	\frac{\mathrm{d}}{\mathrm{d}t} C_{ii}^{zz} &= \gammam \left( \frac{N}{4} - 2 C_{ii}^{zz} + S_i^z \right) + \gamma \left[ \cosh(2r) (C_{ii}^{xx} + C_{ii}^{yy} - 2 C_{ii}^{zz}) + S_i^z \right] ~, \\
	\frac{\mathrm{d}}{\mathrm{d}t} C_{12}^{xx} &= - (\gammam + 4 \gammaz) C_{12}^{xx} + \gamma \left[ C_{12}^{xx} (S_1^z + S_2^z) - \sinh(2r) S_1^z S_2^z - \cosh(2r) C_{12}^{xx} - \sinh(2r) C_{12}^{zz} \right] ~, \\
	\frac{\mathrm{d}}{\mathrm{d}t} C_{12}^{yx} &= - (\gammam + 4 \gammaz) C_{12}^{yx} + \gamma \left[ C_{12}^{yx} + ( S_1^z + S_2^z ) - \cosh(2r) C_{12}^{yx} \right] ~, \\
	\frac{\mathrm{d}}{\mathrm{d}t} C_{12}^{yy} &= - (\gammam + 4 \gammaz) C_{12}^{yy} + \gamma \left[ C_{12}^{yy} (S_1^z + S_2^z) + \sinh(2r) S_1^z S_2^z - \cosh(2r) C_{12}^{yy} + \sinh(2r) C_{12}^{zz} \right] ~, \\
	\frac{\mathrm{d}}{\mathrm{d}t} C_{12}^{zz} &= -2 \gammam C_{12}^{zz} - \gamma \left[ 2 \cosh(2r) C_{12}^{zz} + \sinh(2r) (C_{12}^{xx} - C_{12}^{yy}) \right] ~.
    \label{eq:SM:MFTC12zz}
\end{align}
From these moments and (co)variances, the two-mode operators of interest can be obtained as follows.
\begin{align}
    \langle \hat{X}_{\pm}^2 \rangle &= C_{11}^{xx} + (S_1^x)^2 + C_{22}^{xx} + (S_2^x)^2 \pm 2 \left( C_{12}^{xx} + S_1^x S_2^x \right) ~, \\
    \langle \hat{Y}_{\pm}^2 \rangle &= C_{11}^{yy} + (S_1^y)^2 + C_{22}^{yy} + (S_2^y)^2 \pm 2 \left( C_{12}^{yy} + S_1^y S_2^y \right) ~.
\end{align}
To calculate steady-state observables, we numerically integrate Eqs.~\eqref{eq:SM:MFTSaz} to~\eqref{eq:SM:MFTC12zz} until the norm of the vector of first moments $S_i^\alpha$ and (co)variances $C_{ii'}^{\alpha\beta}$ changes less than $10^{-6}$ between two successive time steps. 
\begin{figure}
    \centering
    \includegraphics[width=\textwidth]{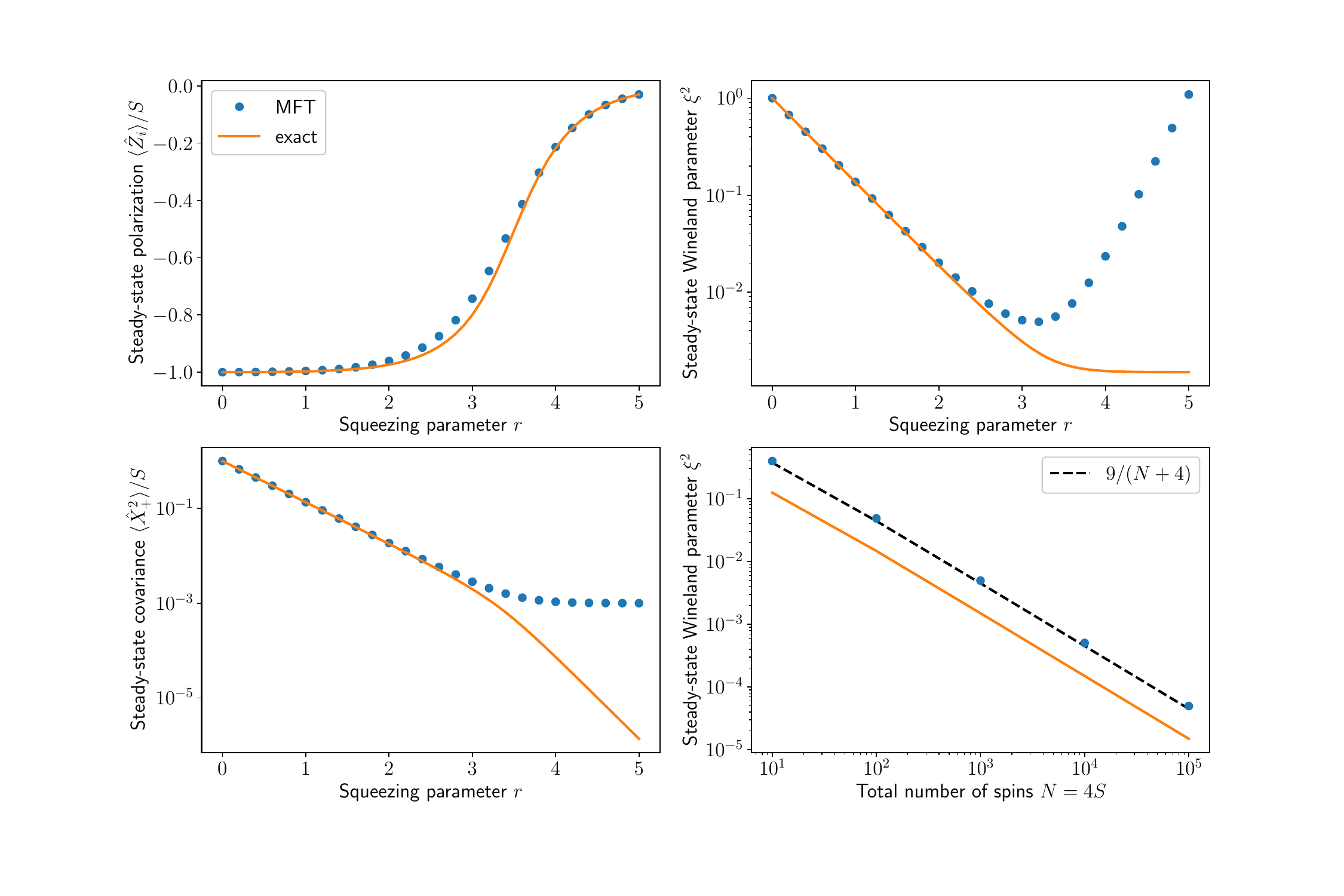}
    \caption{
        Comparison between mean-field theory simulations based on Eqs.~\eqref{eq:SM:MFTSaz} to~\eqref{eq:SM:MFTC12zz} and the exact results given by Eq.~\eqref{eq_ExactResultsSS}.
        Parameters are $N/2=1000$ and $\gammam=\gammaz=0$.
    }
    \label{fig:SM:Benchmark_MFT}
\end{figure}
Figure~\ref{fig:SM:Benchmark_MFT} compares these steady-state results with some exact predictions for the steady-state $\ket{\psi_G(r)}$ with no unwanted dissipation $\gamma_{-}$ or $\gamma_z$, such as,
\begin{equation}
\label{eq_ExactResultsSS}
\begin{aligned}
\langle \hat{Z}_{i=1,2}\rangle &= \frac{f_+ S + \sinh(r)^2[\tanh^{4S}(r)-1]}{f_-},\\
\langle\hat{X}_{+}^2\rangle &=\langle\hat{Y}_{-}^2\rangle = -\frac{e^{-2r}}{2}\left[(2S+1)\frac{f_{+}}{f_{-}}+\cosh(2r)\right],
\end{aligned}
\end{equation}
using the main-text definition $f_{\pm} =\tanh^{4S+2}(r)\pm 1$.

For small squeezing parameter $e^{2r} \ll N$, mean-field theory and the exact results agree very well. 
For large squeezing parameter $e^{2r} \gg N$, mean-field theory is not expected to be an accurate description of the highly non-Gaussian state of the spin system. 
Nevertheless, the decay of the signal $\langle \hat{O}_i^z \rangle$ is reproduced quite well, but the steady-state covariance saturates at a finite value instead of decreasing to zero with increasing squeezing parameter. 
As a consequence, the Wineland parameter obtained from mean-field theory has a minimum value at a finite value of $r$, which is a factor of $3$ larger than the exact steady-state value of the Wineland parameter in the limit $r \to \infty$.
The Heisenberg-like scaling of the minimum Wineland parameter as a function of $N$, however, is reproduced well by mean-field theory.

The results shown in Fig.~3(a) of the main text have been obtained by minimizing the Wineland parameter (obtained from mean-field theory for $\gammaz=0$ and $\gammam > 0$) over the squeezing strength $r$ for different values of the collective cooperativity $C_\mathrm{rel} = N \gamma/\gammam$. 
The dotted lines indicate the mean-field-theory steady-state Wineland parameter obtained for $\gammaz = \gammam = 0$, which is a factor of $3$ larger than the exact result due to the breakdown of mean-field theory for large values of the squeezing parameter $r$.

\begin{figure}
    \centering
    \includegraphics[width=0.48\textwidth]{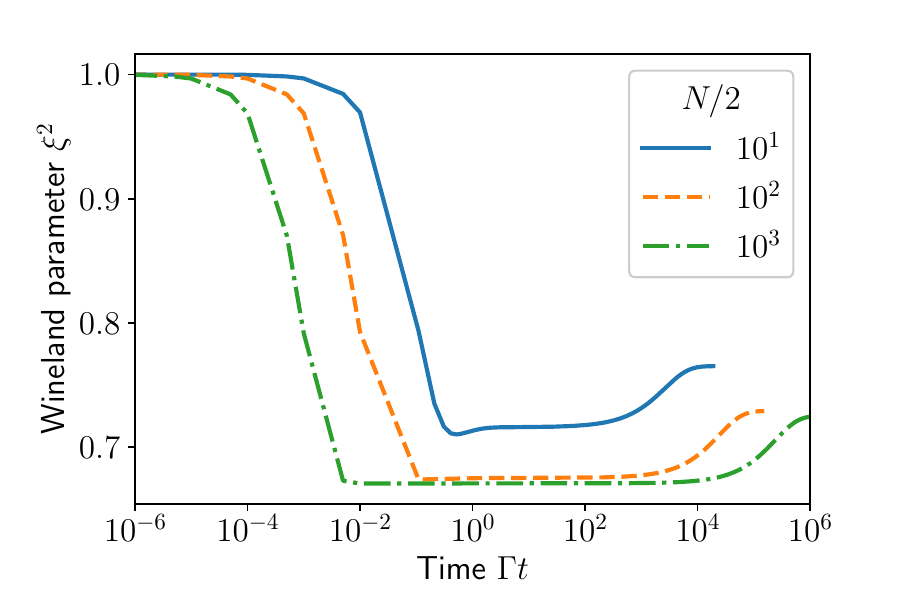}
    \caption{
        Mean-field simulation of the slow timescale emerging in the relaxation dynamics in the presence of weak collective dephasing.
        Parameters are $r=0.2$ and $\gammaz=0.001 \gamma$. 
        Each subsystem is initialized in a coherent spin state polarized along the $-z$ direction. 
    }
    \label{fig:SM:Slow_timescale}
\end{figure}

For local dephasing, $\gammam = 0$ but $\gammaz > 0$, it is known that a slow timescale emerges in the dissipative stabilization of a single-mode spin-squeezed state~\cite{groszkowski2022reservoir}.
A similar effect occurs in the case of dissipative two-mode spin squeezing, as shown in Fig.~\ref{fig:SM:Slow_timescale}.
Starting from a coherent state polarized along the $-z$ direction, the two subensembles relax into a highly spin-squeezed ``prethermal'' state on a timescale $\propto 1/N \gamma$. 
On a much longer timescale $\propto N / \gammaz$, dephasing leads to a reduction of spin squeezing such that the Wineland parameter increases towards its steady-state value. 
Since these timescales are separated by orders of magnitude in $1/\gamma$ for large spin number $N$, it is reasonable to assume that precision metrology will be performed with the highly spin-squeezed transient state. 
Therefore, the results shown in Fig.3(b) of the main text have been obtained by minimizing the Wineland parameter (obtained from mean-field theory for $\gammam=0$ and $\gammaz > 0$) over the evolution time $t$ and the squeezing strength $r$ for different values of the collective cooperativity $C_\phi = N \gamma/\gammaz$.

\clearpage
\section{Alternate non-Gaussian two-mode states}

While the main text focuses on spins and bosonic analogues, we can also consider systems with more exotic annihilation operators $\hat{O}_i$. One such example is parity-restricted bosonic modes, for which the annihilation operators remove pairs of excitations, e.g. $\hat{O}_{i} = \hat{a}_{i} \hat{a}_{i}$ for bosonic lowering operators $\hat{a}_i$~\cite{braunstein1987generalizedSqueezing}, which is of interest to circuit-QED platforms~\cite{leghtas2015yaleTwoPhoton, reglade2024twoPhoton}. For this choice $o(m) = \sqrt{2m(2m-1)}$ and $m_{\mathrm{max}} = \infty$. In this case the squeezed quadrature QFI takes the simple form of
\begin{equation}
\mathcal{Q}_{\mathrm{max}}= 4(1-e^{-2r}+e^{4r}) \sim e^{4r}.
\end{equation}
This scaling is in line with the average photon number of the state, which would also scale as $\sim e^{4r}$, but provides a far more non-trivial example of a system that can nonetheless be engineered for sensing. More concretely, one of the correlated system operators $\hat{X}_{-}$ reads,
\begin{equation}
\hat{X}_{-} = \frac{1}{2}\left[ \hat{a}_{1}^{\dagger}\hat{a}_{1}^{\dagger} + \hat{a}_{1}\hat{a}_{1}-\hat{a}_{2}^{\dagger}\hat{a}_{2}^{\dagger}-\hat{a}_{2}\hat{a}_{2}\right].
\end{equation}
The scheme we describe can thus be used for entanglement-enhanced sensing of parametric driving addressing two separate bosonic modes, e.g. one can test how well-correlated the amplitudes of the driving are for the different modes.

Another non-trivial example with a \textit{finite} Hilbert space is bosonic modes with a hard-core cutoff, enforced for instance by strong non-linear interactions. In this case we can consider the annihilation operators to be conventional bosonic operators $\hat{O}_i = \hat{a}_i$ with the usual matrix elements $o(m)=\sqrt{m}$, but assume finite $m_{\mathrm{max}}< \infty$. For this example, we find an optimal QFI of
\begin{equation}
\begin{aligned}
\mathcal{N}_{Q} e^{2r} &=\frac{2e^{2r}\left[1-\left(1+m_{\mathrm{max}}\text{sech}(2r)\right)\tanh^{2m_{\mathrm{max}}}(r)\right]}{1-\tanh^{2m_{\mathrm{max}}+2}(r)}, \\
\text{lim}_{r \to \infty} \mathcal{N}_{Q} e^{2r}&=\mathcal{Q}_{\mathrm{max}} = 4 m_{\mathrm{max}}.
\end{aligned}
\end{equation}
It is interesting to observe that this QFI scales less favorably with Hilbert space size than the case of spins, which instead features a quadratic scaling $\sim m_{\mathrm{max}}^2$, indicating that the structure of the annihilation operator matrix elements $o(m)$ plays a key role in the ultimate (maximally-squeezed) sensitivity.


\end{document}